\documentclass[aps,prd,longbibliography,reprint,twocolumn,amsmath,amssymb,amsfonts,showpacs,footnote,superscriptaddress]{revtex4-2}

\usepackage[T1]{fontenc}
\usepackage{lmodern}
\usepackage{amsmath,amssymb,amsfonts,mathrsfs}
\usepackage{bm}
\usepackage{mathtools}
\usepackage{physics}
\usepackage{hyperref}
\usepackage{xcolor}
\usepackage{booktabs}
\usepackage{siunitx}
\usepackage{url}

\usepackage{orcidlink}

\definecolor{navyblue}{rgb}{0.0, 0.0, 0.5}
\definecolor{ferrarired}{rgb}{1.0, 0.11, 0.0}
\definecolor{persianblue}{rgb}{0.11, 0.22, 0.73}
\usepackage{float}

\hypersetup{
	colorlinks=true,
	citecolor=blue,%
	linkcolor=ferrarired,%
	urlcolor=persianblue,%
	filecolor=blue,
	linktoc=page
}

\providecommand{\ii}{\mathrm{i}}

\newcommand{\etal}{\emph{et al.}~}

\begin{document}

	\title{Exceptional points of the Schwarzschild spectrum under local potential perturbation: The Princess and the Pea}
	
	\author{Mohamed \surname{Ould~El~Hadj}\,\orcidlink{0000-0002-8558-7992}}
	\email{med.ouldelhadj@gmail.com}
	\affiliation{No-affiliation}

	\author{Sam R. \surname{Dolan}\,\orcidlink{0000-0002-4672-6523}}
	\email{s.dolan@sheffield.ac.uk}
	\affiliation{Consortium for Fundamental Physics, School of Mathematical and Physical Sciences, University of Sheffield, Hicks Building, Hounsfield Road, Sheffield S3 7RH, United Kingdom}
	
	\date{\today}

\begin{abstract}
Black-hole quasinormal modes (QNMs) may coalesce at exceptional points (EPs) in the frequency spectrum. We study EPs and their phenomenology for gravitational perturbations of a Schwarzschild black hole whose Regge-Wheeler potential is perturbed by a localized delta-function of amplitude \(\epsilon\) at radius \(r_0\). For real \(\epsilon\), we find infinite families of discrete EPs. Each family approaches a distinct unperturbed QNM overtone frequency as $r_0$ grows large. Under perturbation at fixed $r_0$, we find that neighboring QNM overtones migrate in the complex plane to meet at these EPs, with the higher overtone moving further. We derive large-$r_0$ asymptotics, finding that EPs have (asymptotically) equal spacing in the tortoise coordinate, and that EP amplitudes $\epsilon$ can be exceedingly small, decaying exponentially according to a rule governed by the complex frequency of the unperturbed QNM.
Using quadratic approximations around EPs and high-precision numerical methods, we investigate a range of phenomena in the model: square-root splittings, avoided crossings and hyperbolae, hysteresis and geometric phases generated by encircling an EP, adiabatic invariants, deformed Cassini ovals and cusp-like trajectories, as well as the lemniscate geometry of the excitation factors. We then replace the delta-function perturbation by a finite-width Gaussian, verify the persistence of the EP structure, and show that varying the width extends isolated EPs into exceptional lines in parameter space. Finally, we study the response to initial data and find that, although a weak perturbation can substantially reconfigure the QNM spectrum, the corresponding time-domain response remains perturbative. This contrast is captured by a re-interpretation of the tale of The Princess and the Pea.
\end{abstract}

	\maketitle
	
	\tableofcontents
	
%=========================================================
\section{Introduction}
\label{sec:introduction}

With the birth of gravitational-wave astronomy in 2015 \cite{LIGOScientific:2016aoc} came confirmation of a long-standing prediction~\cite{Echeverria:1989hg, Flanagan:1997sx}: the final phase of the gravitational-wave signal from a black hole binary coalescence is characterized by a rapidly damped oscillation, as the composite black hole formed in the merger ``rings down'' to a stationary state. In the standard paradigm, the dominant frequency and decay rate of the late-time ringdown are set by the real and imaginary part of the fundamental ($n=0$) quadrupolar ($\ell = 2$) quasinormal mode of the resulting Kerr black hole \cite{Detweiler:1980gk, Leaver:1985ax}.

In black hole perturbation theory, quasinormal modes (QNMs) correspond to poles of the Green function in the complex-frequency plane \cite{Leaver:1986gd}. In principle, there exists an infinite spectrum of discrete QNMs, labeled by mode ($\ell$, $m$) and overtone ($n \ge 0$) numbers. Measuring the complex frequency $\omega_{\ell m n}$ of a \emph{single} QNM gives access to the final black hole mass and spin \cite{Echeverria:1989hg}. The goal of \emph{Black Hole Spectroscopy} \cite{Dreyer:2003bv, Berti:2005ys, Berti:2025hly}, simply stated, is to measure more than one complex QNM frequency, in order to test the no-hair theorem and general relativity itself. 
The burgeoning interest in (and literature on) QNMs, ringdown and spectroscopy can be tracked through the number of review articles written over the years \cite{Kokkotas:1999bd, Nollert:1999ji, Ferrari:2007dd, Berti:2009kk, Konoplya:2011qq, Berti:2018vdi, Cardoso:2019rvt, Berti:2025hly}. 

The black hole spectroscopy program holds promise, but there are obstacles to overcome. A practical challenge is that only the last part of the signal is accurately described by the QNM expansion of perturbation theory, and the signal decays rapidly below the noise floor of the detector. Even for `loud' signals this means that making unambiguous multi-modal detections is difficult; for example, an analysis by Capano \etal\cite{Capano:2021etf} of GW190521 found evidence for an $\ell = 3$ mode (in addition to $\ell = 2$), but this appears to depend on the modeling assumptions made \cite{LIGOScientific:2020ufj, Siegel:2023lxl}, such as the start time for QNM fitting. There are also theoretical issues, not least of which is the fact that a small perturbation in the black hole potential can create dramatic changes in the QNM spectrum, particularly affecting higher overtones $n > 0$ \cite{Nollert:1996rf, Cheung:2021bol, Jaramillo:2020tuu, Jaramillo:2021tmt, Destounis:2021lum}. Broadly, the sensitivity of QNM spectra to perturbations is the core theme of this work.

One aspect of QNM spectral stability has only recently been appreciated. It is well-known that black holes represent open dissipative systems, since perturbations radiate away to infinity or fall through the event horizon, and hence the associated boundary-value problem for the wave equation is \emph{non-Hermitian}. A typical feature in the spectrum of a non-Hermitian system (with some tuning of parameters) is an \emph{Exceptional Point} (EP): a point at which two or more eigenvalues, and their corresponding eigenvectors, coalesce as the parameters are changed smoothly, leading to a branch-point singularity~\cite{Ding:2022juv}. Near a second-order EP, pairs of eigenvalues show avoided crossings (i.e.~hyperbolic trajectories in the complex-frequency domain). Moreover, around closed loops in parameter space, one eigenvalue of the pair is smoothly transformed into the other, and vice versa. The Exceptional Points of QNM spectra have only recently been studied, through pioneering work by Motohashi \cite{Motohashi:2024fwt} and, in parallel,  Cavalcante, Richartz and da Cunha \cite{Cavalcante:2024swt,Cavalcante:2024kmy}. This has led to a flurry of follow-up work, including Refs.~\cite{Yang:2025dbn,Oshita:2025ibu,PanossoMacedo:2025xnf,Destounis:2025dck,Takahashi:2025uwo,Wu:2025wbp,Cao:2025afs,Cavalcante:2025abr,Nakamoto:2026lyo,Cheng:2026gxu,Kubota:2025hjk,Imafuku:2026rpn,Cavalcante:2026vgr,Kubota:2026hdv}. There is a pleasing symmetry here: the study of coalescing black holes has led to the study of coalescing QNMs.

For an \emph{exact} exceptional point to exist in parameter space, generically one must have a model with at least two dimensionless parameters. Cavalcante \etal \cite{Cavalcante:2024swt} showed that, for a scalar field of mass $\mu$ on a rapidly-rotating Kerr spacetime with spin parameter $a$ and mass $M$, there is an exceptional point at $M \mu \approx 0.3704981$ and $a/M \approx 0.9994660$ for $\ell =2$. The EP is associated with a bifurcation of the spectrum into ``damped'' and ``zero-damping'' modes, in the classification of Refs.~\cite{Yang:2012pj,Yang:2013uba}. Motohashi examined the \emph{one-parameter} model of massless Kerr QNMs, identifying an avoided-crossing in the trajectories of the fifth and sixth overtones as $a/M$ is varied, and the lemniscate of Bernoulli in their excitation factors. These are indicators that the system passes ``near'' to an EP in a larger parameter space. This observation also shed light on an anomaly noted by Onozawa in 1996 \cite{Onozawa:1996ux}: the fifth overtone changes direction at $a \approx 0.9M$ and moves toward an isolated point, instead of heading towards the accumulation point at $2M\omega_{\ell m n} = m$ like the rest of the mode frequencies. 

Essential features of EPs can be understood through reasonably general arguments, as we sketch below. Suppose that, after separation of variables, the system is described by a second-order ODE $u'' + U(\omega, x; p_i) u = 0$ with a set of parameters $p_i$. Assume that this system has a pair of homogeneous solutions $u^\pm(\omega, x; p_i)$ that satisfy physical boundary conditions on the left ($-$) and right ($+$) of an open domain in $x$. The Wronskian of these solutions, $W[u^-, u^+]$, can be thought of as a function of frequency and the parameters, so that $W = W(\omega; p_i)$. The QNM frequencies are frequencies $\omega_n(p_i)$ (with $n\ge0$)  such that the Wronskian is zero, i.e., $W(\omega_n(p_i); p_i) = 0$. An exceptional point is defined by imposing an additional condition: that the partial derivative of $W$ with respect to $\omega$, evaluated at $\omega_n$, is also zero. Since $W$ is complex, the EP condition puts two constraints on a real parameter space, in general. Hence, in a two-parameter model one can (generically) find Exceptional Points, and in three-parameter models one has Exceptional Lines, and so on \cite{Ding:2022juv,Yang:2025dbn,Nakamoto:2026lyo,
Cao:2025afs,Cavalcante:2026vgr}. (This is the general situation, but of course there could be special cases of higher-order EPs or degeneracies in parameter space). 
In the region near the EP at $(\hat{\omega}, \hat{p}_i)$, the leading variation of the Wronskian is quadratic in the frequency displacement, while parameter detunings enter at the same order, and so we may write $\omega = \hat{\omega} + \eta \, \delta \omega$ and $p_i = \hat{p}_i + \eta^2 \, \delta p_i$, where $\eta$ is an order-counting parameter. Expanding the Wronskian around the EP leads to a \emph{quadratic approximation}:
\begin{equation}
W(\omega; p_i) = \eta^2 \left( \frac{1}{2} (\delta \omega)^2 \partial_{\omega \omega} W +
\delta \mathbf{p} \cdot \nabla_{\mathbf{p}} W \right) + O(\eta^{3}) 
\end{equation}
where the derivatives are evaluated at the EP position ($\hat{\omega}, \hat{p}_i$). Hence at leading order, the QNMs (the roots of the Wronskian) near the EP are determined by
\begin{equation}
\delta \omega \approx \sqrt{ \frac{- 2 \, \delta \mathbf{p} \cdot \nabla_{\mathbf{p}} W}{\partial_{\omega \omega} W } } \, .
\end{equation}
This square-root structure lies at the heart of the phenomena we describe below, such as avoided-crossings and hysteresis. 

In this work, we consider the EPs of black holes that are slightly perturbed by environmental effects or, more excitingly, by modifications to general relativity. Barausse \etal \cite{Barausse:2014tra} made a systematic study of a range of  perturbations to black hole potentials. Here we consider specifically `Elephant and Flea' models \cite{Cheung:2021bol}, in which the `elephant' is the unperturbed black hole (Regge-Wheeler) potential $V_0(r_\ast)$, and the `flea' is a localized perturbation, $\delta V(r_\ast)$. In the words of Simon \cite{simon1985semiclassical}, \emph{``The flea does not change the shape of the elephant but it can irritate the elephant enough so that it shifts its weight.''}

Cheung \etal \cite{Cheung:2021bol} considered fleas with three parameters (magnitude $\epsilon$, position $r_0$ (or $a$) and width $\sigma$) and of two species (i.e.~Gaussian or P\"oschl-Teller profiles). While such fleas do not directly correspond with potential perturbations in known physically-motivated scenarios, they are effective at probing the general features of the spectrum under perturbation. Recent work has shown that elephant-and-flea models exhibit phenomena associated with exceptional points/lines, including avoided-crossings \cite{PanossoMacedo:2025xnf,Cao:2025afs}. In the current work, building on Ref.~\cite{Torres:2026uey}, we focus on a simple \emph{two}-parameter model: a delta-function perturbation $(\epsilon, r_0)$ that is essentially equivalent to a Gaussian bump of vanishing width ($\sigma \rightarrow 0$).

An advantage of the simple two-parameter model $(\epsilon, r_0)$ is that the task of locating perturbed QNMs reduces to that of finding roots of $F(\omega; r_0) - \epsilon$, where $F(\omega; r_0)$ is a certain function (that we call the \emph{normalized Wronskian}) that depends solely on the properties of the \emph{unperturbed} system (the elephant). In Ref.~\cite{Torres:2026uey}, to find QNM trajectories for fixed $r_0$ and increasing $\epsilon$, this was cast as a simple dynamical system: $d\omega/d\epsilon = 1/F_\omega(\omega; r_0)$ (where $F_\omega = \partial_\omega F$). Here, we instead take the view that perturbed QNMs are the (real) level sets of the complex function $F(\omega; r_0)$. The simplicity of the two-parameter model enables a thorough investigation of EPs and their spectral properties (Sec.~\ref{sec:results}), as well as the study of the response of the perturbed black hole in the time domain (Sec.~\ref{sec:results-time-domain}).  

Two different aspects of spectral instability of this two-parameter model were noted in Ref.~\cite{Torres:2026uey} (dubbed \emph{linear} and \emph{nonlinear} instability) and one aim of the present work is to shed fresh light on these. In essence, linear instability is the exponential sensitivity of the overtones of the QNM spectrum to small perturbations in the potential; this is due to the fact that $F \sim \exp(-2 i \omega r_{0*})$, for sufficiently large $r_0 \gg M$. Nonlinear instability is the observed breakdown of Taylor-series expansions of QNM trajectories at very small values of $\epsilon$, and rapid changes of direction of the QNM trajectories. This aspect is closely related to the presence of EPs (and their cousins, repelling points \cite{Torres:2026uey}) near unperturbed QNMs.

\section{Formulation}
\label{sec:formulation}

\subsection{Perturbations of the Schwarzschild black hole}
\label{subsec:SchwarzschildPerturbations}

In this subsection, we briefly recall the main ingredients of the Schwarzschild perturbation problem and establish the notation and conventions used throughout the article.

We consider the exterior of a Schwarzschild black hole of mass \(M\), described by the line element
\begin{equation}
	\dd s^2 = -f(r)\,\dd t^2 + f(r)^{-1}\,\dd r^2 + r^2\dd\sigma_2^2.
	\label{eq:SchwarzschildMetric}
\end{equation}
Here \(f(r)=1-2M/r\), while \(\dd\sigma_2^2=\dd\theta^2+\sin^2\theta\,\dd\varphi^2\) is the line element on the unit two-sphere \(S^2\). The Schwarzschild coordinates satisfy \(t\in (-\infty,+\infty)\), \(r\in (2M,+\infty) \), \(\theta\in[0,\pi]\), and \(\varphi\in[0,2\pi]\). We adopt units such that \(G=c=1\) and assume a harmonic time dependence of the form \(e^{-\ii \omega t}\) for the perturbative field.

The tortoise coordinate \(r_*\in (-\infty,+\infty) \) is defined in terms of the radial Schwarzschild coordinate \(r\) by
\begin{equation}
	\frac{\dd r}{\dd r_*} = f(r).
	\label{eq:TortoiseDerivative}
\end{equation}
It is explicitly given by
\begin{equation}
	r_*(r) = r + 2M\ln\left(\frac{r}{2M}-1\right) + r_*^{(0)},
	\label{eq:TortoiseCoordinate}
\end{equation}
where \(r_*^{(0)}\) is an arbitrary additive constant. Unless otherwise specified, we take \(r_*^{(0)}=0\). The function \(r_*=r_*(r)\) provides a bijection from \( (2M,+\infty) \) to \( (-\infty,+\infty) \). Thus, the event horizon and spatial infinity correspond, respectively, to \(r_*\to-\infty\) and \(r_*\to+\infty\).

After separation of variables, the radial mode function \( \phi_{\ell\omega}(r) \) associated with a field of spin \(s\) satisfies the Regge-Wheeler equation
\begin{equation}
	\left[ \frac{\dd^2}{\dd r_*^2} + \omega^2 - V_{\ell s}(r) \right] \phi_{\ell\omega}(r) = 0,
	\label{eq:RW}
\end{equation}
where the effective potential is
\begin{equation}
	V_{\ell s}(r) = f(r) \left[ \frac{\ell(\ell+1)}{r^2} + (1-s^2)\frac{2M}{r^3} \right].
	\label{eq:RWPotential}
\end{equation}
In the following, we retain the spin \(s\) in the formal expressions. The numerical analysis, however, is restricted to gravitational Regge-Wheeler perturbations with \(s=2\), and more particularly to the quadrupolar sector \(\ell=2\).

The two standard homogeneous solutions of Eq.~\eqref{eq:RW} are denoted by \(\phi_{\ell\omega}^{\rm in}\) and \(\phi_{\ell\omega}^{\rm up}\). The mode \(\phi_{\ell\omega}^{\rm in}\) is purely ingoing at the event horizon, while the mode \(\phi_{\ell\omega}^{\rm up}\) is purely outgoing at spatial infinity. Their asymptotic behaviors are defined by
\begin{equation}
	\phi_{\ell\omega}^{\rm in}(r_*) \sim
	\begin{cases}
		e^{-\ii\omega r_*},
		&
		r_*\to-\infty,
		\\[2mm]
		A_{\ell\omega}^{(-)}e^{-\ii\omega r_*} + A_{\ell\omega}^{(+)}e^{+\ii\omega r_*},
		&
		r_*\to+\infty,
	\end{cases}
	\label{eq:InModeAsymptotics}
\end{equation}
and
\begin{equation}
	\phi_{\ell\omega}^{\rm up}(r_*)
	\sim
	\begin{cases}
		B_{\ell\omega}^{(-)}e^{-\ii\omega r_*} + B_{\ell\omega}^{(+)}e^{+\ii\omega r_*},
		&
		r_*\to-\infty,
		\\[2mm]
		e^{+\ii\omega r_*},
		&
		r_*\to+\infty.
	\end{cases}
	\label{eq:UpModeAsymptotics}
\end{equation}
Here, \(A_{\ell\omega}^{(\pm)}\) and \(B_{\ell\omega}^{(\pm)}\) are complex scattering amplitudes. The mode functions and the associated amplitudes, initially defined in a domain where the boundary conditions are well posed, are extended into the complex-frequency plane by analytic continuation.

The Wronskian of the two homogeneous solutions is defined, with respect to the tortoise coordinate, by
\begin{equation}
	W_\ell(\omega)
	\equiv
	W\!\left[ \phi_{\ell\omega}^{\rm in}, \phi_{\ell\omega}^{\rm up} \right]
	= \phi^{\rm in}_{\ell\omega} \left[ \overrightarrow{\frac{\dd }{\dd r_*}} -  \overleftarrow{\frac{\dd}{\dd r_*} }\right]	\phi^{\rm up}_{\ell\omega}.
	\label{eq:WronskianDefinition}
\end{equation}
Because the two functions satisfy the same second-order homogeneous differential equation, \(W_\ell(\omega)\) is independent of \(r_*\). By evaluating it at the event horizon or at spatial infinity, we obtain
\begin{equation}
	W_\ell(\omega) = 2\ii\omega A_{\ell\omega}^{(-)} = 2\ii\omega B_{\ell\omega}^{(+)}.
	\label{eq:WronskianAmplitudes}
\end{equation}

We recall that the quasinormal modes of the Schwarzschild black hole are solutions which are purely ingoing at the event horizon and purely outgoing at spatial infinity. They occur when \(\phi_{\ell\omega}^{\rm in}\) and \(\phi_{\ell\omega}^{\rm up}\) become linearly dependent, i.e., when
\begin{equation}
	W_\ell(\omega)=0.
	\label{eq:SchwarzschildWronskianCondition}
\end{equation}
Equivalently, their complex frequencies are the zeros of the incoming amplitude at spatial infinity,
\begin{equation}
	A_{\ell\omega}^{(-)}=0.
	\label{eq:SchwarzschildQNMCondition}
\end{equation}
The zeros lying in the lower half of the complex-frequency plane correspond to damped Schwarzschild QNMs. We denote their frequencies by
\begin{equation}
	\omega_Q^{(n)}, \qquad n=0,1,2,\ldots,
	\label{eq:SchwarzschildQNMNotation}
\end{equation}
where \(n\) is the overtone index, with \(n=0\) corresponding to the fundamental mode.

%===============================================================================================================

\subsection{Localized perturbations of the potential}
\label{subsec:LocalizedPerturbations}

We now introduce localized perturbations of the Regge-Wheeler potential (the `pea'). We first consider a general perturbation of compact support and briefly recall the first-order framework introduced in Ref.~\cite{Torres:2026uey}. It should be noted that our sign convention for the perturbation strength $\epsilon$ is opposite to that adopted in their analysis. We then specialize to a delta-function perturbation, for which the perturbed resonance condition can be obtained exactly and expressed in terms of a normalized Wronskian.

Let the effective potential be modified according to
\begin{equation}
	V_{\ell s}(r_*) \longrightarrow V_{\ell s}(r_*) + \epsilon\,\delta V(r_*),
	\label{eq:GeneralPotentialPerturbation}
\end{equation}
where \(\epsilon\) measures the perturbation strength and \(\delta V(r_*)\) has compact support in an interval \([r_*^{\rm L},r_*^{\rm R}]\). The corresponding radial equation is
\begin{equation}
	\left[ \frac{\dd^2}{\dd r_*^2} + \omega^2 - V_{\ell s}(r) - \epsilon\,\delta V(r_*) \right] \phi_{\ell\omega}^{(\epsilon)}(r_*) = 0.
	\label{eq:GeneralPerturbedRWEquation}
\end{equation}

Let \(\phi_{\ell\omega}^{\rm in,(\epsilon)}\) denote the solution satisfying the ingoing boundary condition at the event horizon and similarly, let \(\phi_{\ell\omega}^{\rm up,(\epsilon)}\) denote the solution satisfying the outgoing boundary condition at spatial infinity. Their unperturbed limits are \(\phi_{\ell\omega}^{\rm in}\) and \(\phi_{\ell\omega}^{\rm up}\), respectively.

For \(r_*>r_*^{\rm R}\), the first-order construction gives, up to an overall normalization,
\begin{equation}
	\begin{split}
		\phi_{\ell\omega}^{\rm in,(\epsilon)}(r_*)
		&={}
		\phi_{\ell\omega}^{\rm in}(r_*)
		\\
		&+
		\frac{\epsilon}{W_\ell(\omega)}
		\left[
		\mathcal V_{\ell\omega}^{(+)} \phi_{\ell\omega}^{\rm up}(r_*) - \mathcal V_{\ell\omega}^{(-)} \phi_{\ell\omega}^{\rm in}(r_*)
		\right] + O(\epsilon^2).
	\end{split}
	\label{eq:PerturbedInSolutionGeneral}
\end{equation}
With the outgoing normalization fixed by Eq.~\eqref{eq:UpModeAsymptotics}, the second solution is simply

$
	\phi_{\ell\omega}^{\rm up,(\epsilon)}(r_*) = \phi_{\ell\omega}^{\rm up}(r_*)
$

for \(r_*>r_*^{\rm R}\).

The coefficient multiplying the incoming contribution is
\begin{equation}
	\mathcal V_{\ell\omega}^{(-)} =
	\int_{r_*^{\rm L}}^{r_*^{\rm R}} \delta V(r_*') \phi_{\ell\omega}^{\rm in}(r_*') \phi_{\ell\omega}^{\rm up}(r_*') \,\dd r_*' .
	\label{eq:VMinusDefinition}
\end{equation}
It follows that the Wronskian of the perturbed solutions is
\begin{equation}
	W_\ell^{(\epsilon)}(\omega) = W_\ell(\omega) - \epsilon\, \mathcal V_{\ell\omega}^{(-)} + O(\epsilon^2).
	\label{eq:GeneralPerturbedWronskian}
\end{equation}

The resonance condition \(W_\ell^{(\epsilon)}(\omega)=0\) can therefore be written, to first order in \(\epsilon\), as
\begin{equation}
	F_{\delta V}(\omega)-\epsilon = O(\epsilon^2),
	\label{eq:GeneralNormalizedSpectralCondition}
\end{equation}
where
\begin{equation}
	F_{\delta V}(\omega) =
	\frac{W_\ell(\omega)}{\mathcal V_{\ell\omega}^{(-)}}.
	\label{eq:GeneralNormalizedWronskian}
\end{equation}
Thus, at the perturbative level, the deformation of the resonance spectrum is determined entirely by the unperturbed radial solutions and their overlap with the localized perturbation through Eq.~\eqref{eq:VMinusDefinition}.

We now specialize to the delta-function model used throughout most of this article. The perturbation is situated at \(r_*=r_{0*}\), where
\(r_{0*}=r_*(r_0)\),  and is defined by
\begin{equation}
	\delta V(r_*) = \delta(r_*-r_{0*}).
	\label{eq:DeltaPotentialDefinition}
\end{equation}
The Regge-Wheeler potential is therefore modified according to
\begin{equation}
	V_{\ell s}(r) \longrightarrow V_{\ell s}(r) + \epsilon\, \delta(r_*-r_{0*}).
	\label{eq:DeltaPotentialPerturbation}
\end{equation}

The perturbed radial function is continuous at \(r_*=r_{0*}\), while its first derivative satisfies the jump condition
\begin{equation}
	\left.
	\frac{\dd\phi_{\ell\omega}^{(\epsilon)}}{\dd r_*} \right|_{r_{0*}^{+}} -
	\left. \frac{\dd\phi_{\ell\omega}^{(\epsilon)}}{\dd r_*} \right|_{r_{0*}^{-}} = \epsilon\, \phi_{\ell\omega}^{(\epsilon)}(r_{0*}).
	\label{eq:DeltaJumpCondition}
\end{equation}

A perturbed resonance is ingoing at the horizon and outgoing at spatial infinity. Its radial function may therefore be written as
\begin{equation}
	\phi_{\ell\omega}^{(\epsilon)}(r_*) =
	\begin{cases}
		\mathcal N_{\rm in}\,
		\phi_{\ell\omega}^{\rm in}(r_*),
		&
		r_*<r_{0*},
		\\[2mm]
		\mathcal N_{\rm up}\,
		\phi_{\ell\omega}^{\rm up}(r_*),
		&
		r_*>r_{0*}.
	\end{cases}
	\label{eq:PiecewisePerturbedResonance}
\end{equation}
Using the continuity and jump conditions at \(r_{0*}\), we obtain the exact resonance condition
\begin{equation}
	\frac{ W_\ell(\omega)}{\phi_{\ell\omega}^{\rm in}(r_{0*})\phi_{\ell\omega}^{\rm up}(r_{0*})} - \epsilon = 0.
	\label{eq:ExactDeltaSpectralCondition}
\end{equation}
In contrast with the general compact perturbation, this expression is exact: no higher-order corrections in \(\epsilon\) are present.

It is therefore natural to introduce the normalized Wronskian
\begin{equation}
	F(\omega;r_0) =
	\frac{W_\ell(\omega)}{\phi_{\ell\omega}^{\rm in}(r_{0*})\phi_{\ell\omega}^{\rm up}(r_{0*})}.
	\label{eq:NormalizedWronskian}
\end{equation}
Using Eq.~\eqref{eq:WronskianAmplitudes}, it may equivalently be written as
\begin{equation}
	F(\omega;r_0) =
	\frac{2\ii\omega A_{\ell\omega}^{(-)}}{\phi_{\ell\omega}^{\rm in}(r_{0*}) \phi_{\ell\omega}^{\rm up}(r_{0*})}.
	\label{eq:NormalizedWronskianAmplitude}
\end{equation}
The perturbed spectral equation then takes the compact form
\begin{equation}
	D(\omega,\epsilon;r_0) \equiv F(\omega;r_0)-\epsilon = 0.
	\label{eq:PerturbedSpectralEquation}
\end{equation}

The function \(F(\omega;r_0)\) is invariant under independent multiplicative rescalings of \(\phi_{\ell\omega}^{\rm in}\) and \(\phi_{\ell\omega}^{\rm up}\). It therefore depends neither on the particular normalization adopted for the homogeneous solutions nor on the radius at which their Wronskian is evaluated.

In the unperturbed limit, \(\epsilon=0\), the spectral condition reduces to
$
	F(\omega;r_0)=0.
$
Provided that the product of radial mode functions does not vanish at \(r_{0*}\), this condition is equivalent to \(W_\ell(\omega)=0\) and hence recovers the usual Schwarzschild QNM spectrum.

The perturbed frequencies may be regarded as trajectories in the complex-frequency plane as the strength \(\epsilon\) is varied. Indeed, differentiating Eq.~\eqref{eq:PerturbedSpectralEquation} at fixed
\(r_0\) gives
\begin{equation}
	\frac{\dd\omega}{\dd\epsilon} = \frac{1}{ F_\omega(\omega;r_0)}.
	\label{eq:SpectralFlowEquation}
\end{equation}
This first-order equation provides the dynamical-system interpretation explored in Ref.~\cite{Torres:2026uey}: the perturbed QNM frequencies migrate along integral curves of the complex vector field \(1/F_\omega\). The singular points of this flow, defined by \(F_\omega=0\), play a central role in the following subsection.

Near an exceptional point, we denote the two local roots of Eq.~\eqref{eq:PerturbedSpectralEquation} by $\omega_+(\epsilon;r_0)$ and $\omega_-(\epsilon;r_0)$. Their identification with particular Schwarzschild QNM frequencies \(\omega_Q^{(n)}\) can only be established by analytically continuing the corresponding branches back to the unperturbed limit
\(\epsilon\to0\).

%===============================================================================================================

\subsection{Repelling points and exceptional points}
\label{subsec:RepellingExceptionalPoints}

We now introduce the repelling points of the spectral flow and establish their relation to exceptional points. Throughout this subsection, the location \(r_0\) of the perturbation is initially regarded as fixed. The perturbed resonance frequencies are the roots of Eq.~\eqref{eq:PerturbedSpectralEquation}. Their migration under variation of the perturbation strength is governed by Eq.~\eqref{eq:SpectralFlowEquation}.
A repelling point is a stationary point of the normalized Wronskian with respect to the complex frequency, defined by
\begin{equation}
	F_\omega(\omega_r;r_0)=0.
	\label{eq:RepellingPointCondition}
\end{equation}
For a simple zero of \(F_\omega\), the spectral-flow vector field \(1/F_\omega\) has a simple pole at \(\omega_r\). Resonance trajectories passing nearby are therefore strongly deflected, motivating the interpretation of \(\omega_r\) as a repelling point of the spectral flow.

For a generic repelling point, the value \(F(\omega_r;r_0)\) is complex. If the perturbation strength is extended to the complex plane, the choice
\begin{equation}
	\epsilon_r = F(\omega_r;r_0)
	\label{eq:ComplexCriticalCoupling_bis}
\end{equation}
places the repelling point on the perturbed spectrum. Indeed, one then has
$
	D(\omega_r,\epsilon_r;r_0) = 0,
$
while the stationary-point condition implies
\begin{equation}
	D_\omega(\omega_r,\epsilon_r;r_0) = 0.
	\label{eq:ComplexStationaryCondition}
\end{equation}
Provided that \(F_{\omega\omega}(\omega_r;r_0)\neq0\), the point \((\omega_r,\epsilon_r)\) is therefore a second-order branch point of the complexified spectral problem.

In the terminology of Ref.~\cite{Torres:2026uey}, a \emph{junction point} is a repelling point that can be reached within the real two-parameter model \((\epsilon,r_0)\). In addition to
the stationary-point condition, this requires
\begin{equation}
	\operatorname{Im} F(\omega_J;r_J) = 0.
	\label{eq:JunctionRealityCondition}
\end{equation}
The associated perturbation strength is then real and is given by
\begin{equation}
	\epsilon_J = \operatorname{Re} F(\omega_J;r_J) .
	\label{eq:JunctionCriticalCoupling}
\end{equation}
At such a point, the real spectral trajectories may reach the branch point and reconnect, rather than merely being deflected in its neighborhood.

We now show that a junction point is a second-order exceptional point of the exact delta-function model. For this purpose, we introduce the exact perturbed Wronskian
\begin{equation}
	\begin{split}
		\mathcal W_\ell(\omega,\epsilon;r_0) ={}& W_\ell(\omega) - \epsilon\, \phi_{\ell\omega}^{\rm in}(r_{0*}) \phi_{\ell\omega}^{\rm up}(r_{0*}) \\
		={}& \mathcal P_\ell(\omega;r_0) D(\omega,\epsilon;r_0),
	\end{split}
	\label{eq:PerturbedWronskianFactorisation}
\end{equation}
where \(\mathcal P_\ell(\omega;r_0) = \phi_{\ell\omega}^{\rm in}(r_{0*}) \phi_{\ell\omega}^{\rm up}(r_{0*})\). The factorization follows directly from the definitions of \(F(\omega;r_0)\) and \(D(\omega,\epsilon;r_0)\).

At a junction point \((\omega_J,\epsilon_J,r_J)\), the spectral equation gives
\begin{equation}
	D(\omega_J,\epsilon_J;r_J) = 0.
	\label{eq:JunctionRootCondition}
\end{equation}
The repelling-point condition gives, independently,
\begin{equation}
	D_\omega(\omega_J,\epsilon_J;r_J) = 0.
	\label{eq:JunctionStationaryCondition}
\end{equation}
The factorization of the perturbed Wronskian therefore implies
\begin{equation}
	\mathcal W_\ell(\omega_J,\epsilon_J;r_J) = 0.
	\label{eq:JunctionWronskianZero}
\end{equation}

Differentiating the factorized expression with respect to the frequency gives
\begin{equation}
	\frac{\partial\mathcal W_\ell}{\partial\omega} = \mathcal P_{\ell,\omega}D + \mathcal P_\ell D_\omega.
	\label{eq:PerturbedWronskianDerivative}
\end{equation}
Both terms vanish at the junction point. Consequently,
\begin{equation}
	\left. \frac{\partial\mathcal W_\ell}{\partial\omega} \right|_{(\omega_J,\epsilon_J,r_J)} = 0.
	\label{eq:JunctionWronskianDerivativeZero}
\end{equation}
The junction point is therefore a multiple zero of the exact perturbed Wronskian.

We assume that the stationary point is non-degenerate, so that
\begin{equation}
	F_{\omega\omega}(\omega_J;r_J) \neq 0.
	\label{eq:SecondOrderNonDegeneracy}
\end{equation}
We also assume that the product of homogeneous mode functions does not vanish at the perturbation location,
\begin{equation}
	\mathcal P_\ell(\omega_J;r_J) \neq 0.
	\label{eq:ModeProductNonzero}
\end{equation}
Differentiating the factorized Wronskian a second time and evaluating it at the junction point then gives
\begin{equation}
	\left. \frac{\partial^2\mathcal W_\ell}{\partial\omega^2} \right|_{(\omega_J,\epsilon_J,r_J)} = \mathcal P_\ell(\omega_J;r_J) F_{\omega\omega}(\omega_J;r_J).
	\label{eq:JunctionWronskianSecondDerivative}
\end{equation}
The right-hand side is nonzero. The exact perturbed Wronskian therefore possesses a zero of algebraic multiplicity two.

At the same frequency, the two boundary-condition solutions of the perturbed problem become linearly dependent and define a single resonance mode, up to an overall normalization. The algebraic multiplicity is therefore two, whereas the geometric multiplicity is one. The junction point is consequently a second-order exceptional point.

In the remainder of the article, we denote the junction frequency by \(\omega_{\rm EP}\), its radial location by \(r_{\rm EP}\), and its critical coupling by \(\epsilon_{\rm EP}\). Under the non-degeneracy assumptions stated above, a real exceptional point is characterized by
\begin{equation}
	F_\omega(\omega_{\rm EP};r_{\rm EP}) = 0,
	\label{eq:RealEPStationaryCondition}
\end{equation}
together with
\begin{equation}
	\operatorname{Im} F(\omega_{\rm EP};r_{\rm EP}) = 0.
	\label{eq:RealEPRealityCondition}
\end{equation}
Its critical coupling is
\begin{equation}
	\epsilon_{\rm EP} = \operatorname{Re} F(\omega_{\rm EP};r_{\rm EP}) ,
	\label{eq:RealEPCriticalCoupling}
\end{equation}
and its second-order character requires
\begin{equation}
	F_{\omega\omega}(\omega_{\rm EP};r_{\rm EP}) \neq 0.
	\label{eq:RealEPNonDegeneracyCondition}
\end{equation}

%===============================================================================================================

\section{Methods}
\label{sec:methods}

\subsection{The quadratic expansion}
\label{subsec:QuadraticExpansion}

We now describe the local expansion of the spectral equation in the vicinity of a repelling point and, more particularly, of an exceptional point. This expansion provides a common framework for the square-root splittings, avoided crossings, hyperbolic trajectories and spectral monodromy discussed below.

The perturbed resonance frequencies are determined by Eq.~\eqref{eq:PerturbedSpectralEquation}. Let \(\omega_r\) be a repelling point associated with a fixed perturbation location \(r_0\), and hence satisfying Eq.~\eqref{eq:RepellingPointCondition}.
In the neighborhood of this point, the normalized Wronskian admits the expansion
\begin{equation}
	\begin{split}
		F(\omega;r_0+\delta r_0)
		={}&
		F(\omega_r;r_0) + \frac{1}{2} F_{\omega\omega}(\omega_r;r_0) (\omega-\omega_r)^2
		\\
		& + F_{r_0}(\omega_r;r_0)\delta r_0 \\
		&+ O\!\left( (\omega-\omega_r)^3, (\omega-\omega_r)\delta r_0, \delta r_0^2
		\right).
	\end{split}
	\label{eq:GeneralQuadraticExpansion}
\end{equation}
The term linear in \(\omega-\omega_r\) is absent as a consequence of Eq.~\eqref{eq:RepellingPointCondition}.

For a generic repelling point, we write \(F(\omega_r;r_0)=\epsilon_c(r_0)+\ii\eta(r_0)\), where \(\epsilon_c(r_0)=\operatorname{Re}F(\omega_r;r_0)\) and
\(\eta(r_0)=\operatorname{Im}F(\omega_r;r_0)\). We then vary the real perturbation strength according to
\begin{equation}
	\epsilon = \epsilon_c(r_0) + \Delta\epsilon.
	\label{eq:LocalEpsilonVariation}
\end{equation}
At fixed \(r_0\), the spectral equation becomes, to quadratic order,
\begin{equation}
	0 \simeq \frac{1}{2} F_{\omega\omega}^{r} (\omega-\omega_r)^2 - \Delta\epsilon + \ii\eta,
	\label{eq:QuadraticEquationRepellingPoint}
\end{equation}
where \(F_{\omega\omega}^{r}=F_{\omega\omega}(\omega_r;r_0)\). The two local roots are therefore
\begin{equation}
	\omega_\pm \simeq \omega_r \pm
	\left[ \frac{ 2(\Delta\epsilon-\ii\eta)}{F_{\omega\omega}^{r} } \right]^{1/2}.
	\label{eq:LocalRootsRepellingPoint}
\end{equation}

This expression describes the generic avoided crossing generated when the image \(F(\omega_r;r_0)\) of the repelling point does not lie on the real axis. Equivalently, the associated branch point lies away from the real-\(\epsilon\) slice. The local geometry becomes more transparent after introducing the complex coordinate
\begin{equation}
	z = \left( \frac{F_{\omega\omega}^{r}}{2} \right)^{1/2} (\omega-\omega_r).
	\label{eq:LocalCoordinateZ}
\end{equation}
Equation~\eqref{eq:QuadraticEquationRepellingPoint} then reduces to
\begin{equation}
	z^2 = \Delta\epsilon-\ii\eta.
	\label{eq:LocalNormalFormRepellingPoint}
\end{equation}
Writing \(z=X+\ii Y\) and equating the real parts gives
\begin{equation}
	X^2-Y^2 = \Delta\epsilon.
	\label{eq:LocalHyperbolaRealPart}
\end{equation}
Equating the imaginary parts gives
\begin{equation}
	2XY = -\eta.
	\label{eq:LocalHyperbolaImaginaryPart}
\end{equation}
Thus, as the real parameter \(\Delta\epsilon\) is varied, the resonance roots locally follow hyperbolic trajectories. The magnitude \(|\eta|\) controls the opening of the avoided crossing. The argument of \(F_{\omega\omega}^{r}\) determines the orientation of the hyperbola in the complex-frequency plane.

We now specialize to a real exceptional point \((\omega_{\rm EP},\epsilon_{\rm EP},r_{\rm EP})\). At this point, \(F_\omega(\omega_{\rm EP};r_{\rm EP})=0\) and \(F(\omega_{\rm EP};r_{\rm EP})=\epsilon_{\rm EP}\), with \(\epsilon_{\rm EP}\in\mathbb{R}\). We introduce the local variables \(\delta\epsilon=\epsilon-\epsilon_{\rm EP}\) and \(\delta r_0=r_0-r_{\rm EP}\). The spectral equation then takes the local form
\begin{equation}
	0 \simeq \frac{1}{2} F_{\omega\omega}^{\rm EP} (\omega-\omega_{\rm EP})^2 + F_{r_0}^{\rm EP}\delta r_0 - \delta\epsilon,
	\label{eq:QuadraticNormalFormEP}
\end{equation}
where \(F_{\omega\omega}^{\rm EP} = F_{\omega\omega}(\omega_{\rm EP};r_{\rm EP})\) and \(F_{r_0}^{\rm EP} = F_{r_0}(\omega_{\rm EP};r_{\rm EP})\).

It is useful to introduce the effective complex discriminant
\begin{equation}
	\zeta = \delta\epsilon - F_{r_0}^{\rm EP}\delta r_0.
	\label{eq:EffectiveDiscriminant}
\end{equation}
The two local branches are then
\begin{equation}
	\omega_\pm \simeq \omega_{\rm EP} \pm
	\left[ \frac{2\zeta} {F_{\omega\omega}^{\rm EP}} \right]^{1/2}.
	\label{eq:GeneralSquareRootUnfolding}
\end{equation}
This is the universal square-root normal form of a second-order exceptional point.

At fixed perturbation location, one has \(\delta r_0=0\). The local branches therefore become
\begin{equation}
	\omega_\pm \simeq \omega_{\rm EP} \pm \left[ \frac{2\delta\epsilon} {F_{\omega\omega}^{\rm EP}} \right]^{1/2}.
	\label{eq:EpsilonSquareRootUnfolding}
\end{equation}
Their separation is consequently
\begin{equation}
	\left| \omega_+-\omega_- \right| \simeq 2 \left| \frac{2} {F_{\omega\omega}^{\rm EP}} \right|^{1/2} |\delta\epsilon|^{1/2}.
	\label{eq:EpsilonSplittingPrediction}
\end{equation}

Conversely, at fixed critical coupling one has \(\delta\epsilon=0\). The two branches then satisfy
\begin{equation}
	\omega_\pm \simeq \omega_{\rm EP} \pm
	\left[ -\frac{2F_{r_0}^{\rm EP}}{F_{\omega\omega}^{\rm EP}} \delta r_0 \right]^{1/2}.
	\label{eq:RadialSquareRootUnfolding}
\end{equation}
It follows that
\begin{equation}
	\left| \omega_+-\omega_- \right| \propto |\delta r_0|^{1/2}.
	\label{eq:RadialSplittingPrediction}
\end{equation}
An equivalent radial expansion may be written in terms of the tortoise coordinate displacement \(\delta r_{0*}=r_*(r_0)-r_*(r_{\rm EP})\). In that parametrization,
\(F_{r_0}^{\rm EP}\) is replaced by \(F_{r_{0*}}^{\rm EP}\).

At the exceptional point, \(\eta=0\). Equation \eqref{eq:LocalHyperbolaImaginaryPart} then becomes
\begin{equation}
	XY = 0.
	\label{eq:DegenerateHyperbola}
\end{equation}
As \(\Delta\epsilon\) varies through zero, the hyperbolic trajectories therefore degenerate into the two axes \(X=0\) and \(Y=0\). These axes are orthogonal and give the two limiting crossing directions of the resonance branches. Changing the sign of the real unfolding parameter rotates the square-root splitting by \(90^\circ\) in the local complex coordinate.

The same normal form reveals the two-sheeted topology associated with the exceptional point. Since the local branches depend on \(\sqrt{\zeta}\), analytic continuation around a closed contour for which \(\zeta\) winds once around the origin changes the sign of the square root:
\begin{equation}
	\sqrt{\zeta} \longrightarrow -\sqrt{\zeta}.
	\label{eq:SquareRootSignChange}
\end{equation}
Consequently, \(\omega_+\) is continued into \(\omega_-\), while \(\omega_-\) is continued into \(\omega_+\). A second circuit is required for each branch to return to its initial sheet.

It should be emphasized that the quadratic approximation is intrinsically local. Higher-order derivatives of \(F\) deform the hyperbolic trajectories away from the repelling point and determine the global continuation of the two branches. In particular, the local symmetry
\begin{equation}
	\omega_--\omega_{\rm EP} \simeq - \left( \omega_+-\omega_{\rm EP}\right)
	\label{eq:LocalBranchSymmetry}
\end{equation}
does not determine which unperturbed Schwarzschild overtones are reached when the branches are continued to \(\epsilon=0\). This local analysis does not fix the Schwarzschild endpoints of the two
sheets; their global connectivity will be examined separately in~Sec.~\ref{subsec:SpectralConnectivitySensitivity}.

%======================================================================================================

\subsection{Asymptotics of exceptional points}
\label{subsec:EPAsymptotics}

We now derive the large-distance behavior of the exceptional points generated by a localized perturbation situated far from the peak of the Regge-Wheeler potential. More precisely, we determine the asymptotic location of the repelling points, the spacing of the real exceptional points and the decay of the corresponding critical coupling.

For this purpose, it is convenient to choose the additive constant in the tortoise coordinate such that the light ring \(r=3M\) is located at the origin. We therefore write
\begin{equation}
	r_*(r) = r + 2M\ln\left(\frac{r}{2M}-1\right) - 3M + 2M\ln 2.
	\label{eq:ShiftedTortoiseAsymptotics}
\end{equation}
This choice only fixes the origin of \(r_*\). It modifies the overall phase of the asymptotic expressions derived below, but it does not affect the spacing of the distant exceptional points or the exponential decay rate of their critical couplings.

We label each EP family by the overtone number \(n\) of the unperturbed Schwarzschild QNM on which its repelling-point branch accumulates as \(r_0\to+\infty\). Thus, \(n=0\) labels the family accumulating on \(\omega_Q^{(0)}\), whereas \(n=1\) labels the family accumulating on \(\omega_Q^{(1)}\).

It is often convenient to use a dimensionless version of the tortoise coordinate, \(\mathcal R=r_{0*}/M\), together with dimensionless frequencies \(\widehat{\omega}=2M\omega\) and \(\widehat{\omega}_Q^{(n)}=2M\omega_Q^{(n)}\). The use of \(\mathcal R\) emphasizes that the large-distance expansion is naturally organized by the tortoise coordinate rather than by the Schwarzschild radial coordinate \(r_0\).

We start from the normalized Wronskian defined in Eq.~\eqref{eq:NormalizedWronskian}.
For \(r_{0*}\to+\infty\), the two homogeneous solutions behave as
\begin{equation}
	\phi_{\ell\omega}^{\rm up}(r_{0*}) \simeq e^{+\ii\omega r_{0*}},
	\label{eq:UpLargeDistance}
\end{equation}
and
\begin{equation}
	\phi_{\ell\omega}^{\rm in}(r_{0*}) \simeq A_{\ell\omega}^{(-)} e^{-\ii\omega r_{0*}} + A_{\ell\omega}^{(+)} e^{+\ii\omega r_{0*}}.
	\label{eq:InLargeDistance}
\end{equation}
Consider a repelling-point branch that approaches the Schwarzschild QNM frequency \(\omega_Q^{(n)}\), assuming that this QNM is simple, so that \(A_{\ell\omega}^{(-)}\) has a simple zero at this frequency. The incoming amplitude therefore admits the local expansion
\begin{equation}
	A_{\ell\omega}^{(-)} \simeq \left. \frac{\dd A_{\ell\omega}^{(-)}}{\dd\omega} \right|_{\omega=\omega_Q^{(n)}} \left( \omega-\omega_Q^{(n)} \right).
	\label{eq:IncomingAmplitudeNearQNM}
\end{equation}

Near the QNM and at large \(r_{0*}\), the denominator of Eq.~\eqref{eq:NormalizedWronskian} is dominated by its outgoing contribution. Using Eq.~\eqref{eq:WronskianAmplitudes}, we then obtain
\begin{equation} 
F(\omega;r_0) \simeq \Xi_n\left( \omega-\omega_Q^{(n)} \right) e^{-2\ii\omega r_{0*}},
	\label{eq:FNearQNMAsymptotic}
\end{equation}
where \(\Xi_n\) is independent of \(r_{0*}\), up to subleading large-distance corrections.

The repelling points are determined by Eq.~\eqref{eq:RepellingPointCondition}. Differentiating Eq.~\eqref{eq:FNearQNMAsymptotic} gives
\begin{equation}
	F_\omega(\omega;r_0) \simeq \Xi_n e^{-2\ii\omega r_{0*}}
	\left[1-2\ii r_{0*} \left(\omega-\omega_Q^{(n)}\right) \right].
	\label{eq:FDerivativeAsymptotic}
\end{equation}
It follows that the repelling points move closer to the unperturbed QNM frequencies, according to
\begin{equation}
	\omega_r-\omega_Q^{(n)} \simeq -\frac{\ii}{2r_{0*}}.
	\label{eq:RepellingPointAccumulation}
\end{equation}
Equivalently, in terms of the dimensionless variables,
\begin{equation}
	\left( \widehat{\omega}_r - \widehat{\omega}_Q^{(n)}\right) \mathcal R \longrightarrow -\ii \qquad \text{as} \qquad \mathcal R\to+\infty.
	\label{eq:DimensionlessRepellingAccumulation}
\end{equation}
Thus, each repelling-point branch accumulates on the corresponding Schwarzschild QNM with a universal correction of order \(1/r_{0*}\).

Evaluating Eq.~\eqref{eq:FNearQNMAsymptotic} at the repelling point gives
\begin{equation}
	F(\omega_r;r_0) \simeq -\frac{\ii\Xi_n}{2e\,r_{0*}} e^{-2\ii\omega_Q^{(n)}r_{0*}}.
	\label{eq:FAtRepellingPointAsymptotic}
\end{equation}
Here \(\Xi_n\) is generally complex. Its modulus, together with the constant dimensional factors arising when the result is written in terms of \(\mathcal R\), is absorbed into the positive constant \(\chi_n\) introduced below.

Writing \(\widehat{\omega}_Q^{(n)} =\widehat{\omega}_{R}^{(n)} +\ii\widehat{\omega}_{I}^{(n)}\), with \(\widehat{\omega}_{I}^{(n)}<0\), we find
\begin{equation}
	\left| F(\omega_r;r_0) \right| \simeq \frac{\chi_n}{\mathcal R} \exp\left[ \widehat{\omega}_{I}^{(n)} \mathcal R \right],
	\label{eq:FRepellingMagnitude}
\end{equation}
where \(\chi_n\) is independent of \(\mathcal R\). 

A real exceptional point is obtained when the image of the repelling point lies on the real \(F\)-axis, so that \(\operatorname{Im}F(\omega_r;r_0)=0\). The associated critical coupling is then \(\epsilon_{\rm EP}=\operatorname{Re}F(\omega_r;r_0)\).

At large distance, the first of these conditions becomes a phase quantization condition. Since the phase in Eq.~\eqref{eq:FAtRepellingPointAsymptotic} varies principally as \(-\widehat{\omega}_{R}^{(n)}\mathcal R\), consecutive crossings of the
real \(F\)-axis satisfy
\begin{equation}
	\mathcal R_{k+1}^{(n)} - \mathcal R_k^{(n)} \simeq \frac{\pi}{ \operatorname{Re} \widehat{\omega}_Q^{(n)}}.
	\label{eq:EPAsymptoticSpacing}
\end{equation}
The distant exceptional points therefore become equally spaced in the shifted tortoise coordinate. The phase changes by approximately \(\pi\) between two consecutive crossings, so that the signs of the associated critical couplings alternate asymptotically.

Their magnitudes obey
\begin{equation}
	\left| \epsilon_{{\rm EP}}^{(n,k)}\right| \simeq \frac{\chi_n}{ \mathcal R_k^{(n)}} \exp\left[ \operatorname{Im} \widehat{\omega}_Q^{(n)} \mathcal R_k^{(n)} \right].
	\label{eq:EPThresholdAsymptoticLaw}
\end{equation}
The asymptotic threshold therefore contains both an algebraic factor \(1/\mathcal R_k^{(n)}\) and an exponential factor governed by the imaginary part of the Schwarzschild QNM.

For comparison with the numerical data, it is useful to remove the algebraic prefactor and write
\begin{equation}
	\log\left[\left|\epsilon_{{\rm EP}}^{(n,k)}\right|\mathcal R_k^{(n)}\right]\simeq a_n + \operatorname{Im} \widehat{\omega}_Q^{(n)} \mathcal R_k^{(n)},
	\label{eq:CorrectedThresholdLinearLaw}
\end{equation}
where \(a_n=\log \chi_n\). A linear fit of the left-hand side as a function of \(\mathcal R_k^{(n)}\) should therefore yield the asymptotic slope
\begin{equation}
	b_n^{\rm fit} \simeq \operatorname{Im} \widehat{\omega}_Q^{(n)}.
	\label{eq:CorrectedThresholdSlope}
\end{equation}

Equations~\eqref{eq:DimensionlessRepellingAccumulation}, \eqref{eq:EPAsymptoticSpacing} and \eqref{eq:EPThresholdAsymptoticLaw} constitute the principal large-distance predictions. They show that the real and imaginary parts of the Schwarzschild QNM play distinct roles: the real part determines the asymptotic spacing of the exceptional points, whereas the imaginary part determines the exponential decay of their critical couplings.

%===============================================================================================================

\subsection{Numerical methods}
\label{subsec:NumericalMethods}

We now briefly describe the numerical methods used to construct the normalized Wronskian and to determine the repelling points and exceptional points.

We first compute the homogeneous solutions \(\phi_{\ell\omega}^{\rm in}\) and \(\phi_{\ell\omega}^{\rm up}\), together with the scattering coefficient \(A_{\ell\omega}^{(-)}\), by numerically integrating the Regge-Wheeler equation~\eqref{eq:RW}. The integrations are initialized using Taylor-series expansions near the event horizon and asymptotic expansions with the appropriate ingoing and outgoing behaviors at spatial infinity. The asymptotic series are resummed using Padé approximants.

The integrations are performed using high-precision arithmetic. An adaptive procedure with automatic stiffness switching is employed to maintain numerical stability, particularly at complex frequencies, for which the homogeneous solutions may contain exponentially growing and decaying components. The resulting solutions are evaluated at the location of the localized perturbation and used to construct the normalized Wronskian \(F(\omega;r_0)\) and its frequency derivatives.

For each fixed value of \(r_0\), the repelling points are determined by solving \(F_\omega(\omega_r;r_0)=0\) with a complex Newton--Raphson algorithm. The first point on a given branch is obtained from a suitable initial estimate, typically chosen near an unperturbed Schwarzschild QNM. The branch is then followed by continuation in \(r_0\), using the repelling point found at one location as the initial
estimate at the next. This procedure allows the same branch to be tracked smoothly over an extended radial interval.

Along each repelling-point branch, we record the corresponding value of \(F(\omega_r;r_0)\). The intervals over which \(\operatorname{Im}F(\omega_r;r_0)\) changes sign provide initial brackets for the locations at which the branch crosses the real \(F\)-axis. These crossings give the approximate locations of the real exceptional points.

Each crossing is subsequently determined more accurately using a secant method applied to \(\operatorname{Im}F(\omega_r;r_0)\). At every step, the repelling-point frequency is recomputed with the Newton--Raphson algorithm, ensuring that \(F_\omega(\omega_r;r_0)=0\) remains satisfied along the branch. Once the reality condition has converged, the critical coupling is obtained from
\(\epsilon_{\rm EP}=\operatorname{Re}F(\omega_{\rm EP};r_{\rm EP})\).

Particular care is required for the most distant exceptional points, because both the real and imaginary parts of \(F\) may become extremely small. An absolute condition on \(\operatorname{Im}F\) alone is then insufficient to establish that the reality condition has been accurately satisfied. In addition to requiring a small residual \(\lvert F_\omega\rvert\), we therefore monitor the relative quantity
\(\lvert\operatorname{Im}F\rvert/ \lvert\operatorname{Re}F\rvert\) and require it to be much smaller than unity. This criterion ensures that the small imaginary part is not merely a consequence of the overall exponential decrease of \(F\), but that \(F\) is genuinely real to the numerical accuracy considered. The second-order nature of each degeneracy is finally verified by checking that \(F_{\omega\omega}(\omega_{\rm EP};r_{\rm EP})\) remains nonzero.

All numerical calculations were performed using \emph{Mathematica}.

%===============================================================================================================

\section{Results: the perturbed spectrum}
\label{sec:results}

\subsection{Exceptional points and repelling points}
\label{subsec:EPFamiliesAsymptotics}

We first present the repelling-point branches associated with the lowest Schwarzschild QNMs and identify the exceptional points of the real perturbed spectral problem. We then examine their organization in the large-\(r_0\) regime and compare the numerical results with Eqs.~\eqref{eq:DimensionlessRepellingAccumulation}, \eqref{eq:EPAsymptoticSpacing}, and \eqref{eq:EPThresholdAsymptoticLaw}.

\subsubsection{Branches}

\begin{figure}[htbp]
	\centering
	\includegraphics[width=0.96\linewidth]{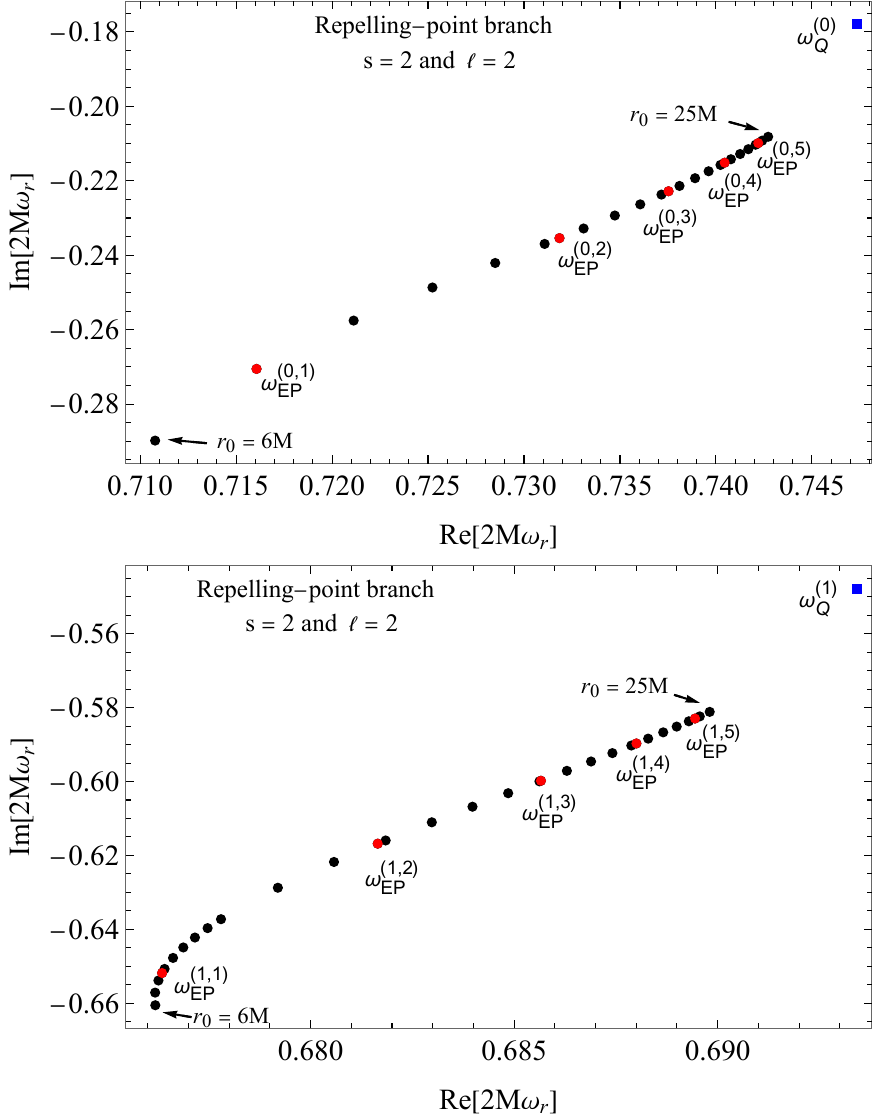}
	\caption{
		Repelling-point branches associated with the first two neighboring pairs of Schwarzschild quasinormal modes
		for \(s=2\) and
		\(\ell=2\).
		The black points show the solutions of
		\(F_\omega(\omega_r;r_0)=0\) as the location \(r_0\) of the
		localized perturbation is varied.
		The blue squares denote the unperturbed Schwarzschild QNM
		frequencies.
		The red points indicate the subset of repelling points for which
		\(\operatorname{Im}F(\omega_r;r_0)=0\).
		At these points, the critical perturbation strength
		\(\epsilon_{\rm EP}
		=\operatorname{Re}F(\omega_r;r_0)\) is real, and the corresponding
		points are exceptional points of the real perturbed spectral
		problem.
	}
	\label{fig:RPbranch}
\end{figure}

Figure~\ref{fig:RPbranch} displays the repelling-point branches obtained by solving \(F_\omega(\omega_r;r_0)=0\) as the perturbation location is varied. The branches are organized around the Schwarzschild QNM frequencies, as predicted by Eq.~\eqref{eq:RepellingPointAccumulation}, and form families associated with distinct unperturbed Schwarzschild QNMs. The first two branches shown correspond to the \(n=0\) and \(n=1\) families, which accumulate on \(\omega_Q^{(0)}\) and \(\omega_Q^{(1)}\), respectively.

Along each continuous branch, the additional condition \(\operatorname{Im}F(\omega_r;r_0)=0\) selects a discrete sequence of points for which the critical coupling \(\epsilon_{\rm EP}=\operatorname{Re}F(\omega_r;r_0)\) is real. These points, shown in red in Fig.~\ref{fig:RPbranch}, are the exceptional points of the real two-parameter problem. The EPs are numbered in order of increasing perturbation radius $r_0$.

The first five exceptional points of the \(n=0\) and \(n=1\) families are summarized in Table~\ref{tab:EPConnectivity}. The table gives their perturbation radii, critical couplings and complex frequencies, and illustrates both the radial ordering used to define the index \(k\) and the rapid decrease of \(\lvert\epsilon_{\rm EP}\rvert\) along the sequences.

\begin{table}[htbp]
	\caption{
		First five exceptional points of the \(n=0\) and \(n=1\) families,
        ordered by increasing perturbation radius \(r_{\rm EP}\).
	}
	\label{tab:EPConnectivity}
	\begin{ruledtabular}
		\begin{tabular}{ccc r@{\,$\times$\,}l c}
			Family & \(k\) &
			\(r_{\rm EP}^{(n,k)}/M\) &
			\multicolumn{2}{c}{\(\epsilon_{\rm EP}^{(n,k)}\)} &
			\(2M\omega_{\rm EP}^{(n,k)}\) \\
			\hline
			
			\(n = 0\) & 1 & 6.98375
			& \(-7.87238\) & \(10^{-2}\)
			& \(0.716068-0.270546\,\ii\) \\
			
			& 2 & 11.35362
			& \(1.61077\) & \(10^{-2}\)
			& \(0.731848-0.235410\,\ii\) \\
			
			& 3 & 15.37436
			& \(-5.17544\) & \(10^{-3}\)
			& \(0.737529-0.222810\,\ii\) \\
			
			& 4 & 19.38346
			& \(1.84934\) & \(10^{-3}\)
			& \(0.740452-0.215144\,\ii\) \\
			
			& 5 & 23.38486
			& \(-7.06709\) & \(10^{-4}\)
			& \(0.742202-0.209884\,\ii\) \\
			
			\hline
			
			\(n = 1\) & 1 & 6.66022
			& \(6.57660\) & \(10^{-3}\)
			& \(0.676351-0.651836\,\ii\) \\
			
			& 2 & 10.83962
			& \(-1.26981\) & \(10^{-4}\)
			& \(0.681649-0.616822\,\ii\) \\
			
			& 3 & 15.05139
			& \(4.95702\) & \(10^{-6}\)
			& \(0.685659-0.599794\,\ii\) \\
			
			& 4 & 19.28905
			& \(-2.53999\) & \(10^{-7}\)
			& \(0.688004-0.589697\,\ii\) \\
			
			& 5 & 23.55514
			& \(1.49143\) & \(10^{-8}\)
			& \(0.689445-0.582948\,\ii\) \\
		\end{tabular}
	\end{ruledtabular}
\end{table}

\subsubsection{Large-distance asymptotics and sensitivity}

We now examine the large-distance behavior of the first two families. For each family, we have constructed the first twenty exceptional points over an extended radial interval. Their locations are expressed in terms of the shifted dimensionless tortoise coordinate $ \mathcal R_k^{(n)} = r_*\!\left(r_{\rm EP}^{(n,k)}\right)/M, $ where \(n=0\) and \(n=1\) label the families accumulating on \(\omega_Q^{(0)}\) and \(\omega_Q^{(1)}\), respectively.

\begin{figure*}[htbp]
	\centering
	\includegraphics[width=0.90\linewidth]{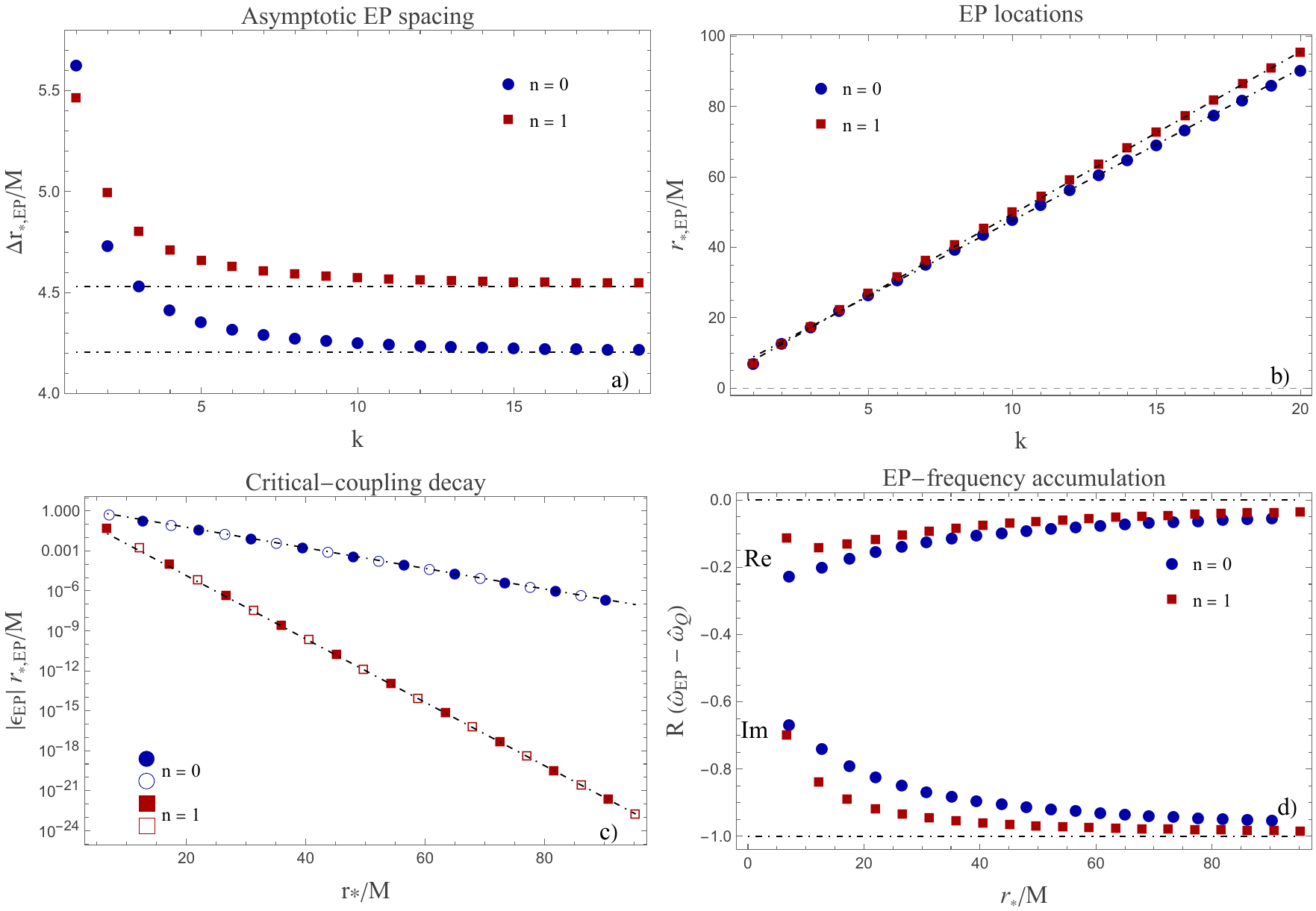}
	\caption{
		Large-distance asymptotics of the \(n=0\) and \(n=1\) exceptional-point families, which accumulate on \(\omega_Q^{(0)}\) and \(\omega_Q^{(1)}\), respectively, shown by blue circles and red squares. The shifted tortoise coordinate is used throughout. In panels (c) and (d), \(r_*\) denotes the tortoise-coordinate location of the corresponding exceptional point; thus \(r_*/M=\mathcal R_k^{(n)}\) for each plotted point. (a) Consecutive EP spacings \(\Delta\mathcal R_k^{(n)} =\mathcal R_{k+1}^{(n)}-\mathcal R_k^{(n)}\). The horizontal dot-dashed lines show the asymptotic predictions \(\pi/\operatorname{Re}\widehat{\omega}_Q^{(n)}\). (b) EP locations as functions of the integer label \(k\). The dot-dashed lines show the corresponding asymptotic linear behavior. (c) Corrected critical-coupling magnitudes \(\lvert\epsilon_{{\rm EP}}^{(n,k)}\rvert r_*/M\), displayed on a logarithmic vertical scale. Filled and open symbols correspond respectively to positive and negative critical couplings. The dot-dashed lines have the predicted asymptotic slopes \(\operatorname{Im}\widehat{\omega}_Q^{(n)}\). (d) Real and imaginary parts of \((r_*/M) [\widehat{\omega}_{{\rm EP}}^{(n,k)} -\widehat{\omega}_Q^{(n)}]\). The horizontal dot-dashed lines mark the predicted limits \(0\) and \(-1\), respectively. The convergence of the two numerical sequences confirms \(\mathcal R_k^{(n)} [\widehat{\omega}_{{\rm EP}}^{(n,k)} -\widehat{\omega}_Q^{(n)}]\rightarrow-\ii\).
	}
	\label{fig:EPAsymptotics}
\end{figure*}

The results are displayed in Fig.~\ref{fig:EPAsymptotics}. Panel (a) shows the consecutive spacings
\begin{equation}
	\Delta\mathcal R_k^{(n)} = \mathcal R_{k+1}^{(n)} - \mathcal R_k^{(n)}.
	\label{eq:NumericalEPSpacing}
\end{equation}
For both families, the spacings approach the constant values predicted in Eq.~\eqref{eq:EPAsymptoticSpacing} from the real parts of the corresponding Schwarzschild QNM frequencies. Panel (b) displays the EP locations themselves. Their approximately linear dependence on \(k\) confirms that the distant exceptional points become equally spaced in the shifted tortoise coordinate. Panel (c) shows the corrected critical-coupling magnitudes \(\lvert\epsilon_{{\rm EP}}^{(n,k)}\rvert\mathcal R_k^{(n)}\). Their exponential decrease is controlled by the imaginary parts of the relevant Schwarzschild QNM frequencies, as anticipated by Eq.~\eqref{eq:EPThresholdAsymptoticLaw}. 

Finally, panel (d) displays the real and imaginary parts of \(\mathcal R_k^{(n)} [\widehat{\omega}_{{\rm EP}}^{(n,k)} -\widehat{\omega}_Q^{(n)}]\). For both families, the real part approaches zero, while the imaginary part approaches \(-1\), providing a direct numerical test of the accumulation law in Eq.~\eqref{eq:DimensionlessRepellingAccumulation}. 

For the \(n=0\) family, the relevant Schwarzschild QNM frequency is
\begin{equation}
	\widehat{\omega}_Q^{(0)} = 2M\omega_Q^{(0)} \simeq 0.7473433688 - 0.1779246314\,\ii .
	\label{eq:QNMFrequency01}
\end{equation}
The predicted asymptotic spacing is therefore
\begin{equation}
	\Delta\mathcal R_{\rm pred}^{(0)} = \frac{\pi}{\operatorname{Re}\widehat{\omega}_Q^{(0)}} \simeq 4.2037.
	\label{eq:PredictedEPSpacing01}
\end{equation}
The numerical spacings displayed in Fig.~\ref{fig:EPAsymptotics}(a) progressively approach this value as \(k\) increases.

The corrected decay fit gives
\begin{equation}
	b_0^{\rm fit} \simeq -0.177755.
   \label{eq:CorrectedSlopeFit01}
\end{equation}
The asymptotic prediction is
\begin{equation}
	\operatorname{Im}\widehat{\omega}_Q^{(0)} \simeq -0.177925.
	\label{eq:PredictedSlope01}
\end{equation}
The relative difference is smaller than \(10^{-3}\).

For the \(n=1\) family, the relevant Schwarzschild QNM frequency is
\begin{equation}
	\widehat{\omega}_Q^{(1)} = 2M\omega_Q^{(1)}
	\simeq 0.6934219938 - 0.5478297506\,\ii.
	\label{eq:QNMFrequency12}
\end{equation}
The corresponding asymptotic spacing is
\begin{equation}
	\Delta\mathcal R_{\rm pred}^{(1)} = \frac{\pi}{ \operatorname{Re}\widehat{\omega}_Q^{(1)} } \simeq 4.5306.
	\label{eq:PredictedEPSpacing12}
\end{equation}
The numerical spacings again converge toward the predicted value.
For the corrected critical-coupling decay, the numerical fit gives
$
	b_1^{\rm fit} \simeq -0.549801,
$
whereas the asymptotic prediction is
$
	\operatorname{Im}\widehat{\omega}_Q^{(1)} \simeq -0.547830.
$
The relative difference is of order \(4\times10^{-3}\).

The distinction between a direct exponential fit and the corrected asymptotic fit is important for interpreting Fig.~\ref{fig:EPAsymptotics}(c). A direct fit of the form \(\lvert\epsilon_{{\rm EP}}^{(n,k)}\rvert \simeq \chi_n^{\rm eff} \exp[-\alpha_n^{\rm eff}\mathcal R_k^{(n)}]\) provides only an effective decay rate, because the asymptotic law also
contains the algebraic factor \(1/\mathcal R_k^{(n)}\). Equivalently,
\begin{equation}
	\log \left| \epsilon_{{\rm EP}}^{(n,k)} \right| \simeq a_n + \operatorname{Im} \widehat{\omega}_Q^{(n)} \mathcal R_k^{(n)} - \log\mathcal R_k^{(n)}.
	\label{eq:LogEpsilonWithAlgebraicCorrection}
\end{equation}
A straight-line fit of \(\log\lvert\epsilon_{\rm EP}\rvert\) alone therefore absorbs part of the slowly varying term \(-\log\mathcal R\) into the fitted slope. The corrected quantity \(\log(\lvert\epsilon_{\rm EP}\rvert\mathcal R)\), used in Fig.~\ref{fig:EPAsymptotics}(c), provides the appropriate numerical test of the asymptotic prediction.

The same asymptotic analysis predicts the accumulation of the EP frequencies according to
\begin{equation}
	\mathcal R_k^{(n)} \left[ \widehat{\omega}_{{\rm EP}}^{(n,k)} -\widehat{\omega}_Q^{(n)} \right] \longrightarrow -\ii.
	\label{eq:EPFrequencyAccumulationTest}
\end{equation}
As shown in Fig.~\ref{fig:EPAsymptotics}(d), the real parts of the two numerical sequences approach zero, while their imaginary parts approach \(-1\) as \(k\) increases. This provides an independent frequency-space confirmation that each exceptional-point family accumulates on the corresponding unperturbed Schwarzschild QNM as \(r_{*,{\rm EP}}\rightarrow+\infty\).

The asymptotic results also provide a quantitative measure of overtone fragility. The decay rate of the EP threshold is controlled by \(-\operatorname{Im}\widehat{\omega}_Q^{(n)}\). For the first two
families,
\begin{equation}
	-\operatorname{Im}\widehat{\omega}_Q^{(1)} \simeq 0.547830 > 0.177925 \simeq -\operatorname{Im}\widehat{\omega}_Q^{(0)}.
	\label{eq:DampingHierarchyCorrected}
\end{equation}
The critical couplings of the \(n=1\) family therefore decrease much more rapidly than those of the \(n=0\) family.

At comparable large values of \(\mathcal R\), the ratio of the two thresholds behaves as
\begin{equation}
	\frac{ \left| \epsilon_{\rm EP}^{(1)} \right|}{\left|\epsilon_{\rm EP}^{(0)}\right|} \sim \frac{\chi_1}{\chi_0} \exp\left[
	\left( \operatorname{Im}\widehat{\omega}_Q^{(1)} - \operatorname{Im}\widehat{\omega}_Q^{(0)} \right) \mathcal R \right].
	\label{eq:ThresholdRatioFragility}
\end{equation}
The difference between the two asymptotic exponents is
\begin{equation}
	\operatorname{Im}\widehat{\omega}_Q^{(1)} - \operatorname{Im}\widehat{\omega}_Q^{(0)} \simeq -0.369905.
	\label{eq:DampingDifference}
\end{equation}
This ratio therefore decreases exponentially with \(\mathcal R\), up to the family-dependent prefactor $\chi_1/\chi_0$. Exceptional points involving the more strongly damped overtones consequently occur at much smaller perturbation strengths.

The Schwarzschild QNMs remain simple resonances of the unperturbed problem, while the exceptional-point sequences accumulate toward them in the enlarged parameter space \((\omega,\epsilon,r_0)\).

The localized perturbation therefore reveals a multi-sheeted spectral geometry in which discrete exceptional-point sequences accumulate on successive Schwarzschild QNMs. The global connectivity of the associated resonance branches is examined separately in Sec.~\ref{subsec:SpectralConnectivitySensitivity}. This hierarchy provides an explicit realization, for the present localized perturbation family, of the enhanced spectral sensitivity of Schwarzschild overtones identified from pseudospectral analyses~\cite{Jaramillo:2020tuu,Cao:2025afs}.

\subsubsection{A fiducial exceptional point}
Throughout the remainder of the article, we shall frequently use as a reference the weak-coupling exceptional point \({\rm EP}_{0}^{(4)}\). In the numerical convention \(2M=1\),
its location is
\begin{equation}
	\frac{r_{\rm EP}^{(0,4)}}{M} \simeq 19.383465.
	\label{eq:MainEPRadius}
\end{equation}
Its complex frequency is
\begin{equation}
	\omega_{\rm EP}^{(0,4)} \simeq 0.7404519 - 0.215144\,\ii .
	\label{eq:MainEPFrequency}
\end{equation}
The normalized Wronskian is real (within numerical accuracy) at this point and satisfies
\begin{equation}
	F\!\left( \omega_{\rm EP}^{(0,4)}; r_{\rm EP}^{(0,4)}\right)\simeq 0.00184934.
	\label{eq:MainEPFValue}
\end{equation}
The corresponding critical coupling is therefore
\begin{equation}
	\epsilon_{\rm EP}^{(0,4)} \simeq 1.849\times10^{-3}.
	\label{eq:MainEPCoupling}
\end{equation}
Numerically, \(F_\omega(\omega_{\rm EP}^{(0,4)}; r_{\rm EP}^{(0,4)})\simeq0\), whereas \(F_{\omega\omega}(\omega_{\rm EP}^{(0,4)}; r_{\rm EP}^{(0,4)})\neq0\). These conditions confirm that \({\rm EP}_{0}^{(4)}\) is a second-order exceptional point.

%===============================================================================================================

\subsection{Avoided crossings and hyperbolae}
\label{subsec:AvoidedCrossingsHyperbolae}

We now examine the local spectral geometry surrounding a second-order exceptional point. The two real parameters \(\epsilon\) and \(r_0\) span a two-dimensional parameter plane. Let us consider two complementary one-dimensional paths through the associated local two-sheeted spectral surface. Along the first path, the perturbation strength $\epsilon$ is varied at fixed location $r_0$; along the second, the location is varied at fixed perturbation strength. When the fixed parameter takes its critical value, the path intersects the exceptional point and the two resonance branches coalesce. A displacement from that critical value causes the path to miss the branch point, thereby unfolding the coalescence into an avoided crossing~\cite{Dias:2022oqm,Motohashi:2024fwt,Cavalcante:2024swt}.

Throughout this subsection, the numerical trajectories are compared with the local forms \eqref{eq:LocalRootsRepellingPoint} and \eqref{eq:GeneralSquareRootUnfolding}. We use \({\rm EP}_{0}^{(4)}\) as a representative exceptional point of the \(n=0\) family.

\subsubsection{Variation of the perturbation strength}
\label{subsubsec:EpsilonVariation}

We first vary the real perturbation strength \(\epsilon\) at a fixed location \(r_0\) close to \(r_{\rm EP}^{(0,4)}\). We write \(r_0=r_{\rm EP}^{(0,4)}+\delta r_0\) and vary the coupling according to
\begin{equation}
	\epsilon=\epsilon_c(r_0)+\Delta\epsilon,
	\label{eq:CouplingVariationNearRP}
\end{equation}
where \(\epsilon_c(r_0)=\operatorname{Re} F(\omega_r(r_0);r_0)\), and \(\omega_r(r_0)\) denotes the repelling point satisfying \(F_\omega(\omega_r(r_0);r_0)=0\).

At \(\delta r_0=0\), the repelling point coincides with the exceptional point and its image under \(F\) lies on the real axis. In this case, \(\epsilon_c=\epsilon_{\rm EP}^{(0,4)}\), and the real \(\epsilon\)-trajectory passes directly through the degeneracy. The two resonance roots approach one another, coalesce at \(\omega_{\rm EP}^{(0,4)}\), and leave the exceptional point along the second local direction as the sign of \(\Delta\epsilon\) is changed.

For \(\delta r_0\neq0\), the corresponding repelling point no longer satisfies the reality condition. Its image may be written as
\begin{equation}
	F\!\left(\omega_r(r_0);r_0\right) = \epsilon_c(r_0) + \ii\eta(r_0),
	\label{eq:ComplexRepellingPointImage}
\end{equation}
where \(\eta(r_0)=\operatorname{Im}F(\omega_r(r_0);r_0)\neq0\). The associated branch point is consequently displaced away from the real-\(\epsilon\) axis. The two roots approach one another but do not coalesce; instead, they turn away and form an avoided crossing.

\begin{figure}[htbp]
	\centering
	\includegraphics[width=0.97\linewidth]{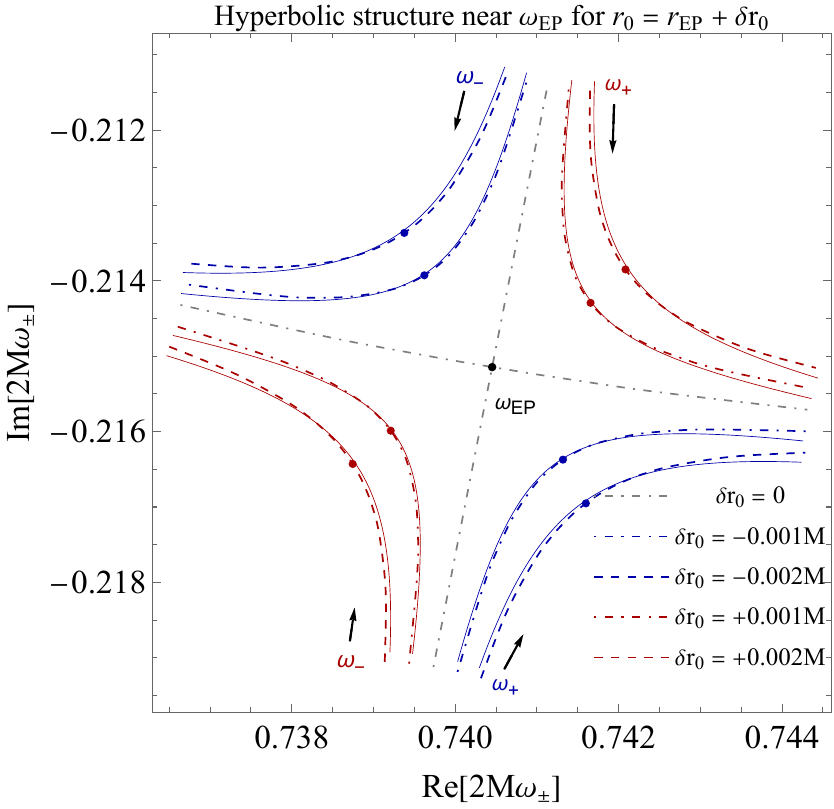}
	\caption{
		Comparison between the full numerical trajectories and the local hyperbolic approximation near \({\rm EP}_{0}^{(4)}\). The perturbation location is written as \(r_0=r_{\rm EP}^{(0,4)}+\delta r_0\), with \(\delta r_0=0\), \(\delta r_0=\pm 10^{-3}M\), and \(\delta r_0=\pm 2\times10^{-3}M\). The gray dot-dashed curves show the limiting crossing directions at the exceptional point, \(\delta r_0=0\). The blue curves correspond to negative radial offsets, \(\delta r_0<0\), while the red curves correspond to positive radial offsets, \(\delta r_0>0\). For each sign, the dot-dashed
		curves denote \(|\delta r_0|=10^{-3}M\), and the dashed curves denote \(|\delta r_0|=2\times10^{-3}M\). The black point marks the exceptional frequency \(\omega_{\rm EP}\), while
		the filled colored points indicate the numerical roots at \(\Delta\epsilon=0\). The arrows show the direction of increasing \(\Delta\epsilon\). For each nonzero value of \(\delta r_0\), the solid red and blue curves represent the local quadratic prediction \(\omega_\pm^{\rm hyp}=\omega_r(r_0)\pm \left[2(\Delta\epsilon-\ii\eta)/ F_{\omega\omega}^{r}\right]^{1/2}\),
		with \(\eta=\operatorname{Im}F(\omega_r(r_0);r_0)\) and \(F_{\omega\omega}^{r} = F_{\omega\omega}(\omega_r(r_0);r_0)\).
	}
	\label{fig:HyperbolicApproximation}
\end{figure}

Figure~\ref{fig:HyperbolicApproximation} displays the crossing at \(\delta r_0=0\), together with the avoided crossings obtained for positive and negative radial offsets. The latter bend in opposite directions on the two sides of the exceptional-point location. Moreover, their opening increases with \(|\delta r_0|\), consistent with the dependence of the local hyperbolic structure on
\(|\eta(r_0)|\).

The full numerical trajectories agree closely with the quadratic prediction in the immediate neighborhood of the corresponding repelling points. The agreement is particularly accurate for \(|\delta r_0|=10^{-3}M\), for which the roots remain within the domain controlled by the local expansion. For the larger offsets, small differences become visible near the ends of the trajectories, where higher-order terms in the frequency expansion of \(F\) begin to contribute.

The square-root character of the crossing can also be tested directly from the separation of the two roots at \(r_0=r_{\rm EP}^{(0,4)}\). A fit using positive and negative values of \(\Delta\epsilon\) gives
\begin{equation}
	\left| \omega_+ - \omega_- \right| \propto |\Delta\epsilon|^{0.5002},
	\label{eq:EpsilonSplittingFitResults}
\end{equation}
in excellent agreement with the exponent \(1/2\) predicted by the quadratic expansion. The agreement between the fitted splitting and the local prediction improves as \(|\Delta\epsilon|\) is reduced, as expected from the local nature of the approximation.

Figure~\ref{fig:HyperbolicApproximation} also makes explicit the continuous relation between the crossing and the neighboring avoided crossings. The crossing is recovered as \(r_0\) approaches \(r_{\rm EP}^{(0,4)}\), or equivalently as \(\eta(r_0)\) tends to zero. In this limit, the opening of the local hyperbola vanishes and its two asymptotic directions become the two crossing directions at the exceptional point.

It should be noted that \(\omega_+\) and \(\omega_-\) are local labels for the two resonance sheets in the neighborhood of the exceptional point. Their identification with particular Schwarzschild overtones requires a global continuation to the unperturbed limit. This global connectivity is examined in Sec.~\ref{subsec:SpectralConnectivitySensitivity}.

\subsubsection{Variation of the perturbation location}
\label{subsubsec:R0Variation}

We now consider a complementary path through the parameter space. The perturbation strength is held fixed close to its exceptional-point value,
\begin{equation}
	\epsilon = \epsilon_{\rm EP}^{(0,4)} + \delta\epsilon,
	\label{eq:CouplingOffsetR0Variation}
\end{equation}
while the perturbation location is varied according to
\begin{equation}
	r_0 = r_{\rm EP}^{(0,4)} + \delta r_0.
	\label{eq:R0VariationNearEP}
\end{equation}

When \(\delta\epsilon=0\), the radial trajectory passes directly through the exceptional point. The two resonance branches coalesce at \(\delta r_0=0\), where \(r_0=r_{\rm EP}^{(0,4)}\), and separate again as the perturbation is moved through its critical location. This is the radial counterpart of the coupling-driven crossing discussed above.

For \(\delta\epsilon\neq0\), the radial path no longer intersects the branch point. The two roots then undergo an avoided crossing as \(r_0\) passes through the neighborhood of \(r_{\rm EP}^{(0,4)}\). Positive and negative coupling offsets select opposite sides of the local exceptional-point geometry and therefore produce avoided crossings with different orientations in the complex-frequency plane.

\begin{figure}[htbp]
	\centering
	\includegraphics[width=0.97\linewidth]{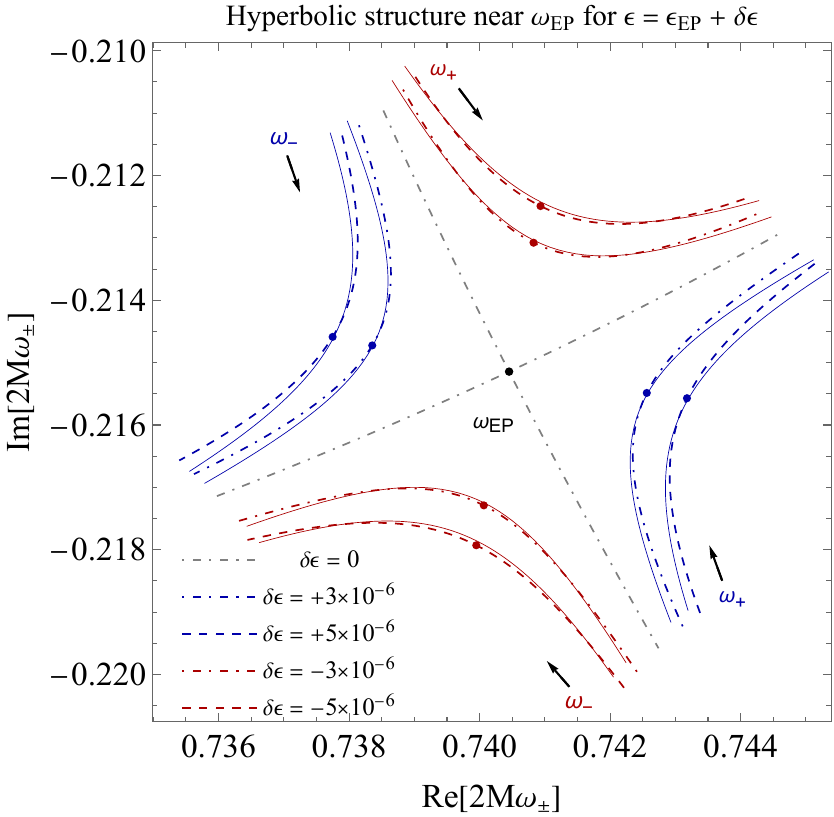}
	\caption{ Comparison between the full numerical trajectories and the local inverse-hyperbolic approximation near \({\rm EP}_{0}^{(4)}\). The coupling is written as \(\epsilon=\epsilon_{\rm EP}^{(0,4)}+\delta\epsilon\), with \(\delta\epsilon=0\), \(\delta\epsilon=+3\times10^{-6}\), \(\delta\epsilon=+5\times10^{-6}\), \(\delta\epsilon=-3\times10^{-6}\), and \(\delta\epsilon=-5\times10^{-6}\). The perturbation location is varied as \(r_0=r_{\rm EP}^{(0,4)}+\delta r_0\). The gray dot-dashed curves show the limiting crossing directions at the exceptional point, \(\delta\epsilon=0\). The blue curves correspond to positive coupling offsets, \(\delta\epsilon>0\), while the red curves correspond to negative coupling offsets, \(\delta\epsilon<0\). For each sign, the dot-dashed curves denote the smaller value of \(|\delta\epsilon|\), and the dashed curves denote the larger value of \(|\delta\epsilon|\).
	The black point marks the exceptional frequency \(\omega_{\rm EP}\), while the filled points indicate the numerical roots at \(\delta r_0=0\). The arrows show the direction of increasing \(\delta r_0\). For each nonzero value of \(\delta\epsilon\), the solid red and blue curves represent the local quadratic prediction
	\(\omega_\pm^{\rm inv} = \omega_{\rm EP} \pm \left[ \frac{2\left(\delta\epsilon - F_{r_0}^{\rm EP}\delta r_0\right)}{F_{\omega\omega}^{\rm EP}}\right]^{1/2}.\) Here \(\delta r_0=r_0-r_{\rm EP}^{(0,4)}\).
	}
	\label{fig:InverseHyperbolicApproximation}
\end{figure}

Figure~\ref{fig:InverseHyperbolicApproximation} shows the full numerical trajectories for several positive and negative values of \(\delta\epsilon\), together with the local quadratic approximation. The gray curves represent the limiting crossing obtained at \(\delta\epsilon=0\). The blue and red curves display the avoided crossings generated by positive and negative coupling offsets, respectively.

As in the coupling-driven case, the numerical trajectories are accurately reproduced by the quadratic approximation close to the exceptional point. The opening of the avoided crossing increases with \(|\delta\epsilon|\), whereas its orientation depends on the sign of the coupling offset. Small discrepancies become visible as the roots move farther from \(\omega_{\rm EP}\), reflecting the increasing
contribution of terms beyond quadratic order.

For a quantitative test of the radial unfolding, we introduce the tortoise-coordinate displacement
\begin{equation}
	\delta r_{0*} = r_*(r_0) - r_*\!\left(r_{\rm EP}^{(0,4)}\right).
	\label{eq:TortoiseDisplacementResults}
\end{equation}
A fit of the frequency splitting on both sides of the exceptional point gives
\begin{equation}
	\left| \omega_+ - \omega_-\right|\propto|\delta r_{0*}|^{0.5008},
	\label{eq:RadialSplittingFitResults}
\end{equation}
again in excellent agreement with the square-root exponent \(1/2\). Thus, varying the location of the localized perturbation provides a second numerical realization of the same universal unfolding.

A further relation between the two parameter variations is visible by comparing Figs.~\ref{fig:HyperbolicApproximation} and \ref{fig:InverseHyperbolicApproximation}. The radial unfolding is rotated relative to the coupling-driven unfolding in the complex-frequency plane. From the local quadratic normal form, the relative angle is
\begin{equation}
	\Delta\theta = \frac{1}{2} \arg\!\left(-F_{r_0}^{\rm EP}\right).
	\label{eq:RelativeUnfoldingRotation}
\end{equation}
For \({\rm EP}_{0}^{(4)}\), using \(F_{r_0}^{\rm EP}\simeq-9.10\times10^{-4}-2.90\times10^{-3}\ii\), we obtain \(\Delta\theta\simeq36.3^\circ\), consistent with the relative orientation observed in the two figures.

The two figures represent complementary real paths through the same local two-sheeted spectral surface. When the fixed parameter takes its critical value, variation of the second parameter carries the two resonance roots directly through the branch point, where they coalesce. When the fixed parameter is displaced from its critical value, the selected path misses the degeneracy and the coalescence is
unfolded into an avoided crossing. The resulting hyperbolic trajectories are therefore two complementary manifestations of the universal square-root unfolding of a second-order exceptional point. Although illustrated here using \({\rm EP}_{0}^{(4)}\) only, the same local behavior has also been verified for exceptional points of the \(n=1\) family.

%===================================================================================================

\subsection{Hysteresis, geometric phase and Riemann sheets}
\label{subsec:HysteresisGeometricPhaseRiemannSheets}

The square-root structure of Eq.~\eqref{eq:GeneralSquareRootUnfolding}, together with the sign change \eqref{eq:SquareRootSignChange}, implies that the two local resonance branches belong to different sheets of a Riemann surface joined at the exceptional point~\cite{Ding:2022juv,Cavalcante:2024swt}. We now examine this structure numerically by continuing the roots along closed paths surrounding \({\rm EP}_{0}^{(4)}\). We first encircle the exceptional point in the complex \(\epsilon\)-plane at fixed \(r_0\), and then consider closed loops involving only the two real parameters \(\epsilon\) and \(r_0\).

Along each contour, the roots are followed by numerical continuation, using the root obtained at one point as the initial estimate at the next. A closed circuit in parameter space need not return an analytically continued resonance to its initial branch. This provides a direct numerical probe of the local monodromy associated with the exceptional point.

We begin by fixing the perturbation location at \(r_0=r_{\rm EP}\) and varying the coupling along the complex contour
\begin{equation}
	\epsilon(\theta) = \epsilon_{\rm EP} + \rho e^{\ii\theta}, \qquad 0\leq\theta\leq2\pi,
	\label{eq:EncirclingEpsilon}
\end{equation}
where \(\rho\) is chosen sufficiently small for the contour to remain within the neighborhood controlled by the quadratic expansion. Since
\(\delta r_0=0\), the local discriminant reduces to \(\zeta=\epsilon-\epsilon_{\rm EP}\), and the two frequencies behave as
\begin{equation}
	\omega_\pm-\omega_{\rm EP} \simeq \pm \left( \frac{2\zeta}{F_{\omega\omega}^{\rm EP}}\right)^{1/2}.
	\label{eq:LocalRiemannSquareRoot}
\end{equation}
A single circuit around \(\epsilon_{\rm EP}\) therefore changes the sign of the square root and exchanges the two local branches. In particular,
\begin{equation}
	\omega_+(2\pi) = \omega_-(0), \qquad \omega_-(2\pi) = \omega_+(0),
	\label{eq:ComplexEpsilonSheetExchange}
\end{equation}
up to the numerical accuracy of the continuation. A second circuit is required for each analytically continued root to return to its initial value and initial sheet.

\begin{figure*}[htbp]
	\centering
	\includegraphics[width=0.98\linewidth]{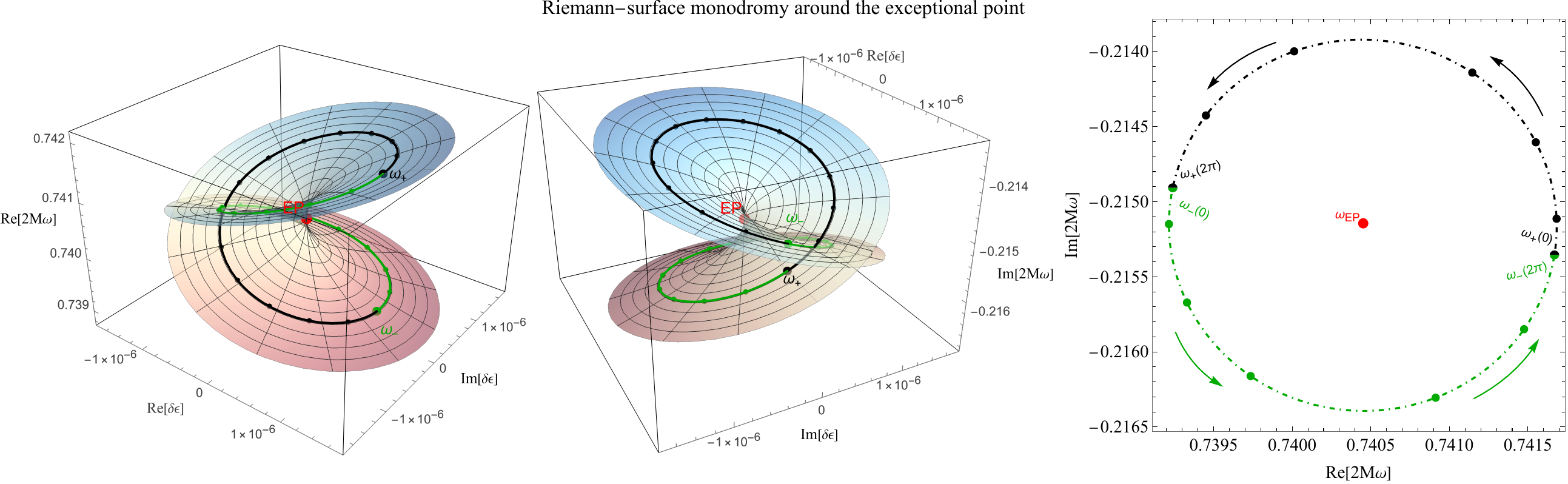}
	\caption{
		Encircling of the exceptional point in the complex perturbation plane. The perturbation strength is varied as \(\epsilon(\theta)=\epsilon_{\rm EP}+\rho e^{\ii\theta}\). The left and middle panels show the trajectory lifted onto the local two-sheeted Riemann surface, with \(\operatorname{Re}(2M\omega)\) and \(\operatorname{Im}(2M\omega)\) used as vertical coordinates, respectively. The right panel shows the corresponding projection onto the complex-frequency plane. The black trajectory follows one sheet during the first circuit, \(0\leq\theta\leq2\pi\), and terminates on the companion sheet. The green trajectory continues from this endpoint during the second circuit. The two pieces therefore display the sheet exchange after one circuit and the
        return to the initial sheet after two circuits, characteristic of the square-root monodromy of a second-order exceptional point.
	}
	\label{fig:EncirclingEP}
\end{figure*}

Figure~\ref{fig:EncirclingEP} provides a geometric representation of this continuation. The left and middle panels show the trajectory lifted onto the local two-sheeted Riemann surface, using respectively the real and imaginary parts of the frequency as vertical coordinates. The right panel shows the projection of the same continuation onto the complex-frequency plane.

Although the perturbation strength returns to its initial value after one circuit, the analytically continued resonance terminates on the companion sheet. This branch-memory effect is a form of spectral hysteresis: a closed path in parameter space returns the spectral equation to its initial form, while analytic continuation carries an individual resonance onto a different branch. The effect is topological and reflects the square-root branch point enclosed by the contour.

The same sheet exchange can be produced without complexifying the perturbation strength. We now consider a closed loop in the real \((r_0,\epsilon)\) parameter plane centered on \({\rm EP}_{0}^{(4)}\). We write
\[
r_0=r_{\rm EP}+\delta r_0,
\qquad
\epsilon=\epsilon_{\rm EP}+\delta\epsilon,
\]
and first consider the elliptical contour
\begin{equation}
	\delta r_0 = a\sin\theta, \qquad \delta\epsilon = b\cos\theta, \qquad 0\leq\theta\leq2\pi.
	\label{eq:RealParameterLoop}
\end{equation}

According to the quadratic expansion, the local unfolding is governed by the effective complex discriminant introduced in Eq.~\eqref{eq:EffectiveDiscriminant},
\[
\zeta = \delta\epsilon - F_{r_0}^{\rm EP}\delta r_0.
\]
Writing
\begin{equation}
	F_{r_0}^{\rm EP} = \mu+\ii\nu,
\end{equation}
the contour \eqref{eq:RealParameterLoop} is mapped to
\begin{equation}
	\zeta(\theta) = b\cos\theta - \mu a\sin\theta - \ii\nu a\sin\theta.
	\label{eq:DiscriminantEllipse}
\end{equation}
For nonzero \(a\) and \(b\), and since \(\nu\neq0\), this mapping defines a nondegenerate ellipse centered on \(\zeta=0\). The contour therefore winds once around the branch point, with its orientation determining the sign of the winding number.

More generally, the winding of an arbitrary closed contour \(\gamma\) is measured by
\begin{equation}
	N_\gamma = \frac{1}{2\pi\ii} \oint_\gamma \frac{d\zeta}{\zeta}.
	\label{eq:EPWindingNumber}
\end{equation}
Analytic continuation of the square root along this contour gives
\begin{equation}
	\sqrt{\zeta} \longrightarrow e^{\ii\pi N_\gamma}\sqrt{\zeta}.
	\label{eq:SquareRootMonodromy}
\end{equation}
Consequently, an odd winding number changes the sign of the local square-root coordinate and exchanges the two resonance sheets, whereas an even winding number restores the initial sheet.

At the level of the resonance frequencies, the same winding also determines the geometric phase of the local spectral displacement. From Eq.~\eqref{eq:LocalRiemannSquareRoot},
\begin{equation}
	\Delta\arg\!\left(\omega-\omega_{\rm EP}\right) = \frac{1}{2}\Delta\arg\zeta = \pi N_\gamma.
	\label{eq:SpectralGeometricPhase}
\end{equation}
Thus the geometric information considered here is carried directly by the complex QNM frequencies: a single winding of the discriminant produces a phase change of \(\pi\) in the local square-root coordinate and simultaneously transports the resonance onto the companion sheet. Such sheet exchange and geometric-phase behavior are characteristic of exceptional-point topology and have also been investigated in black-hole QNM spectra~\cite{Cavalcante:2024swt,Cavalcante:2025abr,Cao:2025afs,Cavalcante:2026vgr}.

A particularly transparent real-parameter contour is obtained by choosing
\begin{equation}
	\delta r_0 = -\frac{\rho}{\nu}\sin\theta, \qquad \delta\epsilon = \rho\cos\theta - \frac{\mu\rho}{\nu}\sin\theta.
	\label{eq:CircularDiscriminantLoop}
\end{equation}
Equivalently, writing \(\rho=-\nu a\), this parametrization takes the form \(\delta r_0=a\sin\theta\) and \(\delta\epsilon=-\nu a\cos\theta+\mu a\sin\theta\). For this choice,
\begin{equation}
	\zeta(\theta) = \rho e^{\ii\theta},
\end{equation}
so that a closed path involving only real values of \(r_0\) and \(\epsilon\) is mapped exactly onto a circle surrounding the branch point in the complex \(\zeta\)-plane.

\begin{figure*}[t]
	\centering
	\includegraphics[width=0.98\linewidth]{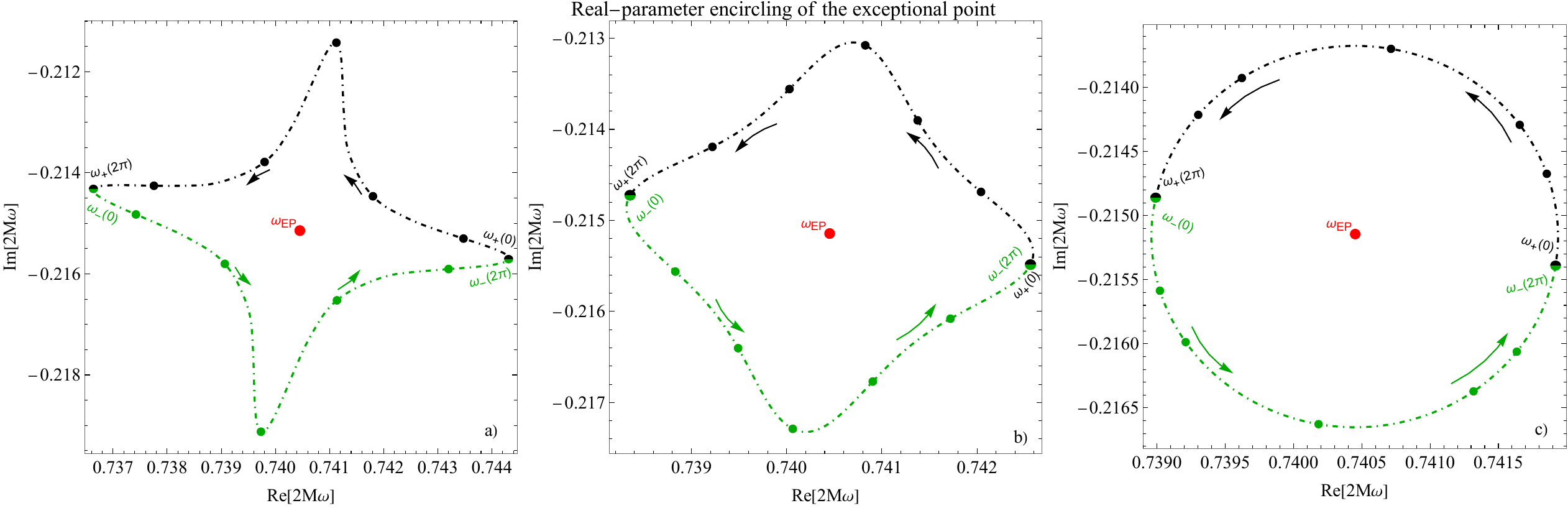}
	\caption{
		Real-parameter encircling of the exceptional point \({\rm EP}_{0}^{(4)}\). The two resonance branches are followed along closed loops in the real \((\delta r_0,\delta\epsilon)\)
		parameter plane, with \(r_0=r_{\rm EP}+\delta r_0\) and \(\epsilon=\epsilon_{\rm EP}+\delta\epsilon\). The black and green curves show the two analytically continued
		branches in the complex-frequency plane, the red point marks \(\omega_{\rm EP}\), and the arrows indicate increasing \(\theta\). In panels (a) and (b), the loops are of the form \(\delta r_0=a\sin\theta\), \(\delta\epsilon=b\cos\theta\), with different relative amplitudes, producing respectively a more star-like trajectory and an intermediate loop. In panel (c), the real parameters are chosen so that the effective discriminant \(\zeta=\delta\epsilon-F_{r_0}^{\rm EP}\delta r_0\) describes a circle around \(\zeta=0\), yielding a nearly circular spectral
		trajectory. In all three cases, one complete circuit exchanges the two branches, \(\omega_+(2\pi)\simeq\omega_-(0)\) and \(\omega_-(2\pi)\simeq\omega_+(0)\), demonstrating the
		square-root monodromy of the exceptional point using only real physical parameters.
	}
	\label{fig:RealParameterEncircling}
\end{figure*}

Figure~\ref{fig:RealParameterEncircling} shows the continued frequencies for three representative real loops around \({\rm EP}_{0}^{(4)}\). The first two use elliptical parameter contours with different relative amplitudes, whereas the third is chosen according to Eq.~\eqref{eq:CircularDiscriminantLoop}. Their projections in the complex-frequency plane have markedly different shapes, but in every case the two branches are exchanged after one complete circuit.

The star-like, intermediate and nearly circular trajectories are therefore different projections of the same underlying two-sheeted spectral structure. Their detailed shapes depend on the chosen path in parameter space, on the complex coefficient \(F_{r_0}^{\rm EP}\), and on corrections beyond the leading quadratic approximation. The shape of the projected frequency trajectory is therefore not itself a topological invariant. The invariant information is instead carried by the winding of \(\zeta\) around the origin and by the resulting permutation of the two resonance sheets.

The complex-\(\epsilon\) and real-parameter constructions thus probe the same local Riemann-surface topology. In the first case, the branch point is encircled directly in the complex \(\epsilon\)-plane. In the second, a closed contour in the real \((r_0,\epsilon)\) plane is mapped through the effective discriminant onto a complex contour winding around \(\zeta=0\). In both cases, one circuit exchanges the two resonance sheets, whereas two circuits restore the initial branch. The associated half-winding of the local square-root coordinate gives the corresponding geometric phase of the resonance-frequency displacement. Spectral hysteresis, geometric phase and sheet exchange are therefore complementary manifestations of the same local square-root topology of the exceptional point. These phenomena belong to the broader branch-point topology of non-Hermitian exceptional points~\cite{Ding:2022juv,Cavalcante:2024swt,Cavalcante:2025abr,Cao:2025afs,Cavalcante:2026vgr}.

%==================================================================================================================

\subsection{Spectral connectivity and sensitivity}
\label{subsec:SpectralConnectivitySensitivity}

The local square-root structure determines the behavior of the two resonance roots in the immediate neighborhood of an exceptional point. Their identification with the unperturbed Schwarzschild modes, however, requires a global continuation in the perturbation strength. We therefore fix the perturbation location at the relevant exceptional-point value and follow each resonance branch from the coalescence point back to the Schwarzschild limit, \(\epsilon=0\).

\begin{figure*}[htbp]
	\centering
	\includegraphics[width=0.85\linewidth] {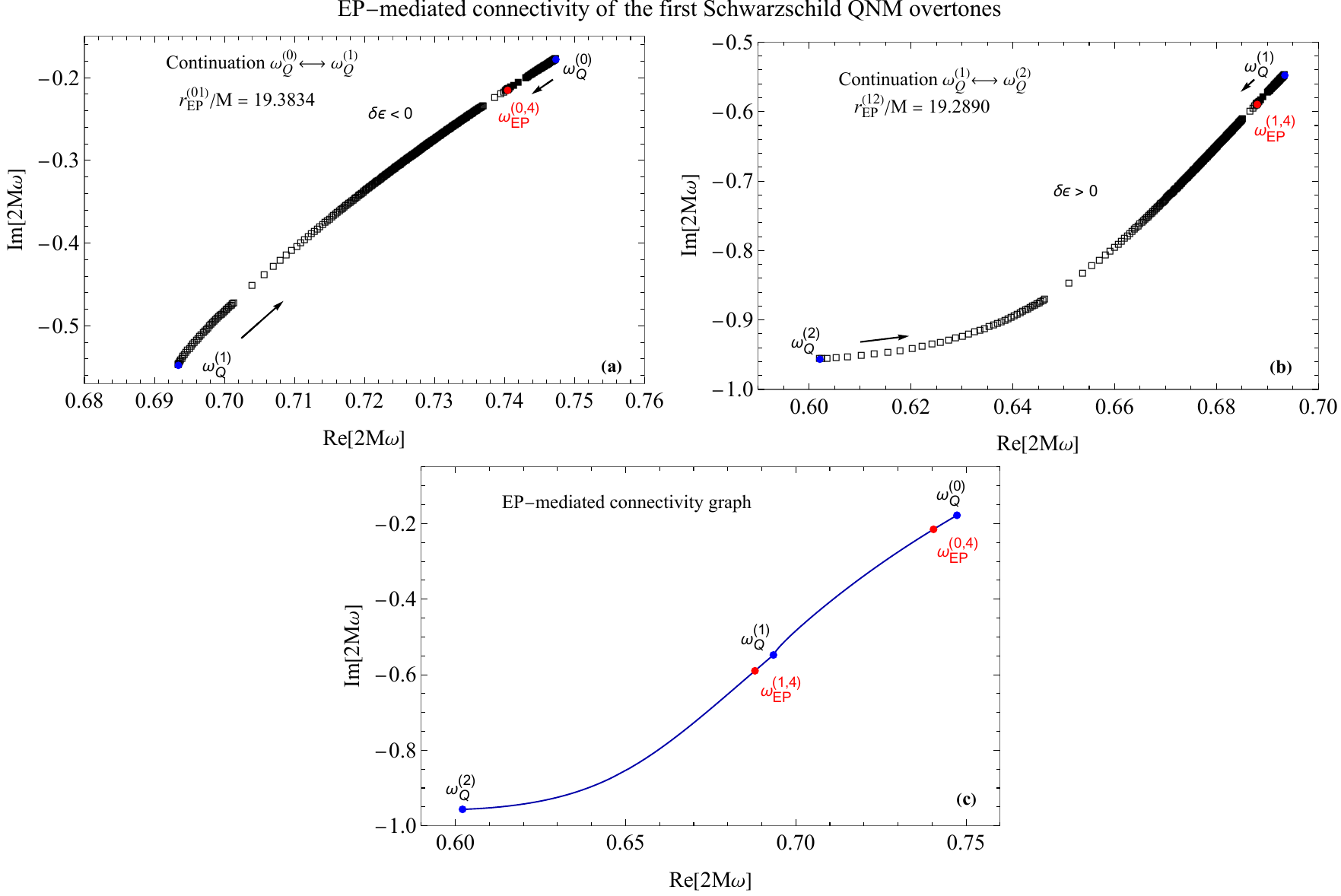}
	\caption{ 
		Exceptional-point-mediated connectivity of the lowest Schwarzschild QNMs. Panel (a) shows the two resonance branches connecting the unperturbed Schwarzschild frequencies \(\omega_Q^{(0)}\) and \(\omega_Q^{(1)}\) through \({\rm EP}_{0}^{(4)}\). Panel (b) shows the corresponding connection between \(\omega_Q^{(1)}\) and \(\omega_Q^{(2)}\) through \({\rm EP}_{1}^{(4)}\). The red points denote the exceptional frequencies, while the blue points denote the unperturbed QNM frequencies reached in the limit \(\epsilon\rightarrow0\). Panel (c) summarizes the resulting spectral connectivity graph. The integers \(0,1,2\) label the unperturbed Schwarzschild overtones. The global continuation establishes that the \(n=0\) family connects \(Q_0\) and \(Q_1\), while the \(n=1\) family connects \(Q_1\) and \(Q_2\).
	}
	\label{fig:EPMediatedSpectralConnectivity}
\end{figure*}

The continuation displayed in Fig.~\ref{fig:EPMediatedSpectralConnectivity} establishes the sequence
\begin{equation}
	\omega_Q^{(0)} \longleftrightarrow \omega_{\rm EP}^{(0,4)} \longleftrightarrow \omega_Q^{(1)} \longleftrightarrow \omega_{\rm EP}^{(1,4)} \longleftrightarrow \omega_Q^{(2)} .
	\label{eq:SpectralConnectivityChain}
\end{equation}
Thus, exceptional points do not merely describe local collisions between two otherwise unidentified resonance roots. They provide global connections between neighboring branches of the Schwarzschild overtone spectrum. Related exceptional-point-mediated connections across overtone spectra have recently been uncovered in the Kerr spectrum~\cite{Cavalcante:2025abr}.

To quantify this spectral migration, we return to the \((n=0)\) family and consider the arclength traveled by each resonance branch in the dimensionless complex-frequency plane.
For the branch connecting the \(n\)-th Schwarzschild mode to \({\rm EP}_{0}^{(k)}\), we introduce the path parameter
\begin{equation}
	\epsilon(\lambda) = \lambda\epsilon_{\rm EP}^{(0,k)}, \qquad 0\leq\lambda\leq1,
\end{equation}
and define
\begin{equation}
	L_n^{(k)} = \int_0^1 \left| \frac{d\widehat{\omega}_n}{d\lambda} \right| d\lambda, \qquad \widehat{\omega}=2M\omega .
	\label{eq:SpectralArclength}
\end{equation}
This definition is independent of the sign of \(\epsilon_{\rm EP}^{(0,k)}\). Numerically, the integral is evaluated as the polygonal arclength of the ordered frequency data,
\begin{equation}
	L_n^{(k)} \simeq \sum_{j=1}^{N-1} \left| \widehat{\omega}_{n,j+1} - \widehat{\omega}_{n,j} \right|.
	\label{eq:PolygonalSpectralArclength}
\end{equation}

For the reference point \({\rm EP}_{0}^{(4)}\), the first-overtone branch travels
\begin{equation}
	L_1^{(4)} \simeq 0.3361 \label{eq:FirstOvertoneTravelK4}
\end{equation}
between \(\epsilon=0\) and \(\epsilon=\epsilon_{\rm EP}^{(0,4)}\). By comparison, the fundamental branch travels only
\[
L_0^{(4)} \simeq 0.03785,
\]
and hence
\begin{equation}
	\frac{L_1^{(4)}}{L_0^{(4)}} \simeq 8.88.
	\label{eq:LengthRatioK4}
\end{equation}
The net endpoint displacement of the first-overtone branch is
\begin{equation}
	\left| \widehat{\omega}_{\rm EP}^{(0,4)} - \widehat{\omega}_Q^{(1)} \right|
	\simeq 0.3360,
	\label{eq:EndpointDisplacementK4}
\end{equation}
consistent with the computed arclength at the quoted precision. The first overtone therefore undergoes a migration in frequency space almost nine times larger than that of the fundamental mode before the two branches coalesce.

\begin{figure}[htbp]
	\centering
	\includegraphics[width=0.97\linewidth]{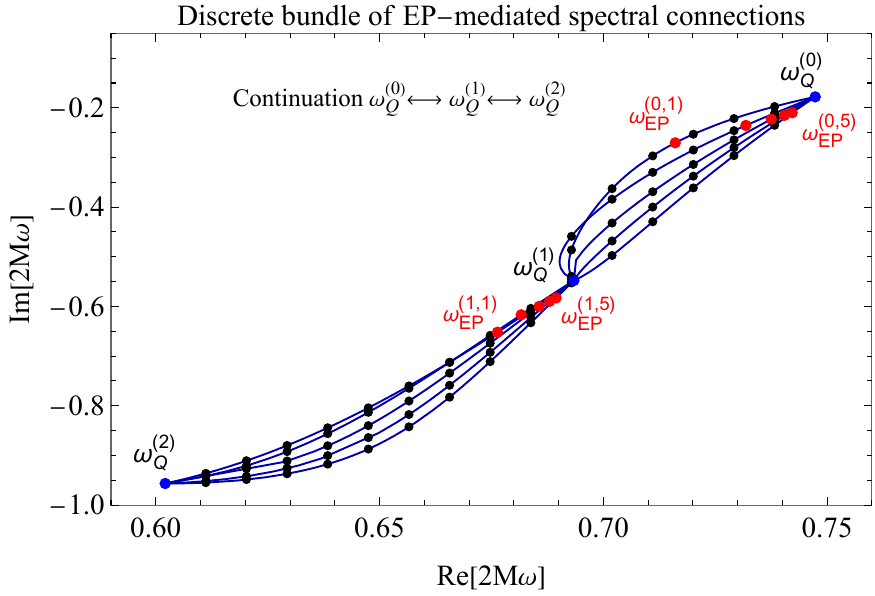}
	\caption{
			Discrete bundles of exceptional-point-mediated spectral connections between the first three Schwarzschild QNM overtones. The upper bundle shows the first five members of the \((n = 0)\) family, while the lower bundle shows the first five members of the \((n = 1)\) family. In both sequences, the signs of the critical couplings alternate, while their magnitudes decrease as the perturbation is moved outward. Each exceptional point provides a distinct spectral bridge between the same pair of neighboring Schwarzschild QNM branches. Taken together, the two families display the connectivity \(\omega_Q^{(0)}\leftrightarrow\omega_Q^{(1)} \leftrightarrow\omega_Q^{(2)}\). Only the first and fifth exceptional points of each family are labeled for clarity.
	}
	\label{fig:DiscreteEPBundle}
\end{figure}

For the \(n = 0\) family shown in the upper bundle of Fig.~\ref{fig:DiscreteEPBundle}, this asymmetry becomes progressively stronger along the sequence. For the five representative exceptional points, the ratio \(L_1^{(k)}/L_0^{(k)}\) increases from approximately \(2.84\) to \(10.55\). For the \(k=5\) connection, for example,
\begin{equation}
	L_0^{(5)} \simeq 0.03237, \qquad L_1^{(5)} \simeq 0.34149.
	\label{eq:SpectralLengthsK5}
\end{equation}

At the same time, the total bridge length,
\begin{equation}
	L_{\rm bridge}^{(k)} = L_0^{(k)}+L_1^{(k)},
	\label{eq:BridgeLength}
\end{equation}
changes by less than one percent across the five displayed connections, decreasing only from approximately \(0.3768\) to \(0.3739\). The total spectral distance bridged by the exceptional point is therefore almost unchanged, while an increasing fraction of that distance is traveled by the first overtone.

The sensitivity is even more pronounced when the migration is normalized by the magnitude of the critical coupling. Across the same five representative exceptional points,
\begin{equation}
	\frac{L_1^{(k)}} {\left|\epsilon_{\rm EP}^{(0,k)}\right|}
\end{equation}
increases from approximately \(3.54\) to \(4.83\times10^2\). Thus, progressively smaller localized perturbations produce a finite spectral migration of the first-overtone branch, corresponding to an increasingly large displacement per unit perturbation strength.

A second manifestation of the spectral sensitivity is the change in orientation of the resonance trajectories as the coupling passes through the exceptional point. We characterize this orientation by the argument of the complex separation vector
\begin{equation}
	\Delta\widehat{\omega}_{\rm sep} = \widehat{\omega}_+ - \widehat{\omega}_- ,
	\label{eq:SeparationVector}
\end{equation}
This quantity characterizes a spectral orientation in the complex-frequency plane and not a physical phase of the perturbing field.

The values \(\epsilon=0\) and \(\epsilon=2\epsilon_{\rm EP}\) lie on opposite sides of the exceptional point and are symmetrically placed about \(\epsilon=\epsilon_{\rm EP}\):
\begin{equation}
	\begin{aligned}
		\epsilon=0
        &\Longrightarrow
		\delta\epsilon=-\epsilon_{\rm EP},
		\\
		\epsilon=2\epsilon_{\rm EP}
		&\Longrightarrow
		\delta\epsilon=+\epsilon_{\rm EP}.
	\end{aligned}
	\label{eq:SymmetricDetunings}
\end{equation}

As the coupling is continued from the Schwarzschild limit through \(\epsilon_{\rm EP}\) and toward \(2\epsilon_{\rm EP}\), the trajectory therefore changes from the negative-\(\delta\epsilon\) direction to the positive-\(\delta\epsilon\) direction.

The quadratic exceptional-point expansion shows that the two limiting directions are orthogonal. Indeed, changing the sign of \(\delta\epsilon\) multiplies the leading square-root displacement by \(i\), up to the choice of sheet, and hence rotates its direction by \(\pi/2\).

This orthogonality is confirmed numerically. At \(\lvert\delta\epsilon\rvert=10^{-7}\), the orientations of the two separation vectors are
\begin{equation}
	\theta_- = 79.736^\circ  \qquad\text{and}\qquad \theta_+ = -10.263^\circ.
\end{equation}
Their relative orientation is therefore
\begin{equation}
	\Delta\theta = 89.999^\circ = \frac{\pi}{2} \quad \text{within numerical accuracy}.
	\label{eq:OrthogonalSpectralDirections}
\end{equation}

A third manifestation of the exceptional-point structure is the failure of a regular Taylor expansion of the QNM frequency in powers of \(\epsilon\). Starting from the Schwarzschild limit, one may formally write
\begin{equation}
	\omega_n(\epsilon) = \omega_Q^{(n)} + c_1^{(n)}\epsilon + c_2^{(n)}\epsilon^2 +\cdots.
	\label{eq:QNMTaylorExpansion}
\end{equation}
Such an expansion may describe the initial displacement of an isolated QNM for sufficiently small \(\epsilon\). In Ref.~\cite{Torres:2026uey}, it was noted that the Taylor expansion provides a poor fit to the trajectory for \(|\epsilon|\gtrsim\epsilon_{\rm nonlin}\), where \(\epsilon_{\rm nonlin}\) is defined as the value of \(\epsilon\) at which the linear and quadratic correction terms in Eq.~\eqref{eq:QNMTaylorExpansion} become equal in magnitude. It is
\begin{equation}
 \epsilon_{nonlin} = \left| \frac{2 (F_\omega)^2}{F_{\omega\omega}} \right|, 
 \label{eq:epsilon-nonlin}
\end{equation}
where the derivatives are evaluated at the unperturbed QNM frequency. In Ref.~\cite{Torres:2026uey} (see Fig.~7) $\epsilon_{nonlin}$ was found to decay exponentially with the overtone number $n$, and this was called \emph{nonlinear instability}. Here we have shown that the critical value $\epsilon_{\rm EP}$ for an EP also has an exponential scaling (Fig.~\ref{fig:EPAsymptotics}c), and that the EP sequences accumulate toward the corresponding unperturbed QNM frequencies (see Fig.~\ref{fig:RPbranch} and Eq.~\eqref{eq:RepellingPointAccumulation}).

The dynamical-systems perspective gives a complementary explanation. At fixed $r_0$, the trajectory is governed by Eq.~\eqref{eq:SpectralFlowEquation}. At the exceptional point,
\begin{equation}
	F_\omega \bigl(\omega_{\rm EP};r_{\rm EP}\bigr) = 0,
\end{equation}
and so \(d\omega/d\epsilon\) diverges. The Taylor expansion constructed about the Schwarzschild limit, Eq.~\eqref{eq:QNMTaylorExpansion}, therefore breaks down as \(\epsilon\rightarrow\epsilon_{\rm EP}\) and cannot describe the coalescence of the two resonance branches. This breakdown is consistent with the square-root branching in Eq.~\eqref{eq:EpsilonSquareRootUnfolding}.

The large distance traveled by the first overtone, the orthogonal reorientation of the spectral trajectories, and the breakdown of the regular Taylor expansion are complementary manifestations of the spectral sensitivity generated by the exceptional point. At the same time, global continuation shows that the exceptional points provide bridges between neighboring Schwarzschild QNM overtones.

%==================================================================================================================

\subsection{Excitation factors and lemniscates}
\label{subsec:ExcitationFactorsLemniscates}

The resonance trajectories provide a direct representation of the exceptional-point structure in the complex-frequency plane. A complementary diagnostic is supplied by the excitation factors
associated with the coalescing poles, which have played a central role in recent studies of resonant QNM excitation and avoided crossings~\cite{Berti:2006wq,Motohashi:2024fwt,Nakamoto:2026lyo,
Kubota:2025hjk}. Near an exceptional point, the two excitation factors become large and acquire opposite leading contributions. Their singular behavior is the residue counterpart of the square-root splitting of the resonance frequencies.

The excitation factor $\mathfrak{B}_j$ of a quasinormal mode of frequency $\omega_j$ is defined in terms of the derivative of the Wronskian \cite{Leaver:1986gd} as
\begin{equation}
\mathfrak{B}_j = \frac{\ii A^{(+)}(\omega_j)}{\left. \partial_\omega W \right|_{\omega_j}}
\end{equation}
This definition can be straightforwardly extended to apply to perturbed resonances.

\subsubsection{Excitation-factor amplification}

For our two-parameter model, the exact perturbed Wronskian factorizes as \(\mathcal W=\mathcal P D\), where \(D=F-\epsilon\) and \(\mathcal P= \phi_{\ell\omega}^{\rm in}(r_{0*}) \phi_{\ell\omega}^{\rm up}(r_{0*})\). At a simple perturbed resonance frequency \(\omega_j\), one has \(D(\omega_j,\epsilon;r_0)=0\). Differentiation with respect to the
frequency therefore gives
\begin{equation}
	\left.
	\frac{\partial \mathcal W}{\partial \omega} \right|_{\omega=\omega_j} = \mathcal P(\omega_j;r_0) F_\omega(\omega_j;r_0).
	\label{eq:WronskianDerivativeAtPole}
\end{equation}
The term proportional to \(\partial_\omega\mathcal P\) vanishes at the pole because it is multiplied by \(D\). Hence the excitation factor associated with the perturbed pole may be written as
\begin{equation}
	\mathfrak{B}_j =
	\frac{\ii A^{(+)}(\omega_j)}{\mathcal P(\omega_j;r_0)F_\omega(\omega_j;r_0)}.
	\label{eq:PerturbedExcitationFactor}
\end{equation}
For the modes considered below, the outgoing amplitude and the prefactor \(\mathcal P\) remain regular and nonzero at the exceptional point. The singular behavior of \(\mathfrak{B}_j\) is therefore governed by
\(1/F_\omega\).

Close to an EP, we can fix \(r_0=r_{\rm EP}\) and set \(\delta\epsilon=\epsilon-\epsilon_{\rm EP}\). The square-root splitting is given by Eq.~\eqref{eq:EpsilonSquareRootUnfolding}. It follows that
\begin{equation}
	F_\omega(\omega_\pm;r_{\rm EP}) \sim \pm
	\left( 2F_{\omega\omega}^{\rm EP} \delta\epsilon \right)^{1/2}.
	\label{eq:FomegaSquareRootExcitation}
\end{equation}
Substitution into Eq.~\eqref{eq:PerturbedExcitationFactor} then yields
\begin{equation}
	\mathfrak{B}_\pm \sim \pm \frac{\ii A^{(+)}(\omega_{\rm EP})}{\mathcal P(\omega_{\rm EP};r_{\rm EP})} \left( 2F_{\omega\omega}^{\rm EP} \delta\epsilon\right)^{-1/2},
	\label{eq:ExcitationInverseSquareRoot}
\end{equation}
where the prefactor is finite and nonzero. Consequently,
\begin{equation}
	\left|\mathfrak{B}_\pm\right| \propto \left|\delta\epsilon\right|^{-1/2}.
	\label{eq:ExcitationScaling}
\end{equation}
The two excitation factors therefore diverge with opposite leading signs as the two resonance frequencies coalesce.

The same behavior may be expressed in terms of the frequency splitting. Writing \(\omega_\pm=\omega_{\rm EP}\pm\delta\omega\), one has \(\Delta\omega=\omega_+-\omega_-=2\delta\omega\). The leading
excitation factors become
\begin{equation}
	\mathfrak{B}_\pm \simeq \pm \frac{\mathcal K_{\rm EP}}{\delta\omega},
	\label{eq:ExcitationInverseFrequency}
\end{equation}
where
\begin{equation}
	\mathcal K_{\rm EP} = \frac{\ii A^{(+)}(\omega_{\rm EP})}{\mathcal P(\omega_{\rm EP};r_{\rm EP})F_{\omega\omega}^{\rm EP}}
	\label{eq:KEPDefinition}
\end{equation}
which is finite.
The excitation-factor difference \(\Delta\mathfrak{B}=\mathfrak{B}_+-\mathfrak{B}_-\) therefore satisfies
\begin{equation}
	\left|\Delta\mathfrak{B}\right| \propto\left|\Delta\omega\right|^{-1}.
	\label{eq:DeltaBInverseSplitting}
\end{equation}
The residue amplification is thus the inverse counterpart of the vanishing frequency separation.

The numerical results for \({\rm EP}_{0}^{(4)}\) confirm these local predictions. A combined log--log fit of the two branches gives
\begin{equation}
	\left|\mathfrak{B}_\pm\right| \propto \left|\delta\epsilon\right|^{-0.5015},
	\label{eq:CombinedExcitationFit}
\end{equation}
in excellent agreement with the theoretical exponent \(-1/2\). Separate fits give exponents \(-0.4896\) and \(-0.5134\) for the two branches. Moreover,
\begin{equation}
	\left|\Delta\mathfrak{B}\right| \propto \left|\Delta\omega\right|^{-0.9934},
	\label{eq:DeltaBNumericalFit}
\end{equation}
consistent with the inverse scaling predicted by Eq.~\eqref{eq:DeltaBInverseSplitting}.

The exceptional point is therefore encoded not only in the trajectories of the resonance frequencies, but also in their residues. The square-root coalescence of the two poles is accompanied by an inverse square-root amplification of their excitation factors.

\subsubsection{Lemniscate structure near avoided crossings}
\label{subsubsec:ExcitationLemniscates}

The preceding analysis concerns the real exceptional point itself, for which \(r_0=r_{\rm EP}\). We now displace the perturbation location slightly away from its critical value. The branch point can then no longer be reached by varying a real coupling, and the divergence of the excitation factors is regularized into a two-lobed trajectory in the complex \(\mathfrak{B}\)-plane.

\begin{figure*}[htbp]
	\centering
	\includegraphics[width=0.65\linewidth]{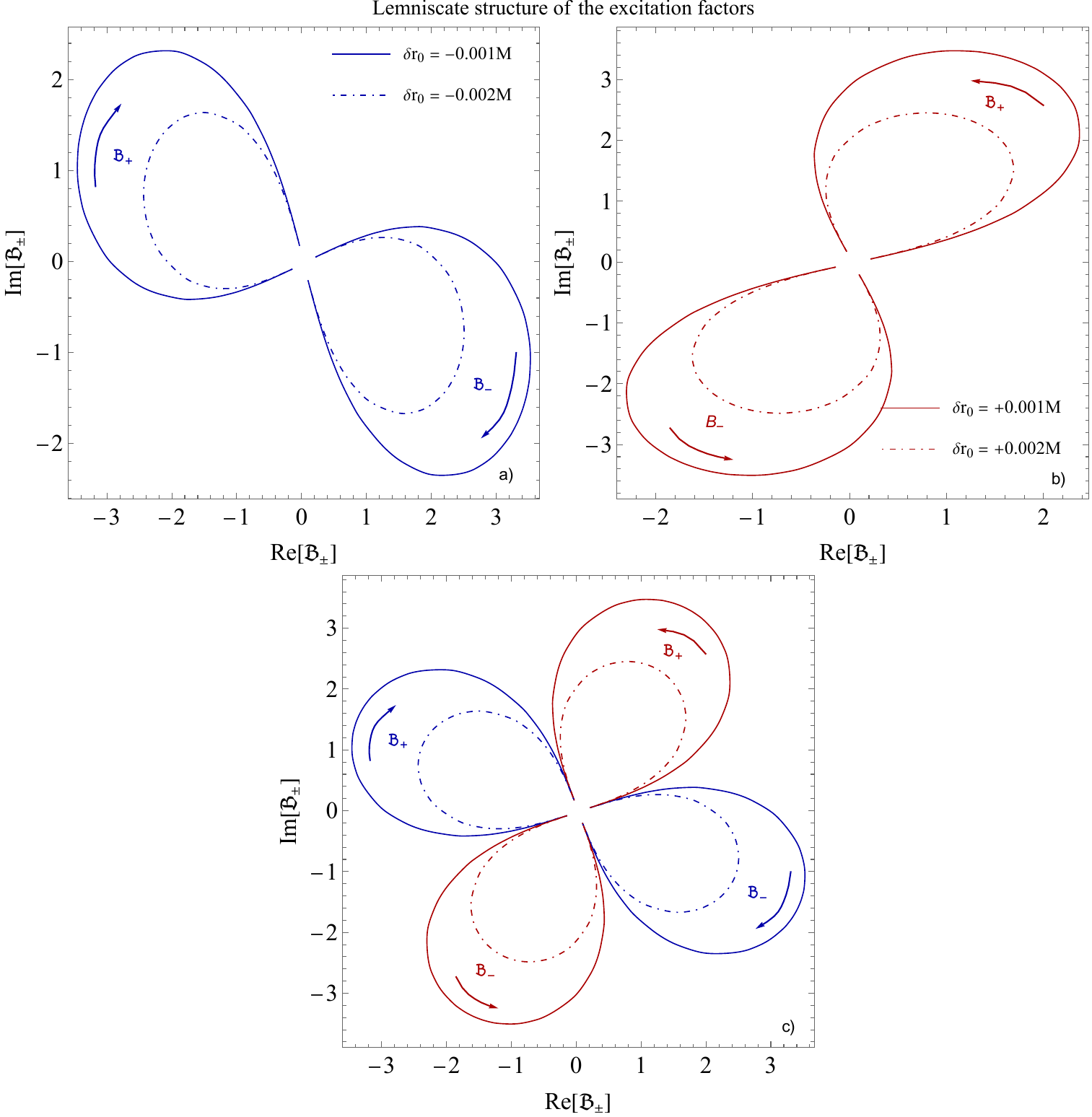}
	\caption{
		Lemniscate structure of the excitation factors near \({\rm EP}_{0}^{(4)}\). The perturbation locations are \(r_0=r_{\rm EP}^{(0,4)}+\delta r_0\), with \(\delta r_0=\pm10^{-3}M\) and \(\delta r_0=\pm2\times10^{-3}M\). For each \(r_0\), the corresponding repelling point \(\omega_r(r_0)\) satisfies \(F_\omega(\omega_r;r_0)=0\), and the real coupling is varied as \(\epsilon=\epsilon_c(r_0)+\Delta\epsilon\), where \(\epsilon_c(r_0)=\operatorname{Re}F(\omega_r;r_0)\). Thus, \(\Delta\epsilon\) is measured from the local avoided-crossing
		center rather than from \(\epsilon_{\rm EP}\). Panels (a) and (b) show the complex trajectories of the full excitation factors for negative and positive radial offsets, respectively. Solid curves correspond to \(\lvert\delta r_0\rvert=10^{-3}M\), and dot-dashed curves to \(\lvert\delta r_0\rvert=2\times10^{-3}M\).
		Panel (c) superposes the two signs of \(\delta r_0\), producing the four-leaf structure. The arrows indicate increasing \(\Delta\epsilon\). The smaller loops farther from the exceptional point reflect the weaker excitation-factor amplification associated with the larger local pole separation.
	}
	\label{fig:BLemniscate}
\end{figure*}

We consider the avoided-crossing configuration described locally by Eq.~\eqref{eq:LocalRootsRepellingPoint}. At a fixed perturbation location \(r_0\) close to \(r_{\rm EP}\), the corresponding repelling point \(\omega_r(r_0)\) satisfies Eq.~\eqref{eq:RepellingPointCondition}.
Away from the real exceptional point, we write \(F(\omega_r;r_0)=\epsilon_c(r_0)+\ii\eta(r_0)\), where \(\epsilon_c(r_0)=\operatorname{Re}F(\omega_r;r_0)\) and \(\eta(r_0)=\operatorname{Im}F(\omega_r;r_0)\neq0\). The real perturbation strength is parametrized as \(\epsilon=\epsilon_c(r_0)+\Delta\epsilon\), with \(\Delta\epsilon\in\mathbb R\).

The local expansion around the repelling point gives
\begin{equation}
	\omega_\pm-\omega_r \simeq \pm
	\left[\frac{2(\Delta\epsilon-\ii\eta)}{F_{\omega\omega}^{r}}\right]^{1/2},
	\label{eq:AvoidedCrossingBranchesExcitation}
\end{equation}
where
\(F_{\omega\omega}^{r} = F_{\omega\omega}(\omega_r;r_0)\). Since \(F_\omega(\omega_\pm;r_0) \simeq F_{\omega\omega}^{r}(\omega_\pm-\omega_r)\), the excitation factors take the local form
\begin{equation}
	\mathfrak{B}_\pm \simeq \pm \mathcal K_r(\Delta\epsilon-\ii\eta)^{-1/2},
	\label{eq:LemniscateExcitationFactor}
\end{equation}
where \(\mathcal K_r\) is a regular complex prefactor. Its modulus rescales the trajectory in the complex \(\mathfrak{B}\)-plane, whereas its argument produces an overall rotation.

The geometrical origin of the two-lobed structure becomes transparent after introducing the locally normalized excitation factor \(b_\pm=\mathfrak{B}_\pm/\mathcal K_r\). Equation~\eqref{eq:LemniscateExcitationFactor} then implies
\begin{equation}
	b_\pm^{-2} \simeq \Delta\epsilon-\ii\eta .
	\label{eq:InverseSquareNormalizedB}
\end{equation}
Writing \(b=U+\ii V\), one finds
\begin{equation}
	b^{-2} = \frac{U^2-V^2}{(U^2+V^2)^2}-\ii\frac{2UV}{(U^2+V^2)^2}.
	\label{eq:InverseSquareCartesianB}
\end{equation}
Because \(\Delta\epsilon\) is real, comparison of the imaginary parts gives
\begin{equation}
	2UV = \eta (U^2+V^2)^2.
	\label{eq:ExcitationLemniscateEquation}
\end{equation}
Introducing polar coordinates \(U=R\cos\varphi\) and \(V=R\sin\varphi\), this becomes
\begin{equation}
	R^2 =\frac{\sin(2\varphi)}{\eta},
	\label{eq:ExcitationLemniscatePolar}
\end{equation}
which is the polar equation of a Bernoulli lemniscate, up to a rotation determined by the sign of \(\eta\) and the overall rotation and rescaling generated by \(\mathcal K_r\).

At the real exceptional point, \(\eta=0\), and the singular behavior of Eq.~\eqref{eq:ExcitationInverseSquareRoot} is recovered. For \(r_0\neq r_{\rm EP}\), the branch point is displaced from the real-\(\epsilon\) axis by the imaginary offset \(\eta\). The divergence is then regularized, and at \(\Delta\epsilon=0\),
\begin{equation}
	\left|\mathfrak{B}_\pm\right| \sim \left|\mathcal K_r\right| \left|\eta\right|^{-1/2}.
	\label{eq:RegularizedExcitationMaximum}
\end{equation}
The excitation factors remain large when \(r_0\) is close to \(r_{\rm EP}\), but they remain finite. The lemniscate is therefore the finite-\(\eta\) remnant of the excitation-factor singularity at the
exceptional point.

Figure~\ref{fig:BLemniscate} confirms the local prediction. On either side of the exceptional point, the two excitation factors trace a two-lobed curve as the real coupling is varied through the avoided crossing. Since \(\eta(r_0)\) changes sign across \(r_{\rm EP}\), the corresponding lemniscates are rotated in opposite directions. Their superposition produces the four-leaf structure displayed in the
combined panel.

The lemniscate geometry therefore provides a residue-space counterpart of the hyperbolic avoided crossings in the frequency plane. Both structures are controlled by the same complex square-root factor:
\((\Delta\epsilon-\ii\eta)^{1/2}\) governs the resonance splitting, whereas its inverse governs the excitation-factor amplification. Related lemniscate structures in QNM excitation have been found in
studies of avoided crossings and nearly coalescing resonances~\cite{Motohashi:2024fwt,Nakamoto:2026lyo}.

%==================================================================================================================================

\subsection{Adiabatic invariants}
\label{subsec:AdiabaticInvariants}

The inverse relation between the excitation-factor difference and the frequency splitting suggests that a finite spectral quantity survives in the exceptional-point limit. We define
\begin{equation}
	\mathcal I(\epsilon) = \Delta\mathfrak{B}(\epsilon)\, \Delta\omega(\epsilon),
	\label{eq:AdiabaticInvariantDefinition}
\end{equation}
where \(\Delta\mathfrak{B}=\mathfrak{B}_+-\mathfrak{B}_-\) and \(\Delta\omega=\omega_+-\omega_-\).

To determine the behavior of \(\mathcal I\) near the exceptional point, we write \(\omega_\pm=\omega_{\rm EP}\pm\delta\omega\), so that \(\Delta\omega=2\delta\omega\). Equation~\eqref{eq:ExcitationInverseFrequency} gives
\begin{equation}
	\Delta\mathfrak{B} \simeq
	\frac{2\mathcal K_{\rm EP}}{\delta\omega}.
	\label{eq:DeltaBInvariant}
\end{equation}
Multiplication by the frequency separation then gives
\begin{equation}
	\mathcal I(\epsilon) = \Delta\mathfrak{B}\,\Delta\omega
	\simeq 4\mathcal K_{\rm EP}.
	\label{eq:AdiabaticInvariantLimit}
\end{equation}
Consequently,
\begin{equation}
	\lim_{\epsilon\rightarrow\epsilon_{\rm EP}} \Delta\mathfrak{B}\,\Delta\omega = \mathcal I_{\rm EP} \equiv 4\mathcal K_{\rm EP},
	\label{eq:AdiabaticInvariantEP}
\end{equation}
The limiting quantity is finite even though \(\Delta\omega\rightarrow0\) and \(\lvert\Delta\mathfrak{B}\rvert\rightarrow\infty\).

The product \(\mathcal I\) is also invariant under an exchange of the two local sheets. Under the relabeling \(+\leftrightarrow-\), both \(\Delta\omega\) and \(\Delta\mathfrak{B}\) change sign, whereas their product is unchanged. The limiting value \(\mathcal I_{\rm EP}\) is therefore independent of the arbitrary local labeling of the two resonance branches.

We refer to \(\mathcal I\) as a local adiabatic invariant in the spectral sense. Under continuous variation of the perturbation parameter through the local exceptional-point regime, the frequency splitting vanishes while the excitation-factor difference diverges, but their product approaches the finite value \(\mathcal I_{\rm EP}\). The term ``adiabatic'' refers here solely to continuous parameter variation and does not imply a time-dependent adiabatic evolution of the black-hole background.

\begin{figure}[htbp]
	\centering
	\includegraphics[width=0.95\linewidth]{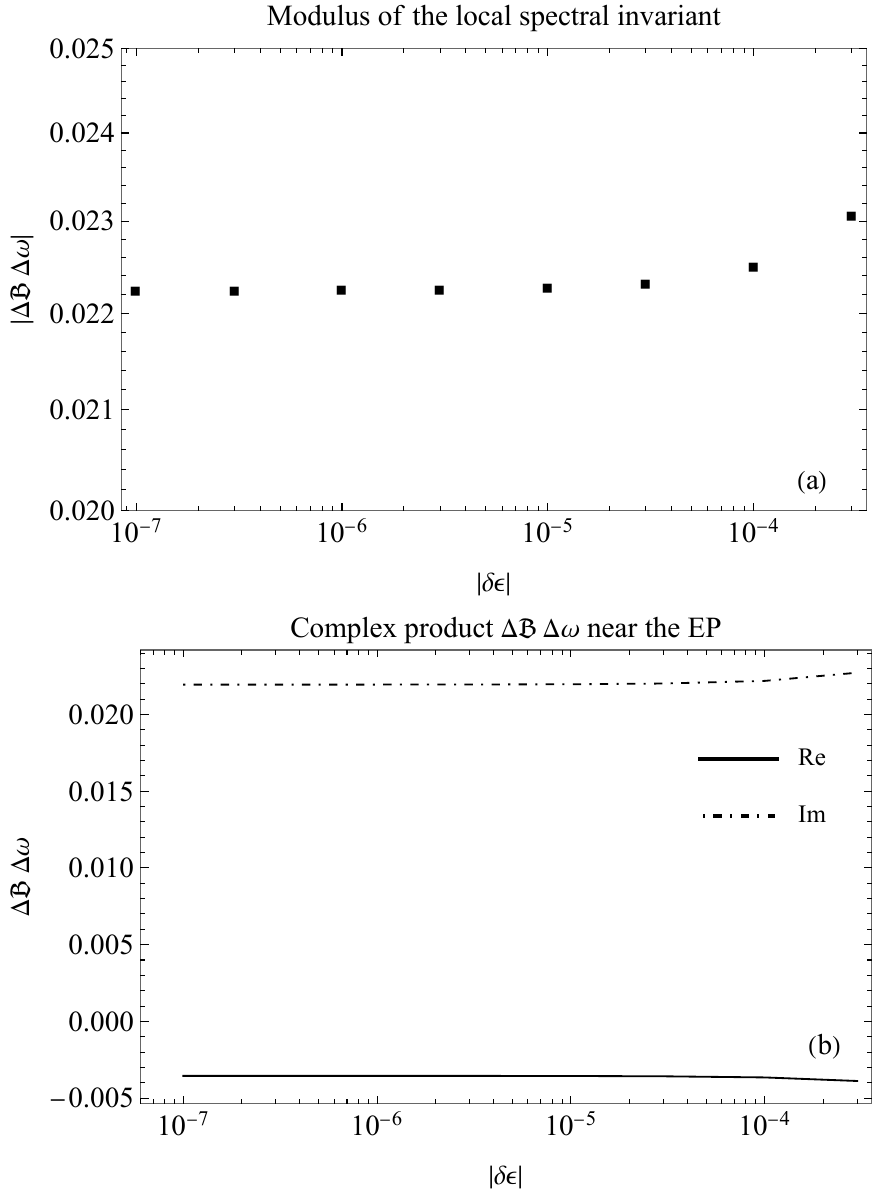}
\caption{
	Local spectral invariant near \({\rm EP}_{0}^{(4)}\), shown as a function of \(\lvert\delta\epsilon\rvert =\lvert\epsilon-\epsilon_{\rm EP}\rvert\) with a logarithmic scale on the horizontal axis. 
	Panel (a) shows the numerical data points for the modulus \(\lvert\Delta\mathfrak{B}\,\Delta\omega\rvert\). Its nearly constant value close to the exceptional point demonstrates that the divergence of the 
	excitation-factor difference compensates the vanishing frequency separation. Panel (b) shows, as dot-dashed curves, the real and imaginary parts of the complex product \(\Delta\mathfrak{B}\,\Delta\omega\). 
	Both approach finite limiting values as \(\lvert\delta\epsilon\rvert\to0\), confirming that the complete complex product, rather than only its modulus, becomes constant in the exceptional-point 
	limit. The residual variation at larger detuning arises from higher-order terms in the resonance function and from the frequency dependence of the regular prefactor.
}
	\label{fig:AdiabaticInvariant}
\end{figure}

The numerical results for \({\rm EP}_{0}^{(4)}\) confirm this prediction. Figure~\ref{fig:AdiabaticInvariant}(a) shows that
\begin{equation}
	\left|\Delta\mathfrak{B}\,\Delta\omega\right| \simeq 2.2\times10^{-2}
	\label{eq:AdiabaticInvariantNumericalModulus}
\end{equation}
close to the exceptional point. The nearly horizontal sequence demonstrates that the divergence of \(\Delta\mathfrak{B}\) compensates the vanishing of \(\Delta\omega\).

Figure~\ref{fig:AdiabaticInvariant}(b) shows that the real and imaginary parts of \(\Delta\mathfrak{B}\,\Delta\omega\) separately approach finite limiting values as \(\lvert\delta\epsilon\rvert\rightarrow0\). It is therefore the complete complex quantity, rather than only its modulus, that becomes constant in the exceptional-point limit. The residual variation at larger detuning is caused by higher-order terms in the resonance function and by the frequency dependence neglected in the leading local approximation.

The approximate constancy of \(\Delta\mathfrak{B}\,\Delta\omega\) expresses a local balance between pole coalescence and residue amplification. The square-root decrease of the frequency separation is compensated, at leading order, by the inverse square-root increase of the excitation-factor difference. The resulting finite complex quantity \(\mathcal I_{\rm EP}\) provides a local spectral descriptor of the coalescing resonance pair that remains well defined in the exceptional-point limit.

This finite combination enters directly into the coalescence limit of the intrinsic two-pole response, Eq.~\eqref{eq:DoublePoleResponse}, where the individually divergent excitation factors combine to produce a finite result. We first turn, however, to the level sets of the spectral function and their Cassini-oval geometry.

%===========================================================================================================

\subsection{Level sets and Cassini ovals}
\label{subsec:LevelSetsCassiniOvals}

The local square-root structure of an exceptional point also appears in the level sets of the spectral function. In the strict quadratic approximation, these level sets are exact Cassini ovals whose foci are the two local resonance roots. The corresponding critical level is a Bernoulli lemniscate passing through the exceptional point. We first describe this local geometry and then compare it with the level sets of the full numerical spectral function.

\subsubsection{Local quadratic geometry}
\label{subsubsec:LocalQuadraticCassini}

At fixed \(r_0=r_{\rm EP}\), we define \(\delta\epsilon=\epsilon-\epsilon_{\rm EP}\). Equation~\eqref{eq:QuadraticNormalFormEP} gives
\begin{equation}
	D_{\rm loc}(\omega,\epsilon) = \frac{1}{2} F_{\omega\omega}^{\rm EP} (\omega-\omega_{\rm EP})^2 - \delta\epsilon.
	\label{eq:LocalSpectralFunctionCassini}
\end{equation}
The local spectral function consequently factorizes as
\begin{equation}
	D_{\rm loc}(\omega,\epsilon) =\frac{1}{2} F_{\omega\omega}^{\rm EP} (\omega-\omega_+^{\rm loc})(\omega-\omega_-^{\rm loc}),
	\label{eq:LocalSpectralFunctionFactorised}
\end{equation}
where the two roots are
\begin{equation}
	\omega_\pm^{\rm loc} = \omega_{\rm EP} \pm\left(\frac{2\delta\epsilon}{F_{\omega\omega}^{\rm EP}}\right)^{1/2}.
	\label{eq:LocalCassiniRoots}
\end{equation}
It follows that the level sets
\begin{equation}
	\left|D_{\rm loc}(\omega,\epsilon)\right|=\tau\left|\delta\epsilon\right|,
	\label{eq:LocalCassiniLevels}
\end{equation}
satisfy
\begin{equation}
	\left|\omega-\omega_+^{\rm loc}\right|\left|\omega-\omega_-^{\rm loc}\right|=\tau\left|\omega_+^{\rm loc}-\omega_{\rm EP}\right|^2.
	\label{eq:CassiniProductEquation}
\end{equation}
These are Cassini ovals with foci \(\omega_+^{\rm loc}\) and \(\omega_-^{\rm loc}\). Since the two foci are symmetrically placed about \(\omega_{\rm EP}\), the critical value \(\tau=1\) gives the Bernoulli lemniscate, which passes through the exceptional point. For \(0<\tau<1\), the level set consists of two disjoint loops surrounding the two foci; at \(\tau=1\), the loops meet at \(\omega_{\rm EP}\); for \(\tau>1\), they merge into a single Cassini oval.

\begin{figure}[htbp]
	\centering
	\includegraphics[width=0.95\linewidth]{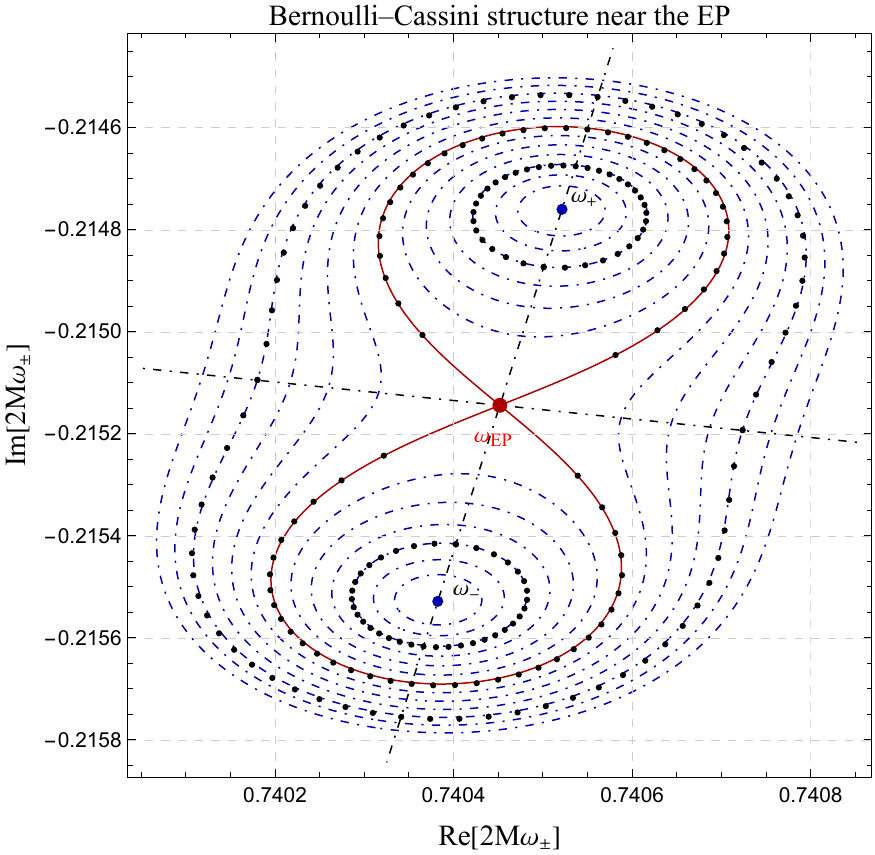}
	\caption{
		Local Cassini-oval geometry generated by the quadratic expansion near \({\rm EP}_{0}^{(4)}\). The perturbation location is fixed at \(r_0=r_{\rm EP}^{(0,4)}\), with \(r_{\rm EP}^{(0,4)}/M\simeq19.383465\), and \(\delta\epsilon=-10^{-7}\). The red point denotes \(\omega_{\rm EP}\). The solid red curve is the critical level \(\lvert D_{\rm loc}\rvert=\lvert\delta\epsilon\rvert\), forming a Bernoulli lemniscate, while the blue dot-dashed curves are neighboring Cassini levels \(\lvert D_{\rm loc}\rvert =\tau\lvert\delta\epsilon\rvert\). The blue points are the analytic local roots and hence the foci of the Cassini ovals; the black markers show the corresponding roots of the full numerical spectral equation. The black dot-dashed lines indicate the two local spectral directions through the exceptional point.
	}
	\label{fig:LocalBernoulliCassini}
\end{figure}

For the detuning considered below, the analytic local roots differ from the roots of the full numerical spectral equation by \(O(10^{-6})\). Figure~\ref{fig:LocalBernoulliCassini} illustrates this local construction for \({\rm EP}_{0}^{(4)}\) at \(\delta\epsilon=-10^{-7}\). At this detuning, the two resonance roots remain within the regime accurately described by the quadratic expansion, which reproduces both their locations and the surrounding level-set geometry on the scale shown.

The two foci in Fig.~\ref{fig:LocalBernoulliCassini} are simply the zeros of \(D_{\rm loc}\) at the chosen detuning, as follows directly from Eq.~\eqref{eq:LocalSpectralFunctionFactorised}.

\subsubsection{Numerical deformation away from the exceptional point}
\label{subsubsec:NumericalCassini}

The exact Cassini geometry follows from the strict quadratic approximation. As the two resonance roots move farther from the exceptional point, cubic and higher-order terms in \(F(\omega;r_{\rm EP})\) progressively deform the level sets. At fixed \(r_0=r_{\rm EP}^{(0,4)}\), we therefore consider the full spectral
function
\begin{equation}
	D(\omega,\epsilon;r_{\rm EP}) = F(\omega;r_{\rm EP}) -\epsilon ,
	\label{eq:FullSpectralFunctionCassini}
\end{equation}
and the corresponding numerical level sets defined by
\begin{equation}
	\left| D\left(\omega,\epsilon_{\rm EP}+\delta\epsilon;r_{\rm EP}\right)\right|=\tau\left|\delta\epsilon\right|.
	\label{eq:FullNumericalCassiniLevels}
\end{equation}

The critical level \(\tau=1\) continues to pass exactly through the exceptional point. Indeed,
\begin{equation}
	D\left(\omega_{\rm EP},\epsilon_{\rm EP}+\delta\epsilon;r_{\rm EP}\right)=-\delta\epsilon ,
	\label{eq:FullSpectralFunctionAtEP}
\end{equation}
and hence \(\omega_{\rm EP}\) belongs identically to the level \(\lvert D\rvert=\lvert\delta\epsilon\rvert\), independently of the higher-order terms in \(F\).

\begin{figure*}[htbp]
	\centering
	\includegraphics[width=0.80\linewidth]{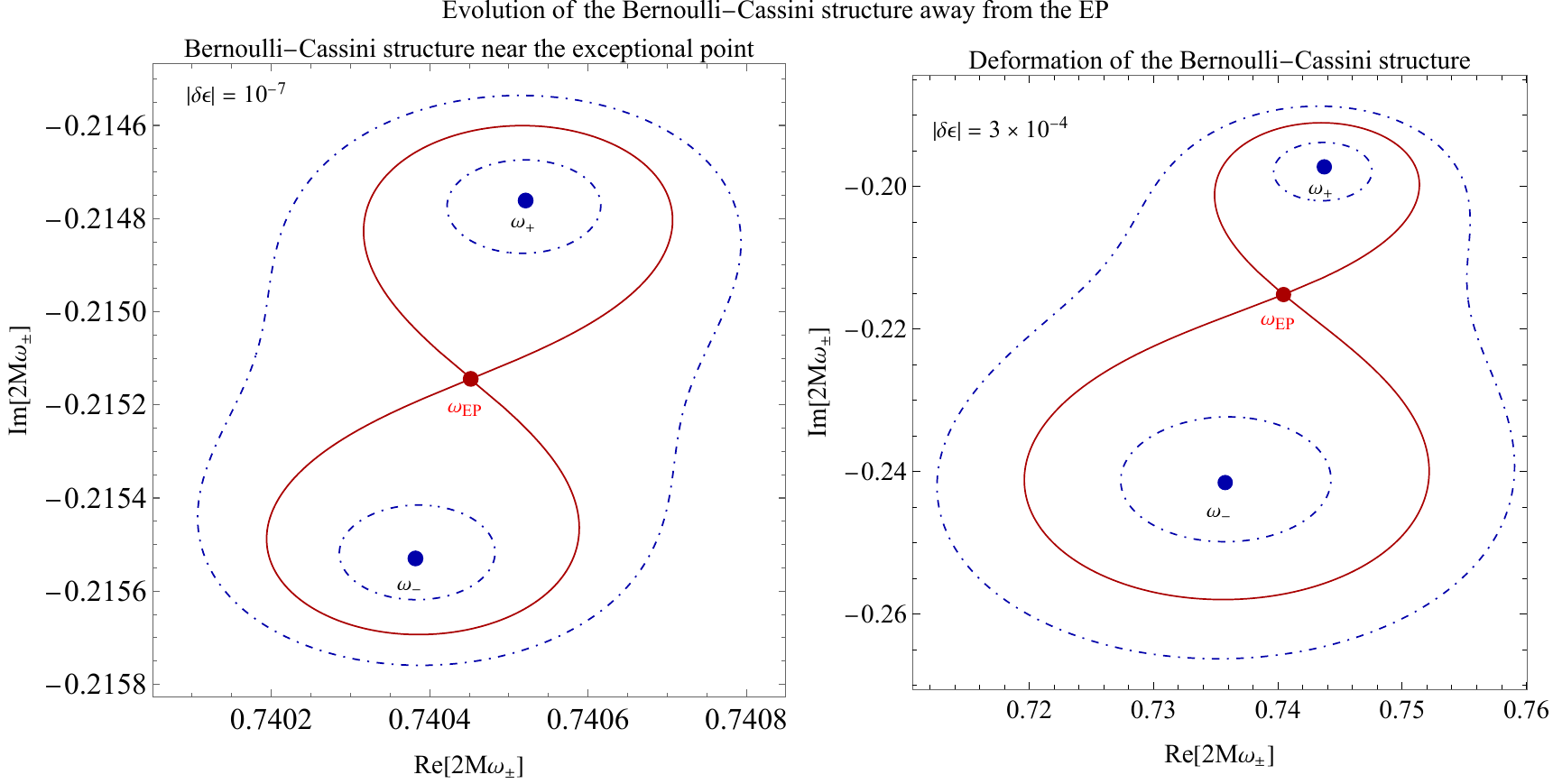}
	\caption{
		Deformation of the local Cassini-oval geometry away from \({\rm EP}_{0}^{(4)}\). The perturbation location is fixed at \(r_0=r_{\rm EP}^{(0,4)}\), with \(r_{\rm EP}^{(0,4)}/M\simeq19.383465\), 
		and \(\epsilon=\epsilon_{\rm EP}^{(0,4)}+\delta\epsilon\). The red point denotes \(\omega_{\rm EP}\), while the blue points denote the two roots \(\omega_\pm\) of the full numerical spectral 
		equation. The solid red curves are the critical levels \(\lvert D\rvert=\lvert\delta\epsilon\rvert\), which pass through \(\omega_{\rm EP}\); they are level sets and not pole trajectories. The 
		blue dot-dashed curves are neighboring levels \(\lvert D\rvert=\tau\lvert\delta\epsilon\rvert\). The left panel corresponds to \(\delta\epsilon=-10^{-7}\), where the quadratic Cassini geometry 
		remains accurate. The right panel corresponds to \(\delta\epsilon=-3\times10^{-4}\), where higher-order terms in the full spectral function produce a pronounced deformation of the lower lobe.
	}
	\label{fig:NumericalBernoulliCassini}
\end{figure*}

Figure~\ref{fig:NumericalBernoulliCassini} illustrates the deformation of this critical geometry as the detuning is increased. For \(\delta\epsilon=-10^{-7}\), the two roots remain close to the exceptional point and the full numerical level sets are nearly indistinguishable from the Cassini ovals of the quadratic model. For \(\delta\epsilon=-3\times10^{-4}\), the roots probe a larger region of
the complex-frequency plane, and the higher-order structure of \(F\) becomes visible through a pronounced deformation of the lower lobe.

The red curves in Fig.~\ref{fig:NumericalBernoulliCassini} are level sets of the full spectral function and should not be
interpreted as trajectories of the resonance poles. At small detuning, the critical level closely follows the local Bernoulli geometry. At larger
detuning, it continues to pass through the exceptional point but is no longer an exact Bernoulli lemniscate. For the detunings shown here, the critical level retains its two-lobed structure, although its two lobes become increasingly asymmetric.

This deformation makes explicit the distinction between local and global spectral geometry. The quadratic approximation imposes the central symmetry
\begin{equation}
	\omega_- - \omega_{\rm EP} \simeq - \left(\omega_+ - \omega_{\rm EP}\right).
	\label{eq:LocalCentralSymmetryCassini}
\end{equation}
The full numerical roots and level sets need not preserve this symmetry away from the exceptional point. Cubic and higher-order terms bend the two branches differently and deform the associated level-set geometry.

The numerical results therefore interpolate between a universal local Cassini geometry and an asymmetric global spectral structure. Close to the exceptional point, the two resonance roots act as the foci of the quadratic Cassini ovals, with the critical level forming an exact Bernoulli lemniscate. Farther from the exceptional point, higher-order terms in the full spectral function progressively deform this local geometry, while the critical level continues to pass exactly through \(\omega_{\rm EP}\).

%=========================================================================================================================

\subsection{Exceptional lines in a three-parameter model}
\label{subsec:ExceptionalLines}

The point-defect model considered in the preceding sections depends on two real parameters: the perturbation location \(r_0\) and its strength \(\epsilon\). Its exceptional points therefore occur at isolated values of \((r_0,\epsilon)\). We now introduce a third parameter by replacing the pointlike defect with a finite-width Gaussian profile. Continuation in the Gaussian width then promotes each isolated point-defect EP to a one-dimensional exceptional line in the enlarged parameter space. Exceptional lines associated with additional continuous parameters have also been identified in related black-hole QNM models~\cite{Cao:2025afs,Nakamoto:2026lyo,Cavalcante:2026vgr}.

We consider the perturbed potential
\begin{equation}
	V_{\ell s}(r) \longrightarrow V_{\ell s}(r) + \epsilon\, \mathcal G_{\sigma}(r_*;r_{s*}),
	\label{eq:GaussianPerturbedPotential}
\end{equation}
where
\begin{equation}
	\mathcal G_{\sigma}(r_*;r_{s*}) = \frac{1}{\sqrt{2\pi}\sigma} \exp\left[ -\frac{(r_*-r_{s*})^2}{2\sigma^2} \right].
	\label{eq:GaussianProfile}
\end{equation}
Here \(r_s\) denotes the center of the perturbation, \(r_{s*}=r_*(r_s)\), and \(\sigma\) is its width in the tortoise coordinate. The profile is normalized to unit integral and satisfies \(\mathcal G_\sigma\rightarrow\delta(r_*-r_{s*})\) in the distributional limit \(\sigma\rightarrow0\). The point-defect model is then recovered by identifying \(r_s=r_0\).

To first order in the perturbation strength \(\epsilon\), and for arbitrary finite width \(\sigma\), the QNM condition may be written in terms of the unperturbed homogeneous solutions as
\begin{equation}
  W_\ell(\omega) - \epsilon \int_{-\infty}^{+\infty} \mathcal G_{\sigma}(r_*;r_{s*}) \phi^{\rm in}_{\ell\omega}(r_*) \phi^{\rm up}_{\ell\omega}(r_*)\, \mathrm{d}r_* = 0.
	\label{eq:GaussianWronskianCondition}
\end{equation}
We define \(g_{\ell\omega} =\phi^{\rm in}_{\ell\omega}\phi^{\rm up}_{\ell\omega}\) and introduce the Gaussian overlap
\begin{equation}
	J_{\sigma}(\omega;r_s) = \int_{-\infty}^{+\infty} \mathcal G_{\sigma}(r_*;r_{s*}) g_{\ell\omega}(r_*)\, \mathrm{d}r_*.
	\label{eq:GaussianOverlap}
\end{equation}
Provided \(J_\sigma(\omega;r_s)\neq0\), the finite-width normalized Wronskian is
\begin{equation}
	F_{\sigma}(\omega;r_s) = \frac{W_\ell(\omega)} {J_{\sigma}(\omega;r_s)} = \frac{2\ii\omega A_{\ell\omega}^{(-)}}{J_{\sigma}(\omega;r_s)}.
	\label{eq:GaussianNormalizedWronskian}
\end{equation}
The corresponding spectral equation takes the same form as in the point-defect problem,
\begin{equation}
	D_{\sigma}(\omega,\epsilon;r_s) = F_{\sigma}(\omega;r_s) - \epsilon = 0.
	\label{eq:GaussianSpectralEquation}
\end{equation}

The point-defect equation is recovered continuously as \(\sigma\rightarrow0\). Indeed,
\begin{equation}
	J_{\sigma}(\omega;r_s) \longrightarrow g_{\ell\omega}(r_{s*}),
\end{equation}
and hence
\begin{equation}
	F_{\sigma}(\omega;r_s) \longrightarrow F(\omega;r_s).
	\label{eq:GaussianFPointlikeLimit}
\end{equation}

At fixed \(\sigma\), a repelling point of the Gaussian model satisfies \(\partial_\omega F_\sigma(\omega_r;r_s)=0\). It becomes an exceptional point for a real perturbation strength when, in addition, \(\operatorname{Im}F_\sigma(\omega_r;r_s)=0\). The corresponding critical coupling is

\begin{equation}
	\epsilon_{\rm EP}^{(n,k)}(\sigma) = \operatorname{Re} F_{\sigma}\left(\omega_{\rm EP}^{(n,k)}(\sigma);r_{s,{\rm EP}}^{(n,k)}(\sigma)\right).
	\label{eq:GaussianCriticalCoupling}
\end{equation}

For all the continued exceptional points considered below, we also verify that \(F_{\sigma,\omega\omega}^{\rm EP}\neq0\), confirming that the degeneracy remains of second order.

For each fixed width, these conditions select isolated EPs in the two-dimensional parameter plane \((r_s,\epsilon)\). Once \(\sigma\) is allowed to vary, the exceptional points form one-dimensional curves in the three-dimensional parameter space \((r_s/M,\epsilon,\sigma/M)\). More precisely, continuation of each point-defect exceptional point \({\rm EP}_{n}^{(k)}\) defines a distinct exceptional line
\begin{equation}
	\mathcal L_n^{(k)} = \left\{ \left( \frac{r_{s,{\rm EP}}^{(n,k)}(\sigma)}{M}, \epsilon_{\rm EP}^{(n,k)}(\sigma), \frac{\sigma}{M} \right) : \sigma\in\Sigma_n^{(k)} \right\},
	\label{eq:ExceptionalLineDefinition}
\end{equation}
where \(\Sigma_n^{(k)}\) denotes the interval over which the continuation is defined. The point-defect exceptional point is recovered at \(\sigma=0\). The associated frequency \(\omega_{\rm EP}^{(n,k)}(\sigma)\) gives the spectral image of \(\mathcal L_n^{(k)}\) in the complex-frequency plane.

The dimensionality of this solution set follows directly from the double-root conditions \(D_\sigma=0\) and \(\partial_\omega D_\sigma=0\). Together, they represent four real equations. The complex frequency, combined with the three real parameters \(r_s\), \(\epsilon\), and \(\sigma\), supplies five real variables. Under the usual non-degeneracy conditions, the solution set is therefore generically one-dimensional.

\begin{figure}[htbp]
	\centering
	\includegraphics[width=0.87\linewidth]{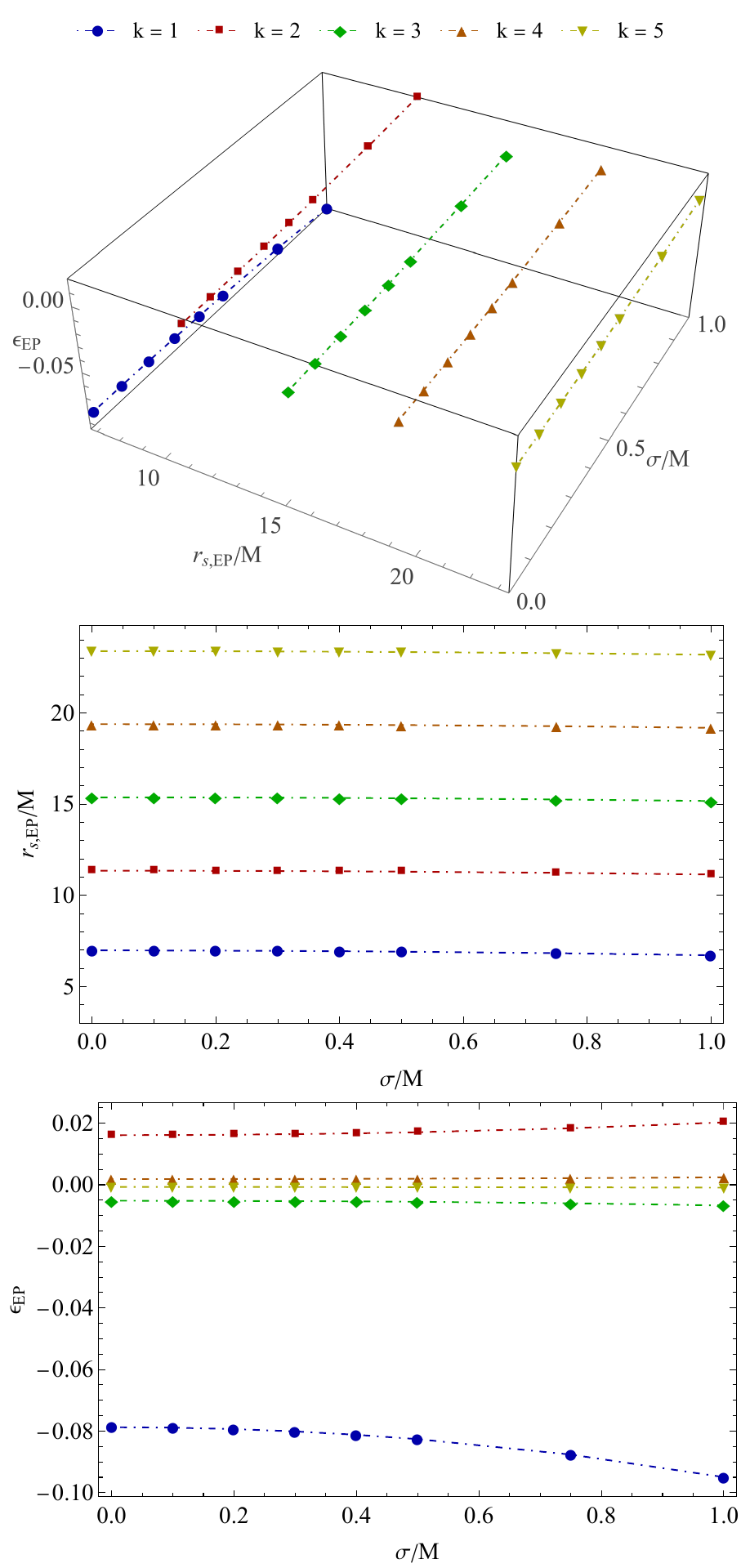}
	\caption{
		Exceptional lines generated by the finite-width Gaussian perturbation. The upper panel shows the continuations of the first five point-defect exceptional points \({\rm EP}_{0}^{(k)}\), \(k=1,\ldots,5\), in the three-dimensional parameter space \((r_s/M,\epsilon,\sigma/M)\). The middle and lower panels show the corresponding projections \(r_{s,\rm EP}^{(0,k)}(\sigma)/M\) and \(\epsilon_{\rm EP}^{(0,k)}(\sigma)\) as functions of \(\sigma/M\). The markers denote the exceptional points computed at \(\sigma/M=0,\,0.1,\,0.2,\,0.3,\,0.4,\,0.5,\,0.75,\) and \(1\). Each point-defect exceptional point evolves smoothly with increasing \(\sigma\), thereby defining a distinct exceptional line \(\mathcal L_0^{(k)}\).
	}
	\label{fig:ExceptionalLinesParameterSpace}
\end{figure}

We have continued the first five point-defect exceptional points \({\rm EP}_{0}^{(k)}\), with \(k=1,\ldots,5\), over the finite-width interval \(0\leq\sigma/M\leq1\). Figure~\ref{fig:ExceptionalLinesParameterSpace} displays the corresponding exceptional lines \(\mathcal L_0^{(k)}\) in the full three-dimensional parameter space, together with their projections onto the \((\sigma,r_s)\) and \((\sigma,\epsilon)\) planes. Each value of \(k\) labels the continuation of a single point-defect exceptional point \({\rm EP}_0^{(k)}\) and therefore defines a distinct exceptional line \(\mathcal L_0^{(k)}\).

Along all five exceptional lines, the exceptional-point location moves inward as the Gaussian width is increased. The radial displacement remains moderate and decreases progressively from the inner to the outer exceptional lines. The absolute variation of the critical coupling is largest for \(k=1\), although its relative variation increases towards the outer exceptional lines.

\begin{figure}[htbp]
	\centering
	\includegraphics[width=0.90\linewidth]{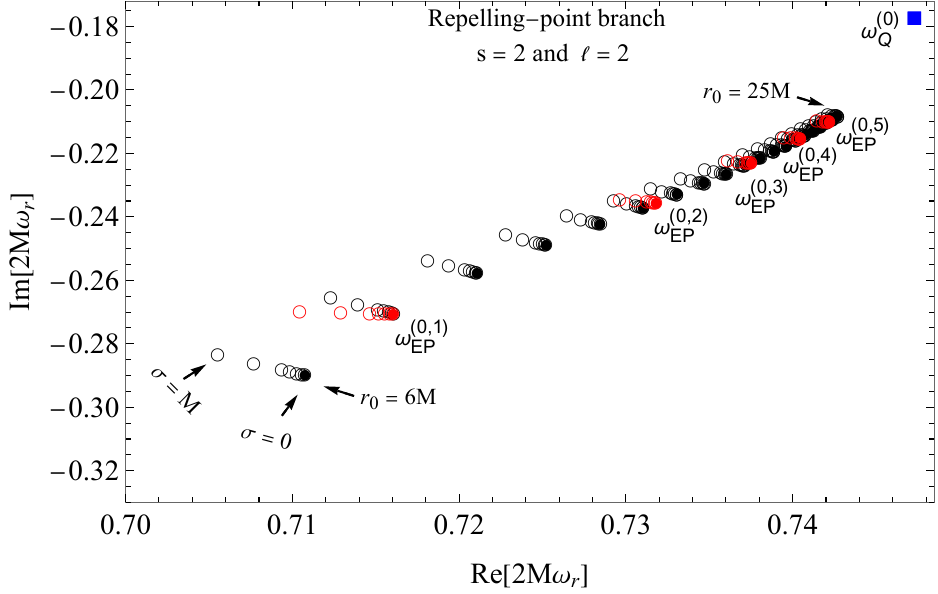}
	\caption{
		Migration of the repelling-point branch under a finite-width Gaussian perturbation of the Regge-Wheeler potential for \(s=2\) and \(\ell=2\). The filled markers show the point-defect limit \(\sigma=0\), while the open markers show the Gaussian repelling points obtained for \(\sigma/M=0.1,\,0.2,\,0.3,\,0.4,\,0.5,\,0.75,\) and \(1\). The branch is followed for perturbation centers in the interval \(6M\leq r_s\leq25M\). Red markers denote the associated exceptional points \({\rm EP}_{0}^{(k)}\) and their finite-width continuations. As the width increases, both the repelling points and the exceptional points migrate smoothly away from their point-defect positions. The displacement is largest along the inner part of the branch, close to \(r_s=6M\), and becomes progressively weaker as the perturbation center moves outward. The figure shows the spectral images of the corresponding exceptional lines in the complex-frequency plane.
	}
	\label{fig:RPBranchGaussian}
\end{figure}

Taken together, the two projection panels determine, for each fixed \(k\), the parametric curve
\begin{equation}
	\sigma \longmapsto \left(\frac{r_{s,{\rm EP}}^{(0,k)}(\sigma)}{M},\epsilon_{\rm EP}^{(0,k)}(\sigma),\frac{\sigma}{M}\right),
\end{equation}
which is precisely the exceptional line \(\mathcal L_0^{(k)}\). The spectral images of these finite-width continuations are shown in Fig.~\ref{fig:RPBranchGaussian}.

Figure~\ref{fig:RPBranchGaussian} shows the spectral images of the finite-width continuations displayed in Fig.~\ref{fig:ExceptionalLinesParameterSpace}, together with the corresponding repelling-point branches. The Gaussian branches deform smoothly from their point-defect limits, and real exceptional points remain present throughout the range of widths considered. For each fixed \(k\), the successive finite-width EPs belong to the same exceptional line. Distinct values of \(k\) label different exceptional lines.

The finite-width displacement varies appreciably along the branch. It is largest near the inner part, around \(r_s=6M\), and becomes progressively weaker as the perturbation center moves outward. This behavior is consistent with the radial variation scale of \(g_{\ell\omega}(r_*)\). At larger \(r_s\), a Gaussian of fixed width samples a region over which this product varies comparatively slowly, so its Gaussian average remains close to the pointwise value \(g_{\ell\omega}(r_{s*})\). Closer to the black hole, the homogeneous mode functions vary more rapidly across the Gaussian support, producing a larger finite-width displacement.

To make explicit the approach to the point-defect limit, we can expand the Gaussian overlap in powers of \(\sigma\), making the assumption that $\sigma$ is small in comparison with the local radial variation scale of \(g_{\ell\omega}\). Since the Gaussian profile is centered and even, its odd moments vanish, yielding
\begin{equation}
	J_{\sigma}(\omega;r_s) = g_{\ell\omega}(r_{s*}) + \frac{\sigma^2}{2} \left. \frac{\mathrm{d}^2g_{\ell\omega}} {\mathrm{d}r_*^2} \right|_{r_*=r_{s*}} + O(\sigma^4).
	\label{eq:GaussianOverlapExpansion}
\end{equation}
Provided \(g_{\ell\omega}(r_{s*})\neq0\), it follows that
\begin{equation}
	F_{\sigma}(\omega;r_s) = F(\omega;r_s)\left[ 1 - \frac{\sigma^2}{2} \frac{ g_{\ell\omega}''(r_{s*})}{g_{\ell\omega}(r_{s*})}+O(\sigma^4)\right].
	\label{eq:GaussianSpectralExpansion}
\end{equation}
The leading finite-width correction is therefore governed by the local curvature of the product of the two homogeneous mode functions. The absence of a term linear in \(\sigma\) follows from the evenness of the Gaussian profile.

At each continued EP for which \(F_{\sigma,\omega\omega}^{\rm EP}\neq0\), the local topology remains that of a second-order exceptional point. At fixed \(\sigma\) and \(r_s=r_{s,{\rm EP}}^{(n,k)}(\sigma)\), expansion around \(\omega_{\rm EP}^{(n,k)}(\sigma)\) and \(\epsilon_{\rm EP}^{(n,k)}(\sigma)\) gives
\begin{equation}
    \begin{split}
	&D_{\sigma}\left( \omega,\epsilon; r_{s,{\rm EP}}^{(n,k)}(\sigma) \right) \simeq {} \\
    &\hspace{+25pt}\frac{1}{2}F_{\sigma,\omega\omega}^{\rm EP} \left[ \omega-\omega_{\rm EP}^{(n,k)}(\sigma) \right]^2 - \left[ \epsilon-\epsilon_{\rm EP}^{(n,k)}(\sigma) \right].
    \end{split}
	\label{eq:GaussianLocalEPExpansion}
\end{equation}
The two local resonance branches therefore satisfy
\begin{equation}
	\omega_\pm^{(n,k)}(\sigma) \simeq \omega_{\rm EP}^{(n,k)}(\sigma) \pm \left[ \frac{ 2\left[ \epsilon-\epsilon_{\rm EP}^{(n,k)}(\sigma) \right] }{ F_{\sigma,\omega\omega}^{\rm EP} } \right]^{1/2}.
	\label{eq:GaussianSquareRootSplitting}
\end{equation}

A finite perturbation width thus shifts the EP locations, frequencies, and critical couplings without altering their local square-root topology. Within this first-order Gaussian extension, the isolated EPs of the point-defect problem persist as continuous exceptional lines in the three-dimensional parameter space.

%====================================================================================================================================

\section{Results: the time domain}
\label{sec:results-time-domain}

\subsection{Exceptional points and the intrinsic time-domain response}
\label{subsec:EPIntrinsicResponse}

The inverse-frequency amplification of the excitation factors, Eq.~\eqref{eq:ExcitationInverseFrequency}, does not imply a divergent time-domain response. The finite product in Eq.~\eqref{eq:AdiabaticInvariantEP} shows that the divergent residues are compensated by the vanishing separation of the two coalescing poles. Their combination therefore provides a natural setting in which to examine the time-domain manifestation of the exceptional point. Related manifestations of nearly coalescing and double-pole QNMs in the time domain have recently been investigated in
Refs.~\cite{PanossoMacedo:2025xnf,Nakamoto:2026lyo,Wu:2026chs,Cheng:2026gxu}.

We first consider the intrinsic ringing associated with the coalescing resonances, without specifying a mechanism by which the perturbation is excited. The construction involves only the resonance frequencies, excitation factors and homogeneous radial solutions, and therefore isolates the QNM pole contribution rather than the complete retarded response.

\subsubsection{Two-pole beating and double-pole limit: intrinsic ringings}
\label{subsubsec:IntrinsicRingings}

\begin{figure*}[htbp]
	\centering
	\includegraphics[width=0.85\textwidth]{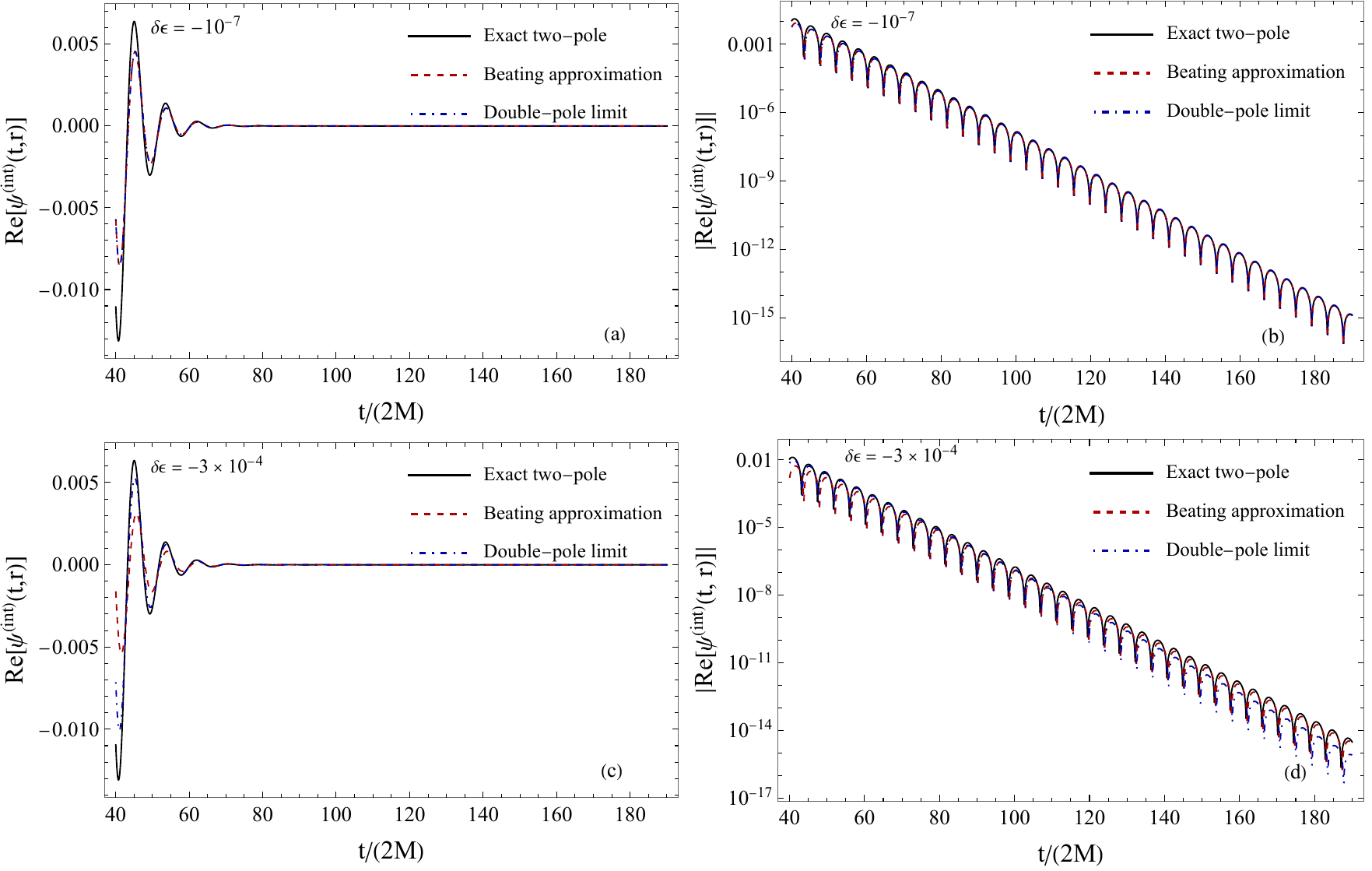}
	\caption{
		Intrinsic two-pole ringing and double-pole limit near \({\rm EP}_{0}^{(4)}\), with \(r=r_{\rm EP}^{(0,4)}\) and \(r'=50M\). Only times beyond the estimated onset \(t_{\rm start}^{\rm int}\simeq r_*(r_{\rm EP}^{(0,4)})+r_*(50M)\) are displayed. The horizontal axis is the physical time \(t/(2M)\). The black solid curves show the direct numerical two-pole ringing, the red dashed curves the local beating approximation, and the blue dot-dashed curves the double-pole limit. The upper panels correspond to \(\delta\epsilon=-10^{-7}\), while the lower panels correspond to \(\delta\epsilon=-3\times10^{-4}\). Panels (a) and (c) show the real part of the complex QNM contribution, whereas panels (b) and (d) show the absolute value of its real part on a logarithmic scale.
	}
	\label{fig:TwoPoleDoublePole}
\end{figure*}

Let the two perturbed resonance frequencies be written as
\begin{equation}
	\begin{aligned}
		\omega_\pm &= \omega_0 \pm \delta\omega, 
	\end{aligned}
	\label{eq:OmegaPmDeltaOmega}
\end{equation}
where
\begin{equation}
        \begin{aligned}
        \omega_0
        &=
        \frac{\omega_++\omega_-}{2},
        &
        \delta\omega
        &=
        \frac{\omega_+-\omega_-}{2}.
        \end{aligned}
	\label{eq:Omega0DeltaOmega}
\end{equation}
The singular excitation-factor behavior is given by Eq.~\eqref{eq:ExcitationInverseFrequency}. Together with Eq.~\eqref{eq:AdiabaticInvariantEP}, it gives the finite relation \(\mathcal K_{\rm EP}=\mathcal I_{\rm EP}/4\), which sets the amplitude of the coalescence limit below.

For a fixed second radial position \(r'\), we define the complex
intrinsic two-pole ringing by
\begin{equation}
	\begin{split}
		\psi_{\rm pair}^{\rm int}(t,r')
		={}& \phantom{+} \,
		\mathfrak{B}_+\, \phi_{\ell\omega_+}^{\rm up}(r_*') e^{-\ii\omega_+t}
		\\
		&+
		\mathfrak{B}_-\, \phi_{\ell\omega_-}^{\rm up}(r_*') e^{-\ii\omega_-t},
	\end{split}
	\label{eq:ExactTwoPoleResponse}
\end{equation}
where \(r_*'=r_*(r')\). 

It is important to recall that QNM expansions of the retarded response do not provide physically relevant results at arbitrarily early times, owing to their exponentially divergent behavior as \(t\) is decreased. One must therefore determine, from physical considerations, the time beyond which the QNM contribution can be meaningfully used. This defines the starting time \(t_{\rm start}\) of the ringing and gives rise to the well-known ``time-shift problem''; see, e.g., Ref.~\cite{Berti:2006wq} and references therein. For massless fields, this starting time admits a simple geometrical estimate. The QNMs are semiclassically associated with the neighborhood of the maximum of the effective potential, which, in the tortoise-coordinate convention adopted here, lies close to \(r_*=0\). If the two radial points are sufficiently far from the black hole, \(r_*,r_*'\gg2M\), one may therefore estimate \(t_{\rm start}\simeq r_*+r_*'\). This is approximately the time required for the signal to propagate from one radial point to the peak of the potential and then from the peak to the second radial point.

To obtain the local exceptional-point form of Eq.~\eqref{eq:ExactTwoPoleResponse}, we expand the outgoing homogeneous solution about the midpoint frequency  \(\omega_0\),
\begin{equation}
	\phi_{\ell\omega_\pm}^{\rm up}(r_*') = \phi_0^{\rm up}(r_*') \pm \delta\omega\, \partial_\omega\phi_0^{\rm up}(r_*') + O(\delta\omega^2),
	\label{eq:PhiUpExpansion}
\end{equation}
where
\begin{equation}
	\phi_0^{\rm up} = \phi_{\ell\omega_0}^{\rm up}, \qquad \text{and} \qquad \partial_\omega\phi_0^{\rm up} = \left. \frac{\partial\phi_{\ell\omega}^{\rm up}} {\partial\omega} \right|_{\omega=\omega_0}.
	\label{eq:PhiUpDerivativeDefinition}
\end{equation}

Using Eqs.~\eqref{eq:ExcitationInverseFrequency} and \eqref{eq:PhiUpExpansion} in Eq.~\eqref{eq:ExactTwoPoleResponse} gives, at leading singular order,
\begin{align}
	\psi_{\rm beat}^{\rm int}(t,r')
	={}&
	2\mathcal K_{\rm EP} e^{-\ii\omega_0t} \left[ \cos(\delta\omega t)\, \partial_\omega\phi_0^{\rm up}(r_*') \right. \nonumber\\
	&\left.
	- \ii t\, \operatorname{sinc}(\delta\omega t)\, \phi_0^{\rm up}(r_*') \right],
	\label{eq:BeatingResponse}
\end{align}
where
\begin{equation}
	\operatorname{sinc}(z)=\frac{\sin z}{z}.
\end{equation}
This expression describes the beating of two nearly coalescing QNMs. The \(1/\delta\omega\) divergences of the individual excitation factors have canceled against the vanishing frequency separation, leaving a finite two-pole ringing. Regular \(O(1)\) contributions to \(\mathfrak{B}_\pm\) generate additional finite terms, but do not alter this cancellation or the characteristic time dependence associated with the coalescence.

In the coalescence regime
\begin{equation}
	\left|\delta\omega\,t\right|\ll1,
	\label{eq:DoublePoleTimeCondition}
\end{equation}
one has \(\cos(\delta\omega t)\simeq1\) and \(\operatorname{sinc}(\delta\omega t)\simeq1\). The leading two-pole ringing therefore reduces to
\begin{equation}
	\psi_{\rm DP}^{\rm int}(t,r') = 2\mathcal K_{\rm EP} e^{-\ii\omega_0t} \left[ \partial_\omega\phi_0^{\rm up}(r_*') - \ii t\,\phi_0^{\rm up}(r_*') \right].
	\label{eq:DoublePoleResponse}
\end{equation}

In the exact coalescence limit, \(\omega_0\rightarrow\omega_{\rm EP}\), this is the singular-residue contribution associated with a second-order pole. More generally, the complete double-pole contribution has the form

\begin{equation}
	\psi_{\rm EP}^{\rm int}(t,r') = \left[ \mathcal A_0(r') + \mathcal A_1(r')\,t \right] e^{-\ii\omega_{\rm EP}t},
	\label{eq:GenericDoublePoleResponse}
\end{equation}
where the coefficient of the polynomial term is fixed by the coalescing residues. The characteristic exceptional-point signature is therefore
$
	t\,e^{-\ii\omega_{\rm EP}t}.
$
This polynomially modified exponential is the characteristic time-domain behavior associated with a second-order pole and is consistent with both general analyses of spectral singularities and recent black-hole QNM studies \cite{Heiss:2010nin,Nakamoto:2026lyo,Wu:2026chs,Cheng:2026gxu,PanossoMacedo:2025xnf}.

For the numerical comparison we approach \({\rm EP}_{0}^{(4)}\) from the negative-detuning side, \(\delta\epsilon<0\). Under continued decrease of \(\epsilon\), the two resonance branches connect continuously to the Schwarzschild fundamental mode \(Q_0\) and first overtone \(Q_1\) at \(\epsilon=0\).

Figure~\ref{fig:TwoPoleDoublePole} compares the direct numerical two-pole ringing with the beating approximation \eqref{eq:BeatingResponse} and the double-pole expression \eqref{eq:DoublePoleResponse}. At \(\delta\epsilon=-10^{-7}\), the beating approximation closely follows the exact two-pole result throughout the displayed interval, while the double-pole form reproduces the response in the regime \(\lvert\delta\omega \, t\rvert\ll1\). At the larger detuning \(\delta\epsilon=-3\times10^{-4}\), the beating expression continues to capture the dominant two-pole behavior, whereas the range over which the double-pole approximation applies is reduced. The numerical results thus display the continuous transition from two nearby simple poles to the characteristic response of a second-order pole.

To place this intrinsic exceptional-point ringing in context, we now compare it with the unperturbed Schwarzschild two-mode contribution formed from the same pair of branches. We define
\begin{equation}
	\begin{split}
    \phantom{+} \, 
		\psi_{\rm Schw}^{01,\rm int}(t,r')
		={}&
		\mathfrak{B}_0^{\rm Schw} \phi_{\ell\omega_Q^{(0)}}^{\rm up}(r_*') e^{-\ii\omega_Q^{(0)}t} \\
		&+
		\mathfrak{B}_1^{\rm Schw} \phi_{\ell\omega_Q^{(1)}}^{\rm up}(r_*') e^{-\ii\omega_Q^{(1)}t},
	\end{split}
	\label{eq:SchwarzschildTwoModeResponse}
\end{equation}
where \(\mathfrak{B}_n^{\rm Schw}\) are the unperturbed Schwarzschild excitation factors and \(\omega_Q^{(0)}\) and \(\omega_Q^{(1)}\) are the fundamental mode and first overtone, respectively.

\begin{figure}[htbp]
	\centering
	\includegraphics[width=0.97\linewidth]{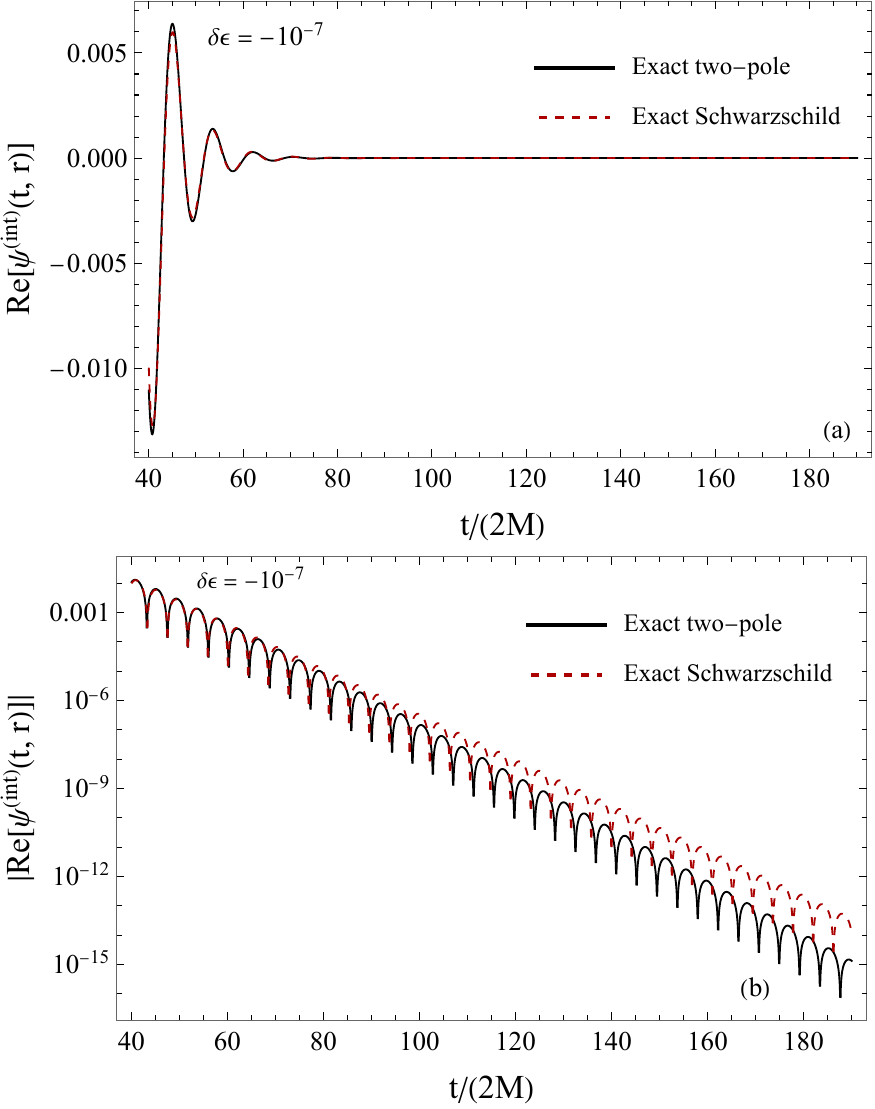}
	\caption{
		Comparison between the intrinsic two-pole ringing near \({\rm EP}_{0}^{(4)}\) and the Schwarzschild \(Q_0+Q_1\) contribution. The outgoing homogeneous solutions are evaluated at \(r'=50M\), and the same time interval \(t\geq t_{\rm start}^{\rm int}\) is used for both curves. The black solid curve shows the perturbed two-pole ringing for \(\delta\epsilon=-10^{-7}\), while the red dashed curve shows the Schwarzschild \(Q_0+Q_1\) reference obtained from the two modes to which the perturbed branches connect continuously as \(\epsilon\rightarrow0\). Panel (a) shows the real part of the complex QNM contribution, and panel (b) its absolute real part on a logarithmic scale.
	}
	\label{fig:SchwVsTwoPole}
\end{figure}

The comparison in Fig.~\ref{fig:SchwVsTwoPole} is not a small-\(\epsilon\) expansion around the Schwarzschild problem. Here the small parameter is \(\delta\epsilon=\epsilon-\epsilon_{\rm EP}\), so that \(\delta\epsilon=-10^{-7}\) describes a configuration extremely close to the exceptional point rather than to \(\epsilon=0\). The two curves instead compare different locations along the same pair of spectral branches.

The responses are initially similar, but progressively dephase as the evolution proceeds. At late times, the Schwarzschild \(Q_0+Q_1\) contribution becomes dominated by the less strongly damped fundamental mode, whereas the perturbed response remains governed by two frequencies lying close to \(\omega_{\rm EP}\).

The exceptional point therefore leaves a finite intrinsic time-domain signature despite the divergence of the individual excitation factors. The vanishing frequency splitting and the opposite residue amplification combine to produce a finite beating signal that approaches the polynomially modified exponential of a second-order pole. The finite quantity \(\mathcal I_{\rm EP}\) defined by Eq.~\eqref{eq:AdiabaticInvariantEP} is precisely the spectral balance underlying this cancellation. We next examine the same coalescence mechanism when a specific excitation process is prescribed.

%=================================================================================================================

\subsection{Exceptional points and the extrinsic time-domain response}
\label{subsec:EPExtrinsicResponse}

The perturbed QNM frequencies determined by Eq.~\eqref{eq:PerturbedSpectralEquation} and the associated excitation factors defined by Eq.~\eqref{eq:PerturbedExcitationFactor} are intrinsic properties of the perturbed spectral problem. Accordingly, the intrinsic two-pole ringing defined by Eq.~\eqref{eq:ExactTwoPoleResponse} does not specify the external mechanism responsible for exciting the system. We therefore examine how the exceptional-point signatures identified at the level of the poles and their residues persist once a definite perturbation is prescribed. The role of resonant QNM excitation and nearly degenerate modes in black-hole ringdown has recently been investigated from several complementary perspectives~\cite{PanossoMacedo:2025xnf,Kubota:2025hjk,Nakamoto:2026lyo,Imafuku:2026rpn}.

A physically motivated perturbation generated, for example, by a particle falling into or plunging towards the black hole would provide a natural setting for such an analysis, but would introduce an additional dynamical problem beyond the scope of the present work. Instead, we consider a simpler initial-value problem generated by Gaussian Cauchy data~\cite{Berti:2006wq,Leaver:1986gd,Andersson:1996cm,Decanini:2014bwa}. This model retains an explicit excitation mechanism while allowing the exceptional-point contribution to be isolated in a controlled manner.

We prescribe
\begin{equation}
	\Psi_\ell(0,r)
	=
	\Psi_0
	\exp\left[
	-\frac{\alpha^2}{(2M)^2}
	\left(r_*(r)-r_*(r_c)\right)^2
	\right],
	\label{eq:GaussianCauchyData}
\end{equation}
with
\begin{equation}
	\partial_t\Psi_\ell(0,r)=0.
	\label{eq:GaussianCauchyVelocity}
\end{equation}
Here \(r_c\) denotes the center of the initial wave packet and \(\alpha\) controls its width. In particular, \(r_c\) should not be confused with the location \(r_0\) of the localized perturbation. For the numerical results below we take \(\Psi_0=1\), \(\alpha=1\), and \(r_c=10M\).

In the standard QNM treatment of an initial-value problem, the pole part of the retarded response is expressed as a sum over the resonances, each weighted by the corresponding excitation coefficient; see, e.g., Ref.~\cite{Decanini:2014bwa}. Restricting this pole contribution to the two resonances \(\omega_\pm\) that coalesce at the exceptional point, we define the complex two-pole contribution to the extrinsic ringing as
\begin{equation}
	\begin{split}
		\psi_{\rm pair}^{\rm ext}(t,r_{\rm obs})
		={}&
		-\ii\omega_+ C_+\, \phi_{\ell\omega_+}^{\rm up}(r_{*{\rm obs}}) e^{-\ii\omega_+t}
		\\
		&-
		\ii\omega_- C_-\, \phi_{\ell\omega_-}^{\rm up}(r_{*{\rm obs}}) e^{-\ii\omega_-t},
	\end{split}
	\label{eq:ExtrinsicTwoPoleResponse}
\end{equation}
where \(r_{*{\rm obs}}=r_*(r_{\rm obs})\). The coefficients \(C_+\) and \(C_-\) are the excitation coefficients associated with the two resonances. Unlike the excitation factors \(\mathfrak{B}_\pm\), which are intrinsic properties of the corresponding poles, they also encode the coupling of the resonances to the prescribed initial data.

For the pointlike perturbation, the resonant radial function has a piecewise form fixed by the QNM boundary conditions and by continuity at the defect. Choosing its normalization to coincide with the ingoing Schwarzschild solution on the horizon side, its value at a perturbed resonance \(\omega_j\) may be written as
\begin{equation}
	\Phi_{\ell j}(r_*)
	=
	\begin{cases}
		\phi_{\ell\omega_j}^{\rm in}(r_*),
		&
		r_*<r_{0*},
		\\[3mm]
		\displaystyle
		\frac{
			\phi_{\ell\omega_j}^{\rm in}(r_{0*})
		}{
			\phi_{\ell\omega_j}^{\rm up}(r_{0*})
		}
		\phi_{\ell\omega_j}^{\rm up}(r_*),
		&
		r_*>r_{0*}.
	\end{cases}
	\label{eq:PerturbedResonantProfile}
\end{equation}
This form follows directly from the matching conditions at the defect: at a resonance the solution satisfying the ingoing horizon condition must be proportional to the outgoing homogeneous solution on the exterior side.

For an observer located outside the defect, the excitation coefficient associated with a simple resonance \(\omega_j\) therefore takes the general form
\begin{equation}
	\begin{split}
		C_j =& \frac{\mathfrak{B}_j}{A^{(+)}(\omega_j)}
		\Bigg[ 
		\int_{-\infty}^{r_{0*}} \Psi_\ell(0,r_*')\, \phi_{\ell\omega_j}^{\rm in}(r_*') \,\dd r_*' \\
		& +
		\frac{ \phi_{\ell\omega_j}^{\rm in}(r_{0*}) }{ \phi_{\ell\omega_j}^{\rm up}(r_{0*}) } \int_{r_{0*}}^{+\infty} \Psi_\ell(0,r_*')\, \phi_{\ell\omega_j}^{\rm up}(r_*') \,\dd r_*'
		\Bigg].
	\end{split}
	\label{eq:ExtrinsicExcitationCoefficient}
\end{equation}
Equation~\eqref{eq:ExtrinsicExcitationCoefficient} applies independently of the position of the initial wave packet relative to the defect. If its effective support lies on the horizon side, the first integral dominates; if it lies on the exterior side, the second one dominates; and if the initial data overlap the defect, both contributions are retained.

It is convenient to introduce the corresponding source factor
\begin{equation}
	\begin{split} 
		\mathcal S(\omega;r_0) &= \frac{1}{A^{(+)}(\omega)}
		\Bigg[
		\int_{-\infty}^{r_{0*}} \Psi_\ell(0,r_*')\, \phi_{\ell\omega}^{\rm in}(r_*') \,\dd r_*' \\
		&+
		\frac{ \phi_{\ell\omega}^{\rm in}(r_{0*}) }{ \phi_{\ell\omega}^{\rm up}(r_{0*}) } \int_{r_{0*}}^{+\infty} \Psi_\ell(0,r_*')\, \phi_{\ell\omega}^{\rm up}(r_*') \,\dd r_*'
		\Bigg],
	\end{split}
	\label{eq:SourceFactor}
\end{equation}
so that, at a perturbed resonance,
\begin{equation}
	C_j = \mathfrak{B}_j\,\mathcal S(\omega_j;r_0).
	\label{eq:CjFactorization}
\end{equation}
In the remainder of this subsection \(r_0\) is fixed, and we suppress its explicit appearance by writing \(\mathcal S(\omega)\equiv\mathcal S(\omega;r_0)\).

For the main numerical configuration considered first below, the Gaussian wave packet is localized well on the horizon side of the defect, \(r_c<r_0\). The second contribution in Eq.~\eqref{eq:SourceFactor} is then negligible over the effective support of the initial data, and one recovers
\begin{equation}
	\mathcal S(\omega) \simeq \frac{1}{A^{(+)}(\omega)} \int_{-\infty}^{+\infty} \Psi_\ell(0,r_*')\, \phi_{\ell\omega}^{\rm in}(r_*') \,\dd r_*'.
	\label{eq:SourceFactorInteriorData}
\end{equation}

The excitation coefficients inherit the local exceptional-point structure of the excitation factors, despite the additional dependence on the prescribed initial data. Indeed, because \(\mathcal S(\omega)\) is regular in a neighborhood of the exceptional point, its values on the two coalescing branches may be expanded as
\begin{equation}
	\mathcal S(\omega_\pm) = \mathcal S_{\rm EP} \pm \delta\omega\,\mathcal S'_{\rm EP} + O(\delta\omega^2),
	\label{eq:SourceFactorExpansion}
\end{equation}
where \(\mathcal S_{\rm EP}=\mathcal S(\omega_{\rm EP})\) and \(\mathcal S'_{\rm EP} =\partial_\omega\mathcal S|_{\omega_{\rm EP}}\). Combining this expansion with Eq.~\eqref{eq:ExcitationInverseFrequency} gives
\begin{equation}
	C_\pm = \pm \frac{ \mathcal K_{\rm EP}\mathcal S_{\rm EP} }{ \delta\omega } + O(1).
	\label{eq:ExtrinsicExcitationEPExpansion}
\end{equation}
For generic initial data with \(\mathcal S_{\rm EP}\neq0\), the excitation coefficients therefore display the same inverse-square-root amplification as the excitation factors,
\begin{equation}
	|C_\pm| \propto |\delta\epsilon|^{-1/2}.
	\label{eq:ExtrinsicExcitationScaling}
\end{equation}
The source factor thus modifies the complex amplitude of the singular contribution without changing its exceptional-point scaling.

The finite balance between the diverging residues and the vanishing frequency separation is inherited as well. Defining \(\Delta C=C_+-C_-\), one obtains
\begin{equation}
	\Delta C \simeq \frac{ 2\mathcal K_{\rm EP}\mathcal S_{\rm EP} }{ \delta\omega },
	\label{eq:DeltaCLeading}
\end{equation}
and hence, using Eq.~\eqref{eq:AdiabaticInvariantEP},
 \begin{equation} 
 	\lim_{\epsilon\rightarrow\epsilon_{\rm EP}} \Delta C\,\Delta\omega = \mathcal S_{\rm EP}\mathcal I_{\rm EP}.
	\label{eq:ExtrinsicInvariantEP}
\end{equation}
Thus, the finite combination \(\Delta\mathfrak{B}\,\Delta\omega\) defined in Eq.~\eqref{eq:AdiabaticInvariantDefinition} persists after the initial data are included, with its limiting value weighted by the source factor evaluated at the exceptional point.

\begin{figure*}[htbp]
	\centering
	\includegraphics[width=0.65\textwidth]{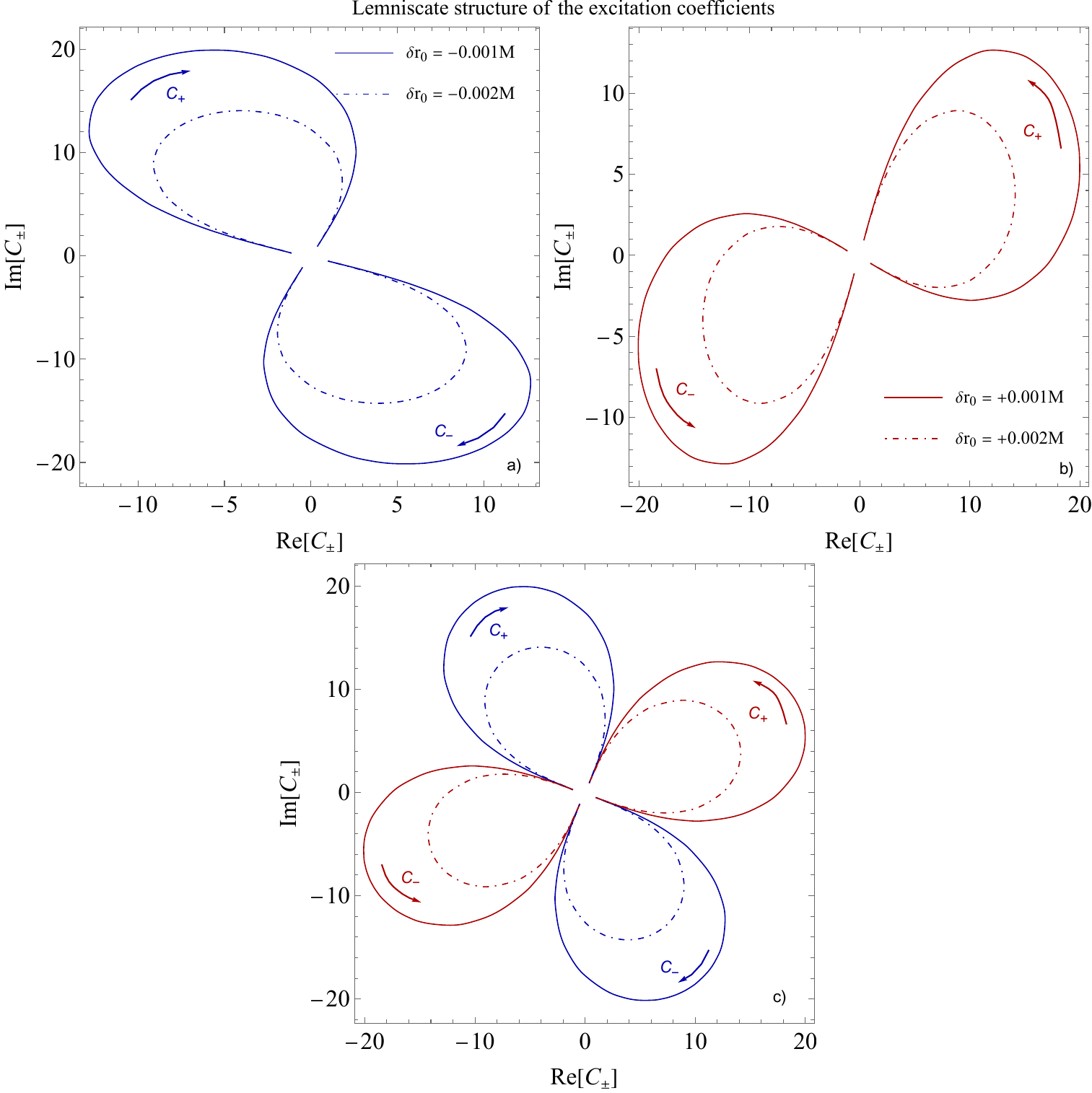}
	\caption{
		Lemniscate structure of the excitation coefficients near \({\rm EP}_{0}^{(4)}\). The perturbation locations are \(r_0=r_{\rm EP}^{(0,4)}+\delta r_0\), with \(\delta r_0=\pm10^{-3}M\) and \(\delta r_0=\pm2\times10^{-3}M\). For each \(r_0\), the corresponding repelling point \(\omega_r(r_0)\) satisfies \(F_\omega(\omega_r;r_0)=0\), and the real coupling is varied as \(\epsilon=\epsilon_c(r_0)+\Delta\epsilon\), where \(\epsilon_c(r_0)=\operatorname{Re}F(\omega_r;r_0)\). Thus, \(\Delta\epsilon\) is measured from the local avoided-crossing center rather than from \(\epsilon_{\rm EP}\). The excitation coefficients are computed for the Gaussian initial data \eqref{eq:GaussianCauchyData}, with \(\Psi_0=1\), \(\alpha=1\), and \(r_c=10M\). Panels (a) and (b) show the complex trajectories of \(C_\pm\) for negative and positive radial offsets, respectively. Solid curves correspond to \(\lvert\delta r_0\rvert=10^{-3}M\), and dot-dashed curves to \(\lvert\delta r_0\rvert=2\times10^{-3}M\). Panel (c) superposes the two signs of \(\delta r_0\), producing the four-leaf structure. The arrows indicate increasing \(\Delta\epsilon\). The source factor produces an overall rescaling and rotation relative to the corresponding excitation-factor trajectories, while preserving their local lemniscate structure.
	}
	\label{fig:CLemniscate}
\end{figure*}

The complex trajectories of the excitation coefficients inherit the same local geometry. Near an avoided crossing, if the source factor varies weakly across the two neighboring resonances, then
\begin{equation}
	C_\pm \simeq \mathcal S(\omega_r)\mathfrak{B}_\pm.
	\label{eq:ExtrinsicLemniscateRelation}
\end{equation}
Multiplication by the finite complex number \(\mathcal S(\omega_r)\) produces an overall rescaling and rotation of the trajectories in the complex plane. The Bernoulli-lemniscate structure expressed by Eqs.~\eqref{eq:ExcitationLemniscateEquation} and \eqref{eq:ExcitationLemniscatePolar} is therefore preserved for \(C_\pm\), up to corrections arising from the frequency dependence of the source factor across the resonance pair.

Figure~\ref{fig:CLemniscate} confirms this inheritance numerically. The two signs of the radial displacement generate complementary two-lobed trajectories, whose superposition produces the same four-leaf structure found for the intrinsic excitation factors. The change in scale and orientation reflects the complex source factor, while the residual deformation is associated with its frequency dependence across the resonance pair.

The exceptional-point structure of the excitation factors is therefore preserved after the Gaussian initial data are taken into account: their singular scaling, finite frequency-residue balance, and local lemniscate geometry are all inherited by the excitation coefficients. This provides the source-dependent counterpart of the resonant excitation structures discussed in Refs.~\cite{Motohashi:2024fwt,Kubota:2025hjk}.

\subsubsection{Two-pole beating and double-pole limit: extrinsic ringings}
\label{subsubsec:ExtrinsicRingings}

\begin{figure*}[htbp]
	\centering
	\includegraphics[width=0.85\textwidth]
	{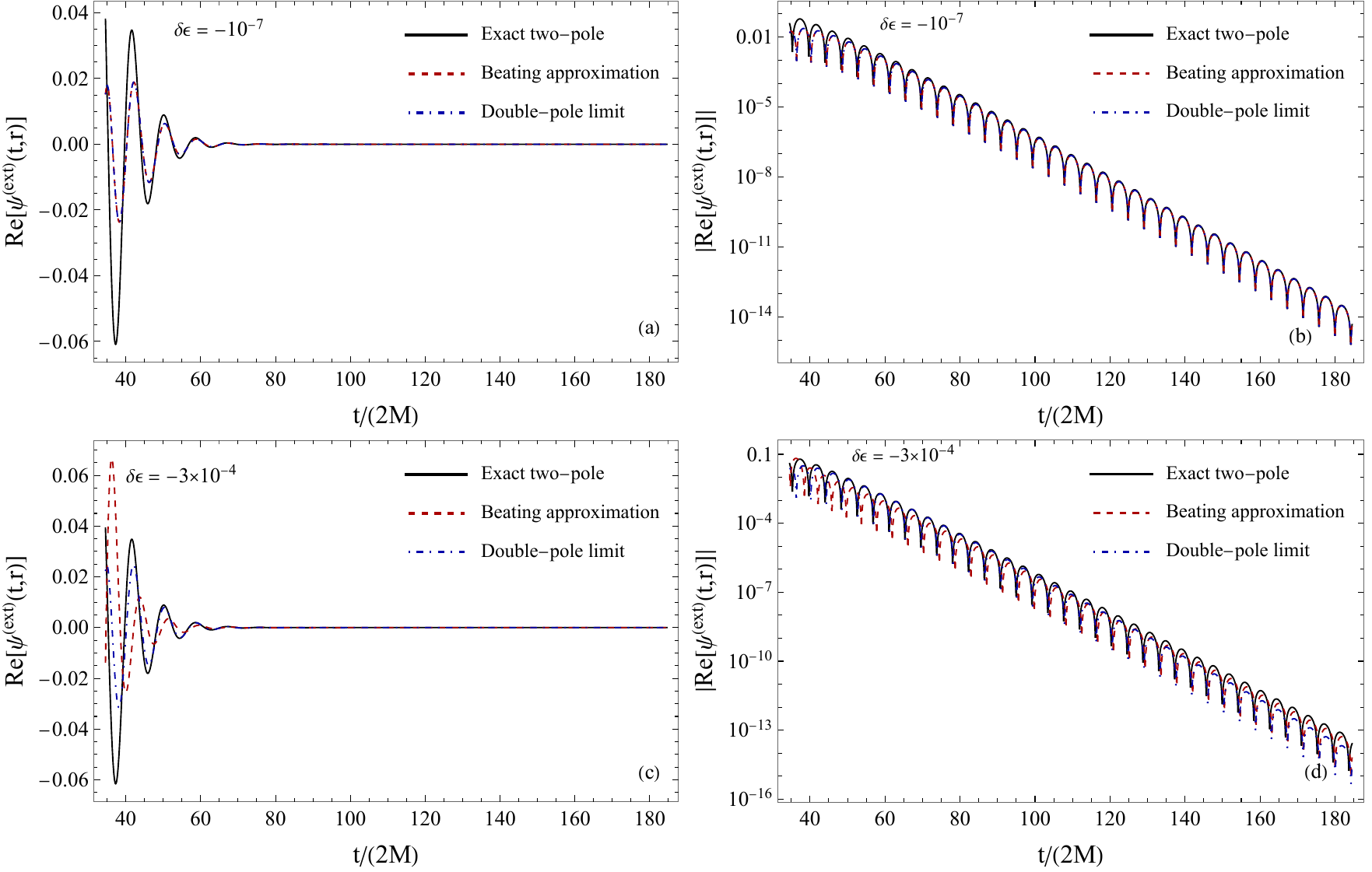}
	\caption{
		Extrinsic two-pole ringing and double-pole limit near \({\rm EP}_{0}^{(4)}\), generated by the Gaussian initial data \eqref{eq:GaussianCauchyData} with \(\Psi_0=1\), \(\alpha=1\), and \(r_c=10M\), and evaluated at \(r_{\rm obs}=50M\). Only times beyond the estimated onset \(t_{\rm start}^{\rm ext}\simeq r_*(10M)+r_*(50M)\) are displayed. The black solid curves show the direct numerical two-pole ringing, the red dashed curves the beating approximation \eqref{eq:ExtrinsicBeatingResponse}, and the blue dot-dashed curves the double-pole limit \eqref{eq:ExtrinsicDoublePoleResponse}. The upper panels correspond to \(\delta\epsilon=-10^{-7}\), and the lower panels to \(\delta\epsilon=-3\times10^{-4}\). Panels (a) and (c) show the real part of the complex QNM contribution, while panels (b) and (d) show the absolute value of its real part on a logarithmic scale.
	}
	\label{fig:ExtrinsicTwoPoleDoublePole}
\end{figure*}

To obtain the local form of the extrinsic ringing near the exceptional point, we introduce the shorthand
\begin{equation}
	q(\omega,r_{\rm obs}) = \mathcal S(\omega) \phi_{\ell\omega}^{\rm up}(r_{*{\rm obs}}).
	\label{eq:ExtrinsicQDefinition}
\end{equation}
The source factor and the outgoing homogeneous solution are regular at the exceptional point, and hence so is \(q\). Expanding about the midpoint frequency \(\omega_0\), we have
\begin{equation}
	q(\omega_\pm,r_{\rm obs}) = q_0 \pm \delta\omega\,q'_0 + O(\delta\omega^2),
	\label{eq:ExtrinsicQExpansion}
\end{equation}
where
\begin{equation}
	q_0=q(\omega_0,r_{\rm obs}), \qquad q'_0= \left. \partial_\omega q(\omega,r_{\rm obs}) \right|_{\omega=\omega_0}.
\end{equation}
Using Eq.~\eqref{eq:ExtrinsicExcitationEPExpansion} in Eq.~\eqref{eq:ExtrinsicTwoPoleResponse}, together with Eq.~\eqref{eq:ExtrinsicQExpansion}, gives the leading two-pole beating response
\begin{align}
	&\psi_{\rm beat}^{\rm ext}(t,r_{\rm obs}) = -2\ii\mathcal K_{\rm EP} e^{-\ii\omega_0t} \Big[ \cos(\delta\omega t) \left(q_0+\omega_0q'_0\right) \nonumber\\
	&\hspace{3.1cm}
	-\ii\omega_0 t\, \operatorname{sinc}(\delta\omega t)\, q_0 \Big].
	\label{eq:ExtrinsicBeatingResponse}
\end{align}

In the coalescence regime \(\lvert\delta\omega\,t\rvert\ll1\), the beating expression reduces to
\begin{equation}
	\psi_{\rm DP}^{\rm ext}(t,r_{\rm obs}) = -2\ii\mathcal K_{\rm EP} e^{-\ii\omega_0t} \left[ q_0+\omega_0q'_0 -\ii\omega_0t\,q_0 \right].
	\label{eq:ExtrinsicDoublePoleResponse}
\end{equation}
In the limit \(\omega_0\rightarrow\omega_{\rm EP}\), the extrinsic response therefore acquires the same characteristic polynomially modified exponential dependence as its intrinsic counterpart,
\begin{equation}
	\psi_{\rm EP}^{\rm ext}(t,r_{\rm obs}) = \left[ \mathcal A_0^{\rm ext}(r_{\rm obs}) + \mathcal A_1^{\rm ext}(r_{\rm obs})\,t \right] e^{-\ii\omega_{\rm EP}t}.
	\label{eq:GenericExtrinsicDoublePoleResponse}
\end{equation}
The characteristic \(t e^{-\ii\omega_{\rm EP}t}\) behavior of the second-order pole thus persists after the coupling to the prescribed initial data is included.

Before comparing the beating and double-pole expressions, Eqs.~\eqref{eq:ExtrinsicBeatingResponse} and \eqref{eq:ExtrinsicDoublePoleResponse}, with the direct numerical two-pole ringing defined in Eq.~\eqref{eq:ExtrinsicTwoPoleResponse}, we recall that the QNM contribution is subject to the time-shift problem and should not be interpreted at arbitrarily early times. For the present Gaussian initial-value problem, a natural estimate of the starting time is obtained from the propagation of the initial disturbance toward the peak of the effective potential and then outward to the observation point. For a sufficiently localized wave packet centered at \(r_c\), and with both \(r_c\) and \(r_{\rm obs}\) sufficiently far from the potential barrier, one may therefore estimate \(t_{\rm start}^{\rm ext}\simeq r_*(r_c)+r_*(r_{\rm obs})\). This is approximately the time required for the initial disturbance to reach the potential peak and for the resulting ringing to propagate to the observer.

Figure~\ref{fig:ExtrinsicTwoPoleDoublePole} compares the direct numerical two-pole ringing with the local beating and double-pole expressions. Very close to the exceptional point, at \(\delta\epsilon=-10^{-7}\), the beating approximation reproduces the two-pole evolution throughout the ringdown regime, and the double-pole expression provides an accurate description over the interval for which \(\lvert\delta\omega t\rvert\ll1\). At the larger detuning \(\delta\epsilon=-3\times10^{-4}\), the finite frequency splitting becomes more apparent. The beating expression continues to capture the dominant oscillatory evolution, whereas the double-pole approximation progressively departs from it as the condition \(\lvert\delta\omega t\rvert\ll1\) becomes more restrictive in time. The same coalescence mechanism found for the intrinsic ringing thus survives the weighting by the Gaussian initial data.

To separate the effect of the exceptional-point spectral structure from that of the chosen initial perturbation, we finally compare the perturbed two-pole ringing with the Schwarzschild \(Q_0+Q_1\) contribution generated by the same Gaussian initial data.

For the unperturbed Schwarzschild modes, the corresponding excitation coefficients are
\begin{equation}
	C_n^{\rm Schw} = \mathfrak{B}_n^{\rm Schw} \frac{1}{A^{(+)}(\omega_Q^{(n)})} \int_{-\infty}^{+\infty} \Psi_\ell(0,r_*')\, \phi_{\ell\omega_Q^{(n)}}^{\rm in}(r_*') \,\dd r_*',
	\label{eq:SchwarzschildExtrinsicCoefficient}
\end{equation}
and the two-mode contribution is
\begin{equation}
	\psi_{\rm Schw}^{01,\rm ext}(t,r_{\rm obs}) = \sum_{n=0}^{1} -\ii\omega_Q^{(n)} C_n^{\rm Schw} \phi_{\ell\omega_Q^{(n)}}^{\rm up}(r_{*{\rm obs}}) e^{-\ii\omega_Q^{(n)}t}.
	\label{eq:SchwarzschildExtrinsicResponse}
\end{equation}

\begin{figure}[htbp]
	\centering
	\includegraphics[width=0.97\linewidth]
	{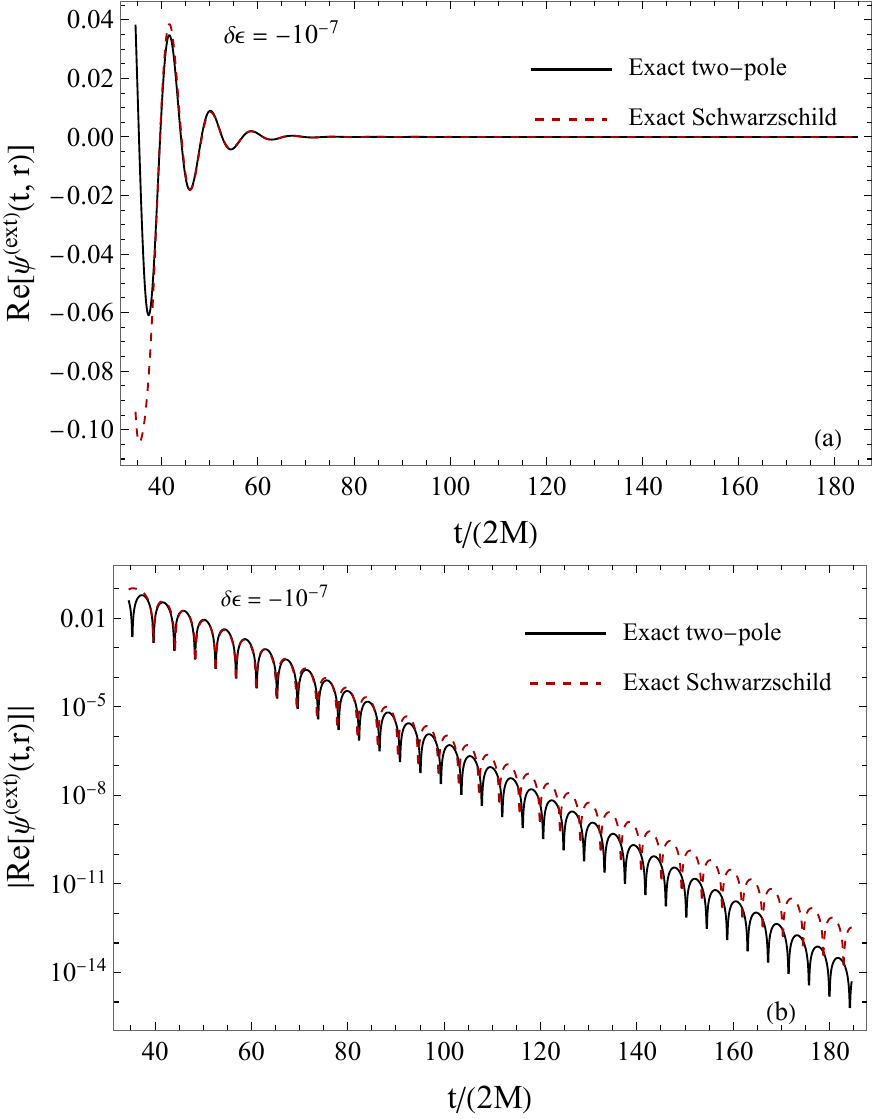}
	\caption{
		Comparison between the extrinsic two-pole ringing near \({\rm EP}_{0}^{(4)}\) and the Schwarzschild \(Q_0+Q_1\) ringing generated by the same Gaussian initial data. We use \(\Psi_0=1\), \(\alpha=1\), \(r_c=10M\), and \(r_{\rm obs}=50M\). The perturbed response is evaluated at \(\delta\epsilon=-10^{-7}\), and the same time interval \(t\geq t_{\rm start}^{\rm ext}\) is used for both curves. The black solid curve shows the perturbed two-pole contribution, while the red dashed curve shows the Schwarzschild \(Q_0+Q_1\) contribution formed from the two unperturbed modes to which the perturbed branches connect continuously as \(\epsilon\rightarrow0\). Panel (a) shows the real part of the complex QNM contribution, and panel (b) the absolute value of its real part on a logarithmic scale.
	}
	\label{fig:SchwVsExtrinsicTwoPole}
\end{figure}

Figure~\ref{fig:SchwVsExtrinsicTwoPole} shows that the two extrinsic ringings are initially very similar despite the substantial reorganization of the underlying spectrum near the exceptional point. The responses progressively dephase, and their late-time decay rates become distinguishable. In the Schwarzschild case, the \(Q_0+Q_1\) contribution eventually becomes dominated by the less strongly damped fundamental mode. The perturbed ringing, by contrast, is governed by the pair of resonances lying close to \(\omega_{\rm EP}\), whose damping rates are comparable. The difference becomes particularly clear on the logarithmic scale.

The exceptional-point amplification of the individual residues is therefore compatible with a finite response to prescribed initial data, in line with the broader distinction between spectral sensitivity and time-domain observability emphasized in recent studies~\cite{Berti:2022xfj,PanossoMacedo:2025xnf, Kubota:2025hjk,Nakamoto:2026lyo,Imafuku:2026rpn}. The singular behavior of the excitation factors is inherited by the excitation coefficients, while the opposite contributions of the coalescing poles combine to produce a regular time-domain ringing with the characteristic double-pole limit.

The comparison with the Schwarzschild \(Q_0+Q_1\) contribution also reveals a distinct late-time effect. The two responses remain close during the early part of the ringdown, but progressively separate as the more strongly damped exceptional-point pair decays. For the configuration considered here, the Schwarzschild response eventually becomes dominated by the fundamental mode, whose damping rate is smaller than that of the resonances lying near \(\omega_{\rm EP}\). The localized perturbation can therefore produce a substantial modification of the late-time QNM response when its parameters place the spectrum close to an exceptional point and the prescribed initial data couple efficiently to the coalescing resonances.

Finally, we examined how the extrinsic response depends on the position of the initial wave packet relative to the localized perturbation. We first placed the Gaussian on the exterior side of the defect at \(r_c=29.4084M\), chosen to be approximately symmetric, in tortoise distance, to the \(r_c=10M\) configuration about the reference defect. In this case the excitation coefficients \(C_\pm\) are enhanced by approximately an order of magnitude and undergo a substantial rotation in the complex plane. Nevertheless, the corresponding two-pole time-domain response remains of the same order of magnitude as in the horizon-side configuration. This illustrates that the magnitude of the raw excitation coefficients alone does not determine the observable ringdown. Indeed, writing \(t=t_{\rm start}^{\rm ext}+\tau\), each modal contribution contains the effective amplitude \(C_j\exp[-\ii\omega_j t_{\rm start}^{\rm ext}]\). Since \(\operatorname{Im}\omega_j<0\), the larger propagation delay associated with the exterior source partially compensates the enhancement of \(\lvert C_j\rvert\). In addition, close to the exceptional point the large individual pole contributions combine coherently into the finite double-pole response derived above.

We also considered the limiting configuration in which the Gaussian is centered directly on the defect, \(r_c=r_0=r_{\rm EP}\), for which both terms in the general source factor \eqref{eq:SourceFactor} contribute. Denoting the first and second integrals on the right-hand side of Eq.~\eqref{eq:SourceFactor} by \(J_<\) and \(J_>\), respectively, we find that the two contributions have comparable magnitudes, with \(\lvert J_>\rvert/\lvert J_<\rvert\simeq1.27\), and a relative phase of approximately \(45^\circ\). Their sum is therefore predominantly constructive rather than canceling:
\begin{equation}
	\frac{\lvert J_<+J_>\rvert} {\lvert J_<\rvert+\lvert J_>\rvert} \simeq0.92.
\end{equation}
Despite this constructive spatial coupling, the extrinsic ringing again remains of the same order of magnitude as for the one-sided source configurations. These checks show that the finite time-domain response near the exceptional point is not tied to a particular placement of the initial data, nor to a cancellation between the two sides of the defect. Rather, it results from the combination of the source-dependent propagation delay with the coherent recombination of the two coalescing resonances.

\section{Discussion and conclusions}
\label{sec:conclusions}

In this work, we have investigated the exceptional-point structure of the Schwarzschild quasinormal-mode spectrum under a localized perturbation at $r=r_0$ of the Regge-Wheeler potential. The exact spectral equation of the delta-function model \eqref{eq:PerturbedSpectralEquation} allowed us to characterize \emph{repelling points} as stationary points of the normalized Wronskian \eqref{eq:NormalizedWronskian} and to identify the subset accessible for real perturbation strength as second-order \emph{exceptional points}. The analysis revealed that the EPs are not isolated spectral accidents. For $r_0 > 3 M$, each QNM frequency is associated with a nearby repelling point, and each RP approaches a QNM frequency as $r_0 \rightarrow +\infty$, according to Eq.~\eqref{eq:RepellingPointAccumulation}. Under perturbation, neighboring Schwarzschild overtone pairs are connected through discrete sequences of exceptional points distributed along corresponding RP branches. For large $r_0$, the organization is controlled directly by the unperturbed QNM spectrum, and the real and imaginary parts of the corresponding Schwarzschild QNM frequency play distinct roles: the former fixes the asymptotic spatial organization of the EP sequence, whereas the latter controls the perturbation strength $\epsilon$ required to reach it. Under perturbation, strongly damped overtones reach exceptional points at exponentially-smaller perturbation strengths. This behavior provides a concrete realization of the broader overtone sensitivity discussed in previous spectral and pseudospectral analyses \cite{Jaramillo:2020tuu,Torres:2026uey,Cao:2025afs}.

Global continuation further shows that each exceptional point in the \(n\)-th family provides a spectral bridge between the neighboring Schwarzschild modes \(Q_n\) and \(Q_{n+1}\). For the lowest modes, the resulting connectivity may be represented schematically as
\[
\omega_Q^{(0)} \longleftrightarrow {\rm EP}_{0}^{(k)} \longleftrightarrow \omega_Q^{(1)} \longleftrightarrow {\rm EP}_{1}^{(k)} \longleftrightarrow \omega_Q^{(2)}.
\]
where each value of \(k\) labels a distinct bridge connecting the same pair of neighboring Schwarzschild modes. Locally, the same structure is manifested through the universal square-root splitting, hyperbolic avoided crossings, the orthogonal reorientation of the spectral branches, and the exchange of the two Riemann sheets upon encircling an EP. The associated branch-point topology and mode permutation are characteristic features of exceptional points in non-Hermitian systems and have also been discussed in black-hole QNM spectra~\cite{Ding:2022juv,Cavalcante:2024swt,Cavalcante:2025abr}.

In Sec.~\ref{subsec:ExceptionalLines}, we showed the persistence of these features when the point defect is replaced by a finite-width Gaussian, with isolated EPs continuing into exceptional lines as the width is varied. This shows that the phenomenon is not specific to the distributional delta-function limit. Related exceptional-line structures have recently been identified in other black-hole QNM settings~\cite{Cao:2025afs,Nakamoto:2026lyo,Cavalcante:2026vgr}.

A complementary manifestation of the EP structure appears in the residues of the resonances \cite{Motohashi:2024fwt}. The excitation factors of the two coalescing modes diverge with opposite leading contributions, with
\(\lvert\mathfrak B_\pm\rvert\propto\lvert\delta\epsilon\rvert^{-1/2}\), or equivalently
\(\lvert\Delta\mathfrak B\rvert\propto\lvert\Delta\omega\rvert^{-1}\). Thus, the amplification of the residues is the inverse counterpart of the vanishing frequency separation. Away from the exact EP, the same square-root structure generates Bernoulli lemniscates in the complex excitation-factor plane (see Fig.~\ref{fig:BLemniscate}), providing the residue-space counterpart of the hyperbolic avoided crossings in the frequency plane. Related lemniscate structures in QNM excitation have also been identified in studies of avoided crossings and nearly coalescing resonances \cite{Motohashi:2024fwt,Nakamoto:2026lyo}.

The apparently singular behavior also contains a finite spectral quantity: \(\mathcal I=\Delta\mathfrak B\,\Delta\omega\) approaches a finite limit that is invariant under exchange of the two local sheets as the poles coalesce. Together with the Cassini-oval structure of the local spectral level sets, these results show that the frequency trajectories, spectral level sets and pole residues are complementary manifestations of the same underlying two-sheeted geometry associated with the exceptional point.

The time-domain analysis clarifies the consequences of this singular spectral structure. For the intrinsic response, the individual \(1/\delta\omega\) divergences of the two excitation factors cancel when the two pole contributions are combined, yielding a finite beating signal that approaches the characteristic double-pole form
\([{\cal A}_0+{\cal A}_1 t]e^{-\ii\omega_{\rm EP}t}\) in the coalescence limit. With prescribed Gaussian initial data, the corresponding excitation coefficients inherit the same EP singularity and lemniscate geometry, but are additionally weighted by a source-dependent factor. Importantly, large changes in the individual coefficients do not translate directly into a comparable amplification of the time-domain ringing. Moving the Gaussian from the horizon side of the defect to the exterior side increases \(\lvert C_\pm\rvert\) by roughly an order of magnitude, while the resulting two-pole response remains of the same order. The larger propagation delay partly compensates the enhancement of the raw coefficients, while the two coalescing pole contributions recombine coherently into a finite response. The configuration \(r_c=r_0=r_{\rm EP}\) provides an independent check: the two spatial contributions to the source factor are comparable and predominantly constructive, yet the resulting ringing again remains of the same order. The finite response near the EP is therefore neither tied to a particular source placement nor produced by a cancellation across the defect.

More broadly, these results illustrate that a dramatic reorganization of the resonance spectrum and a strong amplification of individual residues need not produce a comparably large enhancement of the QNM time-domain response. Related distinctions between spectral sensitivity, resonant excitation and time-domain observability have emerged in recent studies of perturbed and nearly degenerate black-hole QNMs \cite{Berti:2022xfj,Yang:2025dbn,PanossoMacedo:2025xnf,Kubota:2025hjk,Nakamoto:2026lyo,Imafuku:2026rpn}. Going back further, thirty years ago Nollert \cite{Nollert:1996rf} noted that ``quasinormal frequencies [...] seem to be extremely sensitive to very small changes in the underlying potential. The question arises whether -- and how -- it is possible to make any definite statements about the significance of quasinormal modes of black holes at all''. Moreover, the observation that the time-domain response remains robust under perturbation is also not a new one: after studying a range of environmental perturbations, Barausse \etal \cite{Barausse:2014tra} concluded that, in general, ``the [QNM] mode structure can drastically differ from the vacuum case, yet the BH response to external perturbations is unchanged at the time scales relevant for detectors''. Relatedly, there is a creative tension between the local, intuitive view of QNMs as `vibrations' originating near the light-ring \cite{Goebel:1972}, and the global, formal view of QNMs as particular solutions to a (non-Hermitian) boundary-value problem posed in the frequency domain.

To summarise the situation in the preceding paragraphs, let us offer a re-interpretation of the classic tale of \emph{The Princess and the Pea} (H.~C.~Andersen, 1835). In our version, the `pea' is a small potential perturbation, and the tower of mattresses is the black hole potential. The Princess cannot sleep because she somehow detects the presence of the pea, and this marks her out as a lady of higher sensibilities. In our retelling, the disturbed rest of the Princess is not due to any non-perturbative change in the response of the bed to external excitations; instead, it is because the Princess is unduly concerned about the implications of a re-configured QNM spectrum and the exceptional points of the slightly-perturbed system.

Several extensions follow naturally from the present work. A first step would be to examine gravitational perturbations of even parity and determine whether the same RP/EP sequences, asymptotic organization and fragility hierarchy arise in the Zerilli sector. Extending the construction to Kerr would be particularly important for black-hole spectroscopy, introducing the spin and azimuthal dependence of the rotating problem and allowing the localized-perturbation mechanism studied here to be compared with the avoided-crossing and exceptional-point structures of the Kerr spectrum. Another natural direction is provided by complex angular momentum. Quasinormal modes describe resonant poles in the complex-frequency plane, whereas Regge poles organize scattering at fixed real frequency through poles in the complex angular-momentum plane. It would be interesting to determine whether localized perturbations generate analogous coalescences, avoided crossings and multi-sheeted structures in Regge trajectories, and whether these can be related to the QNM connectivity identified here. A further step would be to replace the prescribed Gaussian initial data by a physically motivated dynamical source, such as a plunging particle. This would allow us to determine how the EP-associated resonances are excited by a realistic source and to assess whether the spectral reorganization found here leaves a discernible imprint on the full retarded waveform, rather than only on its isolated QNM pole contribution.

\begin{acknowledgments}
With thanks to Theo Torres for discussions, and for initial numerical evidence of the position of the repelling points at large $r_0$ for the Nariai spacetime with a P\"oschl-Teller potential; and to Rodrigo Panosso Macedo and Hayato Motohashi for email correspondence.
\end{acknowledgments}

\bibliography{refs}

\end{document}